\documentclass[a4paper,fleqn,usenatbib]{mnras}
\pdfoutput=1

\usepackage{newtxtext,newtxmath}
\usepackage[T1]{fontenc}
\usepackage{ae,aecompl}

\usepackage{graphicx}	% Including figure files
\usepackage{amsmath}	% Advanced maths commands
\usepackage{tablefootnote}

\newcommand{\fesc}{$f_{\rm esc}$}
\newcommand{\Tigm}{$T_{\rm IGM}$}
\newcommand{\Tm}{$\langle T_{\rm IGM} \rangle$}
\newcommand{\Lint}{($L_{900}/L_{1500}$)$_{int}$}
\newcommand{\Frat}{($F_{900}/F_{1500}$)$_{obs}$}
\usepackage{mathtools}

\title[The IGM Transmission of LyC Detections]{IGM
  Transmission Bias for $z$ $\geq$ 2.9 Lyman Continuum Detected Galaxies}

\author[R. Bassett et al.]{
R. Bassett$^{1,2}$\thanks{E-mail: rbassett@swin.edu.au (RB)},
E. V. Ryan-Weber$^{1,2}$,
J. Cooke$^{1,2}$,
U. Me\v{s}tri\'{c}$^{1,2}$, 
K. Kakiichi$^{3,4}$,\newauthor
L. Prichard$^{5}$,
M. Rafelski$^{5,6}$,
\\
$^{1}$Centre for Astrophysics and Supercomputing, Swinburne University
of Technology, PO Box 218, Hawthorn VIC 3122, Australia\\
$^{2}$ARC Centre of Excellence for All Sky Astrophysics in 3 Dimensions (ASTRO 3D), Australia\\
$^{3}$Department of Physics, University of California, Santa Barbara,
CA 93106, USA\\
$^{4}$Department of Physics and Astronomy, University College London, London, WC1E 6BT, UK\\
$^{5}$Space Telescope Science Institute, 3700 San Martin Drive,
Baltimore MD 21218, USA\\
$^{6}$Department of Physics \& Astronomy, John Hopkins University,
Baltimore, MD 21218, USA
}

\date{Accepted XXX. Received YYY; in original form ZZZ}

\pubyear{2021}

\begin{document}
\label{firstpage}
\pagerange{\pageref{firstpage}--\pageref{lastpage}}
\maketitle

\begin{abstract}
Understanding the relationship between the underlying escape fraction
of Lyman continuum (LyC) photons ({\fesc}) emitted by galaxies and
measuring the distribution of observed {\fesc} values at high redshift
is fundamental to the interpretation of the reionization process.
In this paper we perform a statistical
exploration of the attenuation of LyC photons by neutral hydrogen in
the intergalactic medium using ensembles of simulated
transmission functions. We show that LyC detected galaxies are more
likely to be found in sightlines 
with higher-than-average transmission of LyC photons. This means that
adopting a mean transmission at a given
redshift leads to an overestimate of the true {\fesc} for LyC
detected galaxies. We note, however, that mean values are appropriate
for {\fesc} estimates of larger parent samples that include LyC non-detected galaxies. We
quantify this IGM transmission bias for LyC detections in photometric
and spectroscopic surveys in the 
literature and show that the bias is stronger for both shallower
observations and for fainter parent samples (i.e. Lyman $\alpha$
emitters versus Lyman break galaxies). We also explore the effects of varying the
underlying probability distribution function (PDF) of {\fesc} on recovered values,
showing that the underlying {\fesc} PDF may depend on sample selection
by comparing with observational surveys. This work represents a first
step in improved interpretation of LyC detections in the context of
understanding {\fesc} from high redshift galaxies.
\end{abstract}


%%%%%%%%%%%%%%%%%%%%%%%%%%%%%%%%%%%%%%%%%%%%%%%%%%

%%%%%%%%%%%%%%%%% BODY OF PAPER %%%%%%%%%%%%%%%%%%

\section{Introduction} \label{sec:intro}

Understanding the details of cosmic reionization, the epoch at
$z\simeq6-10$ during which the hydrogen content of the
intergalactic medium (IGM)
transitioned from neutral to mostly ionized
\citep[e.g.][]{fan06,planck16,greig17,mason18},
is a major goal of the
international astronomical community. The general 
consensus currently favours a picture in which ionizing, or Lyman
continuum (LyC), photons originating from young, massive stars and/or
X-ray binaries and Wolf-Rayet stars in
star-forming galaxies are the primary driver. This picture is
supported by extensive theoretical
\citep[e.g.][]{wise09,yajima11,paardekooper15} and observational 
\citep[e.g.][]{inoue06,ouchi09,robertson15}
efforts. Active galactic nuclei (AGN), though pridigious producers of
LyC emission, are expected to play only a minor role due to their low
number density at $z > 6$ \citep[e.g.][]{hopkins07,parsa18,kakiichi18a}.

Detailed modelling of the reionization process critically requires an
accurate census of the fraction of LyC photons (with respect to
ultraviolet, UV, continuum photons) produced in galaxies that manage to
escape into the IGM, typically referred to as the LyC escape fraction
({\fesc}). The first major challenge in using {\fesc} to understand
reionization is the fact that no LyC photons from galaxies during the
Epoch of Reionization (EoR) will ever reach a telescope due to
absorption from intervening hydrogen. The second is the inherent
faintness of LyC emission from
galaxies \citep[as demonstrated by pioneering works
of][]{giallongo02,fernandez-soto03,inoue05},which is driven largely 
by two key factors. 

The first factor driving the faintness of LyC emission is that {\fesc}
is typically found to be very low 
(or zero) as inferred from the lack of LyC detections in 
\citep[e.g.][]{boutsia11,japelj17,bian20}. This may,
in part, be due to the fact that  
observations of galaxies, and thus, their LyC emission, at high redshift ($z\geq2.9$) are limited
to relatively high stellar mass ($M_{*} \geq 10^{9} M\odot$) galaxies that
are likely to contain significant quantities of neutral hydrogen
\citep[consistent with their high star-formation rates,
SFRs, e.g.][]{steidel01,iwata09,nestor11,grazian16}
that absorbs ionizing radiation before it can enter the IGM and drive
reionization. Indeed, for the small sample of such known
LyC emitting galaxies at $z \gtrsim 2.8$, the observed LyC flux is
relatively faint \citep[e.g.][]{shapley06,micheva17,vanzella18}. Even if
{\fesc} is larger in lower mass galaxies, such 
galaxies are inherently faint
and their LyC emission will likely be at least as difficult to detect as
their higher mass counterparts \citep[apart from the rare cases of
strong gravitational lensing, ][]{bian17,rivera-thorsen19}. The most
straightforward way past this problem is
to perform larger and deeper surveys targeting LyC emission across a
range of redshifts. A variety such surveys are currently
in progress.

The second issue resulting in faint LyC emission is that the IGM
itself contains large
fractions of neutral hydrogen above $z\simeq3$
\citep[e.g.][]{inoue14}. This means that after LyC
escapes from a galaxy it is largely absorbed in the IGM before reaching
Earth. For any individual LyC detection, there is currently no
reliable method for inferring the IGM transmission ({\Tigm}) of LyC
photons for that particular sightline. This is troubling
as observationally {\Tigm} and {\fesc} are degenerate meaning that,
in order to estimate {\fesc}, a value of {\Tigm} must be assumed
that may or may not be appropriate for a given IGM sightline. There is,
however, hope of a way forward as the
differential column density distribution of HI absorption
systems is well constrained
\citep[e.g.][]{meiksin06,becker13,rudie13}, providing a statistical
description of the probability that LyC photons escaping galaxies
will be absorbed by hydrogen in the IGM at a given redshift.

Such a statistical approach to estimate {\Tigm} in a theoretical
context has been explored using Monte Carlo (MC)
simulations for around three decades
\citep[e.g.][]{moller90,bershady99,inoue14}. Similarly, the application
of such MC simulations of {\Tigm} to detections (and non-detections)
of LyC radiation has a long history \citep[e.g.][S18
hereafter]{shapley06,siana07,steidel18} In
general, the most probable value of {\Tigm} at $z > 3$ is zero,
though individual sightlines with {\Tigm} $>$ 0.8 can exist (see
Section \ref{section:TMC}. The
typical probability distribution of {\Tigm} (around $\lambda_{rest} \sim
910$ {\AA}) at $z$ = 2.9-4.0 can be
described as bimodal with a sharp peak at $T_{\rm IGM} = 0.0$ and a
broader, less prominent peak at higher values. Both the location and
prominence of this secondary peak decrease with redshift until
$z\sim$5-6, at which point the presence of high
{\Tigm} sightlines is negligible. The result is that LyC is unlikely
to be observed from galaxies \textit{during} the EoR.

Using knowledge of the probability distribution of {\Tigm} at a
given redshift, astronomers can put forward an estimate of
{\fesc} for LyC detected galaxies. One method is to apply the full
suite of {\Tigm} models to a given observation (or set of
observations), however this typically results in largely unconstrained
{\fesc} values including a large number with the unphysical case of
{\fesc} $>$ 1.0 \citep{shapley16,vanzella16}. Another method is to
assume the mean value of {\Tigm}, $\langle T_{\rm IGM} \rangle$,
among all simulated sightlines thus providing a single {\fesc} value
\citep[S18,][hereafter F19 and
M20]{bassett19,fletcher19,mestric20}. The problem with this second
method 
is that a single statistic belies to complexities of the underlying
{\Tigm} distribution. Indeed, the mean of a bimodal distribution
will be found to lie between the two peaks, and will not fall
among the most likely values. This issue has been highlighted in the
context of Ly$\alpha$ transmission by \citet{byrohl20} who find that
assuming a median or mean transmission curve ``is misleading and
should be interpreted with caution''. 

There exist, however, important observational priors
that can provide more realistic constraints on the most likely value
of {\Tigm} for LyC detected galaxies. First and foremost, the fact
that a galaxy has been detected at LyC wavelengths means that
{\Tigm} for that galaxy \textit{cannot} be zero. This fact
automatically reduces the underlying bimodal {\Tigm} distribution
for all sightlines to a unimodal distribution for sightlines with LyC
detections. In this case, standard statistics such as the mean and
median of {\Tigm} may be more applicable. Secondly, while the
probability distribution function (PDF) of {\Tigm} is routinely
considered, the underlying PDF of {\fesc}
itself, which so far has been left out, may also be important. As we
have stated, low or zero 
{\fesc} values seem to be preferred, which is not reflected in current
{\fesc} calculations. It is possible that the broad behaviour of the
{\fesc} PDF may be inferred through consideration of the detection
rates in LyC surveys (this intriguing idea is explored further in
Section \ref{section:detrate}).  It is likely that a full understanding of the 
underlying {\fesc} PDF of galaxies will require a theoretical
underpinning through the careful analysis of high-resolution,
hydrodynamical simulations employing radiative transfer of ionizing
photons \citep[e.g.][]{trebitsch17,rosdahl18,ma20}. 

In this paper, we explore in detail the probability distributions of
both {\Tigm} and {\fesc} in the context of known LyC surveys at high
redshift. Our goal is to provide a statistically sound framework
within which astronomers can calculate meaningful estimates of {\fesc}
for both individual LyC detections as well as stacked samples. In
particular, we show that both the assumption of the mean {\Tigm}
value and (to a lesser extent) a lack of consideration of the underlying {\fesc} PDF result
in an overestimate of {\fesc} for LyC detected galaxies. Here we
quantify the IGM transmission bias, $T_{\rm bias}$, as $\langle T_{\rm
  det} \rangle - \langle T_{\rm IGM} \rangle$ where $\langle T_{\rm
  det} \rangle$ is the average IGM transmission for LyC detected galaxies
for a given observational detection limit. We note that, although
a transmission value is not inherently an additive quantity, our
definition leads to a roughly redshift independent correction to
$\langle T_{\rm IGM} \rangle$ as opposed to an alternative definition
such as $T_{\rm bias} =  \langle T_{\rm det} \rangle / \langle T_{\rm
  IGM} \rangle $ (see Section \ref{section:results} for further
discussion). 

This paper is
organised as follows: in Section \ref{section:method} we describe our
method of generating simulated IGM sightlines and spectra of mock LyC
emitting galaxies, in Section \ref{section:results} we describe the
results of our various models, in Section \ref{section:discussion} we
explore the implications of our results in the context of past and
on-going LyC surveys, and in Section \ref{section:conclusions} we
provide a brief summary of our findings.

\begin{figure*}
  \includegraphics[width=\textwidth]{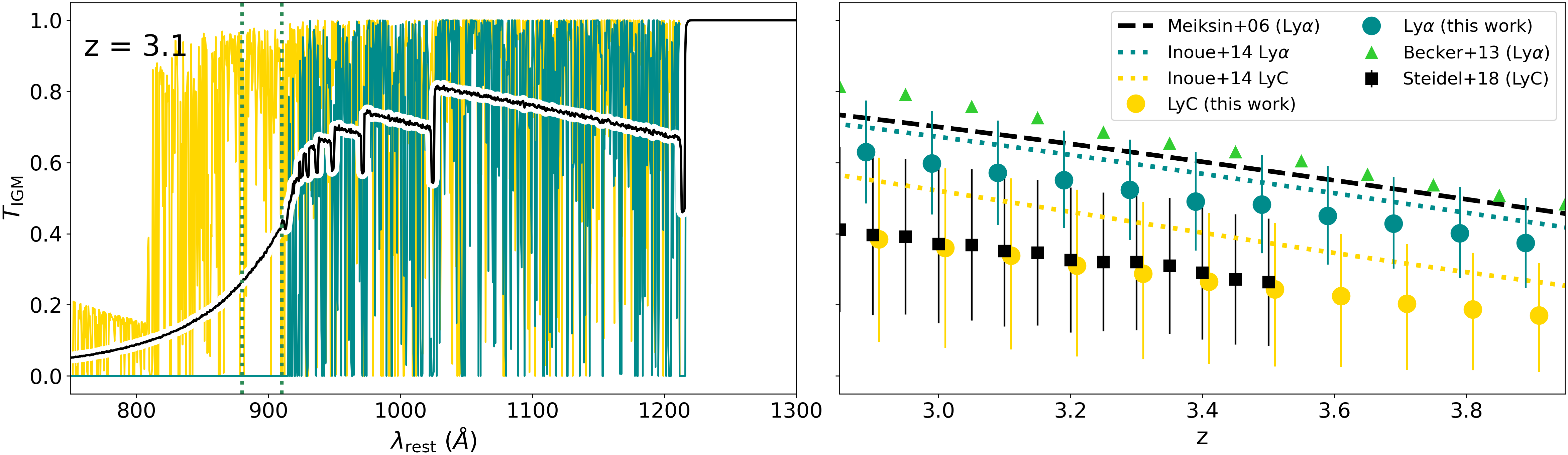}
  \caption{\textit{Left:} Example IGM transmission function for a galaxy
  at $z=3.1$. In gold and cyan are single transmission functions
  with highest and lowest 
  {\Tigm} at 880 $<$ $\lambda$ $<$ 910 {\AA}
  (range indicated by vertical, cyan, dotted lines) among our ensemble of 10,000 transmission functions at $z=3.1$. The black curve shows the average transmission for the
  entire ensemble. \textit{Right:} The mean transmission of Ly$\alpha$
  (1210 $<$ $\lambda$ $<$ 1215 {\AA}, cyan)
and LyC emission (880 $<$ $\lambda$ $<$ 910 {\AA}, gold) as a function of redshift for our simulated
IGM transmission functions. Error bars indicate the range containing
68.1\% of all values about the median in
each bin. We note that mean and median values
differ given the complex, bimodal underlying distribution. Here we also
compare to theoretical and observational work
in the literature from \citet{becker13}, \citet{meiksin06}, 
\citet{inoue14}, and S18.}
  \label{fig:TMCex}
\end{figure*}

\begin{figure}
  \includegraphics[width=\columnwidth]{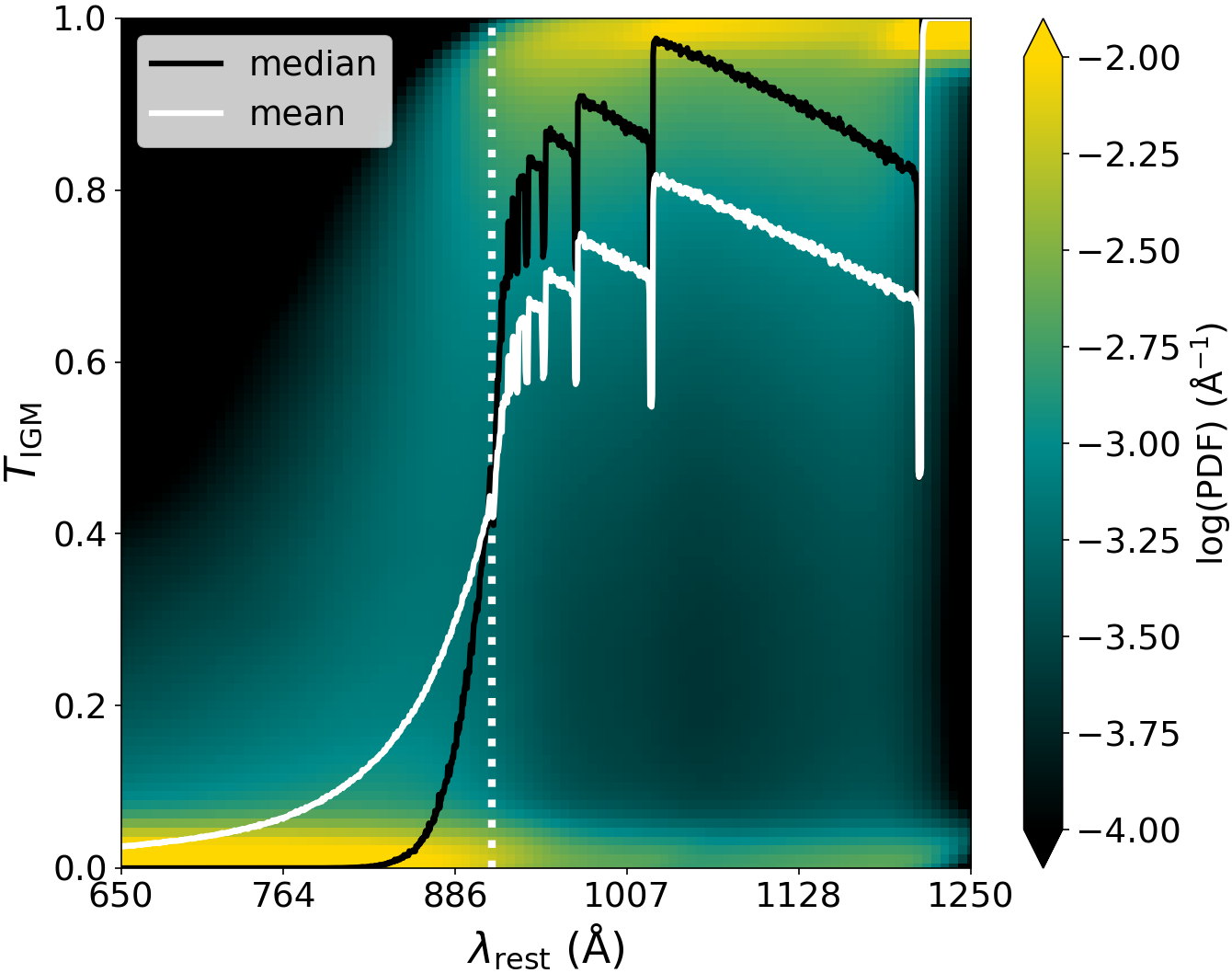}
  \caption{A full statistical description of our 10,000 IGM
    transmission functions at $z=3.1$. The shading represents the
    probability of a given {\Tigm} value at each wavelength with
    probability increasing from black to gold (note the colour scaling is
    logarithmic). Blueward of the Lyman
    limit (911.8 {\AA}, white dotted line) {\Tigm} is strongly peaked at {\Tigm} = 0.
    The behaviour at fixed $\lambda$ shifts from unimodal at the
    shortest wavelengths to bimodal redward of $\sim$880 {\AA}. For illustration
  we show the median and mean {\Tigm} functions in black and
  cyan.}
  \label{fig:2dpdf}
\end{figure}

\section{Simulating LyC Leaking Galaxies} \label{section:method}

In this Section we describe our method of producing mock observations
of LyC flux from high redshift galaxies. There are three primary
ingredients in creating an individual high 
redshift galaxy observation for our simulation: an IGM transmission 
function, {\fesc}, and the input SED model. Our method for
producing an IGM transmission function is described in Section
\ref{section:TMC}. Although the
underlying PDF of {\fesc} for galaxies is
largely unknown, we test two models described in Section
\ref{section:fesc_PDF}. Finally, we take our input SED model from
BPASSv2.1 \citep[described further in Section
\ref{section:mock_spectra}]{eldridge17}, matching the assumed LyC to
non-ionizing UV flux ratio from previous studies. In particular we
compare with results from the Keck Lyman Continuum Survey
(KLCS, S18),
the LymAn Continuum Escape Survey (LACES, F19), and
the ground based photometric work of M20 based on deep 
$u$-band photometry from the Canada France Hawaii Telescope (CFHT)
Large Area U-band Deep Survey \citep[CLAUDS,][]{sawicki19}. 

\subsection{IGM Transmission Functions}\label{section:TMC}

{\Tigm} functions are produced following the method outlined
in S18, Appendix B \footnote{All code for producing IGM transmission curves is open
source and available at https://github.com/robbassett/TAOIST\_M
C.}. We perform a Poisson sampling
of the number of HI absorbers in redshift intervals, $\Delta z$, from
$z=0$ to a specified redshift, $z_{em}$. Following \citet{inoue14} we
select a value of $\Delta z = 5 \times 10^{-5}$, noting however that
deviations from this value would not affect our results. The value of
$z_{em}$ for a given analysis is determined by the redshift of the
galaxy, or sample of galaxies, being considered. In this work we
create suites of 10,000 IGM transmission functions at 10 discrete $z_{em}$
values in the range 2.9-3.9 with $\Delta z_{em} = 0.1$ (we also
explore IGM transmission bias at $z=2.4$ and $z=4.4$ for HST F275W and
F435W observations, respectively, in Section \ref{section:lowz}). 

To generate a single {\Tigm} function at a given $z_{em}$ we
must first produce a random sampling of hydrogen absorption systems in
redshift bins of $\Delta z = 5 \times 10^{-5}$ from $z=0$ to $z_{em}$. This is
achieved assuming a differential HI column density distribution,
$f(N_{\rm HI},X)$, following the prescriptions outlined for the ``IGM+CGM''
model in S18 Appendix B. In each redshift interval we derive the
expected number of absorption systems in each bin of $log(N_{\rm HI})$
(sampled from $log(N_{HI}) = 12.0-21.0$ with $\Delta log(N_{\rm HI}) =
0.1$) as:
\begin{equation}\label{eq:Nabs}
  N_{\rm abs} = \int_{N_{\rm HI,min}}^{N_{\rm HI,max}}\int_{z}^{z+\Delta
    z} N_{\rm HI}^{-\beta}A(1+z)^{\gamma}dN_{\rm HI}dz
\end{equation}
Where $N_{\rm HI,min}$ and $N_{\rm HI,max}$ are the lower and upper
bounds, $\beta$ is the slope of $f(N_{\rm HI},X)$, $A$ is a constant
chosen to match observed $N_{\rm abs}$, and $\gamma$ describes the
redshift evolution of $N_{\rm abs}$. Values for $\beta$, $A$, and $\gamma$ are taken
directly from Table B1 of S18. We assume the presence of absorption
systems is a Poissonian process, thus for each sightline the number of
absorption systems at a given $z$ and $N_{\rm HI}$ is calculated using
{\sc numpy.random.poisson} with $\lambda$ set to
$N_{\rm abs}$.

For each individual absorber in a given observed sightline, we then
apply the transmission
function for LyC photons at $\lambda_{\rm rest} \leq 911.8$ {\AA} and a
transmission for Lyman series forest for photons with $\lambda_{\rm rest}
\geq 911.8$ {\AA}, noting that in this case we are considering the rest
wavelength at the redshift of the absorption system and \textit{not}
the LyC emitting galaxy. For LyC photons we apply the functional form:
\begin{equation}\label{eq:tau_LyC}
  \tau_{\rm HI}^{LyC}(\nu_{\rm rest}) = N_{\rm HI} \sigma_{\rm HI} (\nu_{\rm rest})
\end{equation}
where $\nu_{\rm rest}$ is the photon frequency at the rest frame of a
given absorbtion system and $\sigma_{\rm HI}(\nu_{\rm rest})$ is the
frequency dependent interaction cross section of HI to ionising 
photons given by $\sigma_{L} (\nu_{\rm rest} / \nu_{911.8 \rm{\AA}})^{-3}$. Here $\sigma_{L}$
is a constant with a value of $6.3 \times 10^{-18}$
cm$^{2}$ \citep{osterbrock89}. For Lyman series lines we use the
following for each Lyman transition, $i$ \citep[e.g.][]{inoue08}:
\begin{equation}\label{eq:tau_Lyseries}
  \tau_{i}(\nu_{\rm rest}) = N_{\rm HI} \frac{\sqrt{\pi}e^{2}f_{i}}{m_{e}c \nu_{D}}\phi_{i}(\nu_{\rm rest})
\end{equation}
where $m_{e}$ and $e$ are the electron mass and charge, respectively,
and $c$ is the speed of light. The parameter $f_{i}$ is the oscillator
strength of Lyman transition $i$, which we take from tables provided
with the {\sc VPFIT} package \citep{carswell14}. In our calculation we
include the first 32 Lyman series transitions. $\nu_{D} = \nu_{i}(b/c)$ is the
Doppler broadening of the Lyman line at frequency $\nu_{i}$ where $b$,
the Doppler parameter, is randomly sampled from \citep{hui99}:
\begin{equation}
  h(b) = \frac{4b_{\sigma}^{4}}{b^{5}}e^{-b_{\sigma}^{4}/b^{4}}
\end{equation}
with $b_{\sigma} = 23$ km s$^{-1}$ \citep[e.g.][]{janknecht06}. Finally,
$\phi_{i}(\nu)$, the absorption profile, is taken as the analytic
approximation of the Voigt profile given by
\citet{tepper-garcia06}. Here, as with $f_{i}$, we also sample
$\Gamma_{i}$, the damping constant for transition $i$, from the {\sc
  VPFIT} values. The total optical depth of an individual absorber is
then taken as $\tau(\nu) = \tau_{\rm HI}^{LyC}(\nu_{\rm rest}) + \Sigma
\tau_{i}(\nu_{\rm rest})$, where $\nu$ refers to the observed frame
frequency, $\nu = \nu_{\rm rest} / (1+z)$. The total $\tau(\nu)$ for a given sightline is
the sum of the ensemble of $\tau(\nu)$ for all absorbers in that
sightline.

It is worth mentioning that our transmission curves are produced
as a function of wavelength, rather than frequency, and we employ a
fixed resolution of $\Delta \lambda = 2.2$ {\AA} per pixel in the
observed frame. This choice is motivated by the fact that we compare
extensively with LRIS spectroscopy of S18, who quote a spectral
resolution of 2.18 {\AA} per pixel for their
observations. \citet{inoue08} note that spectral resolution can have
a significant impact on the resultant IGM transmission, we
have tested the effect of increasing the spectral resolution to 0.4
{\AA} per pixel, finding no statistically significant difference
compared to our standard 2.2 {\AA} per pixel transmission curves. 

Throughout this paper we consider values in terms of IGM transmission,
{\Tigm} = $e^{-\tau}$, rather than considering $\tau$ computed as
described above. The reasons being first that the value of {\Tigm} is
typically included in calculations of {\fesc} and second that
{\Tigm} has a dynamic range between 0 and 1, which provides more
intuitive comparisons. We note that throughout this paper the
symbol {\Tigm} may refer to a wavelength dependent transmission
function or a single value at some specified wavelength. We avoid
introducing an explicitly wavelength dependent symbol,
i.e. $T_{\rm IGM}(\lambda)$, as the usage here is consistent with the
conventions in the literature \citep[e.g.][]{inoue08}.

Example IGM transmission functions at $z=3.1$ are shown in Figure
\ref{fig:TMCex}. In the left panel in black we show the mean transmission curve of all
10,000 simulated sight lines at $z=3.1$ while gold and cyan curves show
two individual sight lines having the highest and lowest
$\lambda_{\rm rest} = 910$ {\AA} transmission, respectively. 
At a given redshift the transmission of LyC in the IGM may vary from 0.0
to nearly 1.0. In the right panel we show the redshift evolution of
the mean Ly$\alpha$ and LyC transmission predicted by {\sc TAOIST-MC}
in comparison with observational and theoretical estimates from the
literature. In all cases, our model agrees, within errors, with previously
reported results. 

We note that our measurements are systematically
lower than some previous results, which can be attributed to the
inclusion of the circumgalactic medium component introduced in
S18. Furthermore, a single statistic (such as the mean) belies the
complexity of the underlying {\Tigm} distribution as shown in Figure
\ref{fig:2dpdf}. Thus, we do not
place a large emphasis on differences between the average values
of {\Tigm} between different studies. For theoretical {\Tigm}
functions this behaviour may be, in part, attributed to the exact form of the 
differential $N_{\rm HI}$ distribution assumed and the details of the
implementation. For example, \citet{inoue08} and S18 assume different
behaviours for the exponent $\beta$ of $f(N_{\rm HI},X)$ producing
different relative numbers of low and high $N_{\rm HI}$ systems. These
differences will affect the {\Tigm} of LyC and Ly$\alpha$
differently and will appear as complex systematic offsets between the mean
{\Tigm} at a given redshift between the two implementations. It
should also be mentioned that, to our knowledge, no study employing MC
simulations of IGM transmission curves have accounted for the effects
of HI clustering, which may further alter the mean
{\Tigm} curve \citep[see, however,][who demonstrate Ly$\alpha$ may
be more attenuated from galaxies in high density
environments]{kakiichi18b}. 

Differences in the behaviour between theoretical {\Tigm}
implementations are only apparent from the mean
transmission curves while individual IGM transmission curves
are likely indistinguishable. The implications regarding the
statistical behaviour of IGM sightline ensembles, however, is
precisely the topic of this paper. As we will repeat, the absolute
values of quantities calculated throughout
will be imprinted with the assumptions regarding our $N_{\rm HI}$
distribution sampling and may change slightly if different
implementations are used. Thus, it is key to keep in mind that the
absolute results are for our implementation only. Qualitatively,
however our results are independent of the various input parameters. 

\begin{figure}
  \includegraphics[width=\columnwidth]{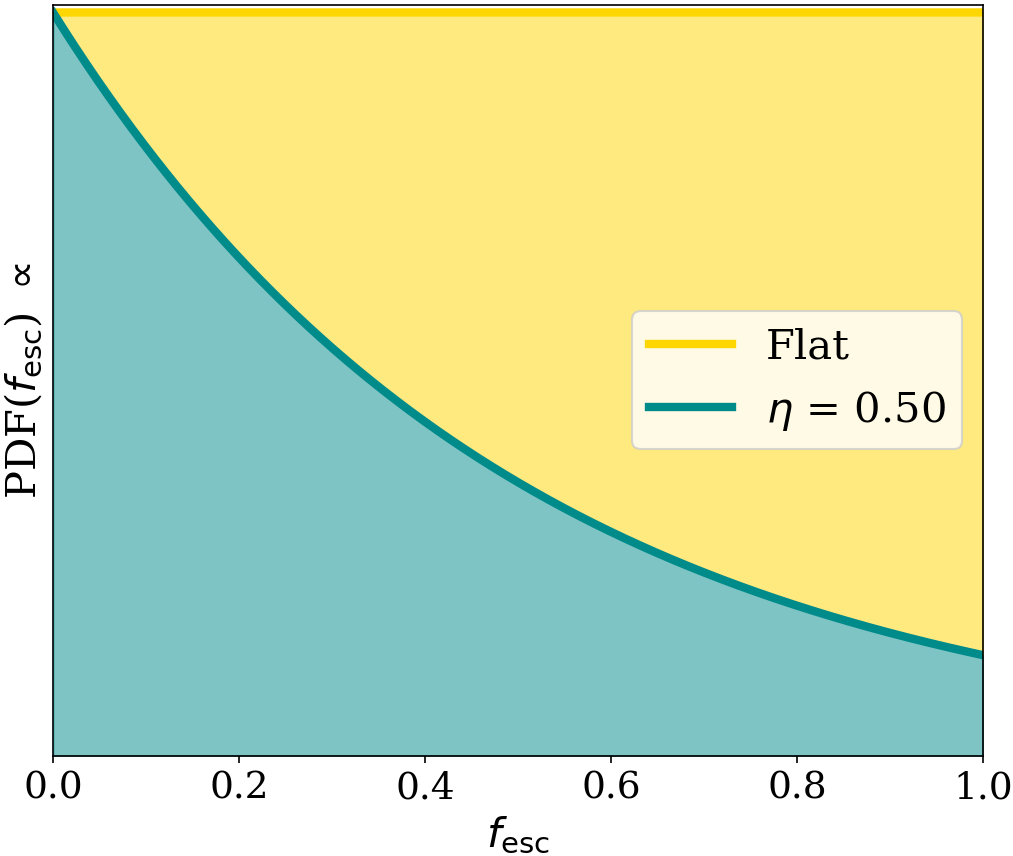}
  \caption{A comparison of the two {\fesc} PDFs used in this
    work. The ``Flat'' distribution represents the case of no assumed
    prior when calculating {\fesc} and is representative of most
    studies in the literature. The alternative explored here is an
    exponentially declining models of the form PDF $\propto$
    $e^{-1/\eta}$, here shown with $\eta$ = 0.50.}
  \label{fig:fesc_pdfs}
\end{figure}

\subsection{$f_{\rm esc}$ Distribution Functions}\label{section:fesc_PDF}

One of the key unknowns in this study is the distribution function of
$f_{\rm esc}$ for galaxies at $z \geq 2.9$. While quantifying
$f_{\rm esc}$ from galaxies has been a long standing goal in the astrophysics
of reionization, this parameter remains elusive. In a broad sense, a
number of studies have estimated the average $f_{\rm esc}$ required
for all galaxies in order to match the constraints on the timing of
reionization, finding values in the range 0.05 $<$ $\langle
f_{\rm esc} \rangle$
$<$ 0.20 \citep[e.g.][]{bouwens15,robertson15,finkelstein19}. From
hydrodynamical
simulations of individual galaxies employing full radiative transfer,
however, the likelihood that all galaxies will have a constant and/or
single valued $f_{\rm esc}$ over their lifetime seems vanishingly
small \citep[][see also Section \ref{section:3dv2d} for a brief discussion of
the 3D versus line-of-sight {\fesc} values]{kimm14,paardekooper15}. 

Given the lack of strong constraints on {\fesc} from the literature, for
the fiducial model of our analysis, presented in Section
\ref{section:IGMbias}, we simply uniformly apply values of {\fesc}
between 0.0 and 1.0 to 
our mock spectra. This allows for mock spectra with the highest
possible LyC flux for a given IGM sightline, representing the most
likely galaxies to be detected in a LyC survey. As such, the results
of our fiducial model should be interpreted as the \textit{minimum}
level of {\Tigm} bias expected for LyC detected galaxies.

It seems most likely that allowing extremely high {\fesc} is unrealistic
for the vast majority of
real galaxies \citep[e.g.][]{vanzella10,siana15,japelj17}. In
Section \ref{section:fescdist} we 
test the effects on our measured {\Tigm} bias of applying an additional, more
realistic {\fesc} distribution, to our simulations. For this test, we assume an
exponentially declining {\fesc} PDF, i.e. $P(f_{\rm esc}) \propto
e^{-1/\eta}$, resulting in a model with the most probable value of
{\fesc} being zero. For our exponentially declining {\fesc} PDF we
choose a value of $\eta = 0.5$, which is motivated by the observed detection
rates of KLCS (S18, see Section \ref{section:detrate}). We illustrate
the relative PDF shapes of our fiducial and exponentially declining models in Figure
\ref{fig:fesc_pdfs} for clarity.

\subsection{Producing Mock Galaxy Spectra}\label{section:mock_spectra}

\begin{figure}
  \includegraphics[width=\columnwidth]{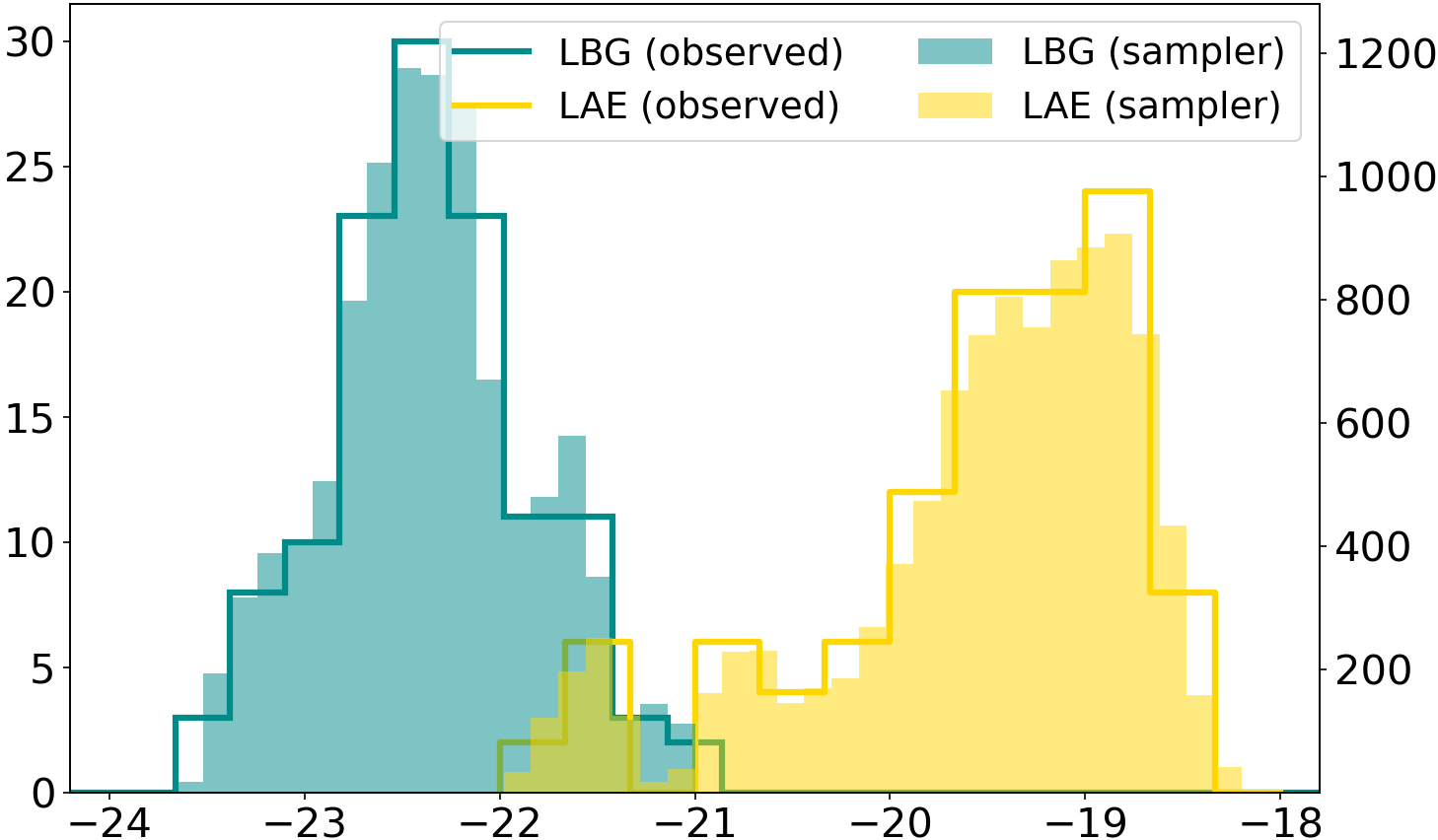}
  \caption{Input 1500 {\AA} absolute magnitude distributions for LBG
    (cyan) and LAE (gold) samples. Open histograms represent LRIS 1500
  {\AA} fluxes from S18 and UV magnitudes from ground-based imaging
  reported in F19 for LBGs and LAEs, respectively. Filled histograms
  represent one realisation of 10,000 sampled values for our mock
  galaxy spectra produced using {\sc cdf\_sampler.py} (see footnote 2)
  with the open histograms as inputs. Values on the left y-axis refer
  to observed samples (open histograms) and on the right y-axis
  refer to mock samples (closed histograms), noting in the latter case
  these values are based on an arbitrarily selected ``parent sample'' size.}
  \label{fig:UVfluxes}
\end{figure}

\begin{figure*}
  \includegraphics[width=\textwidth]{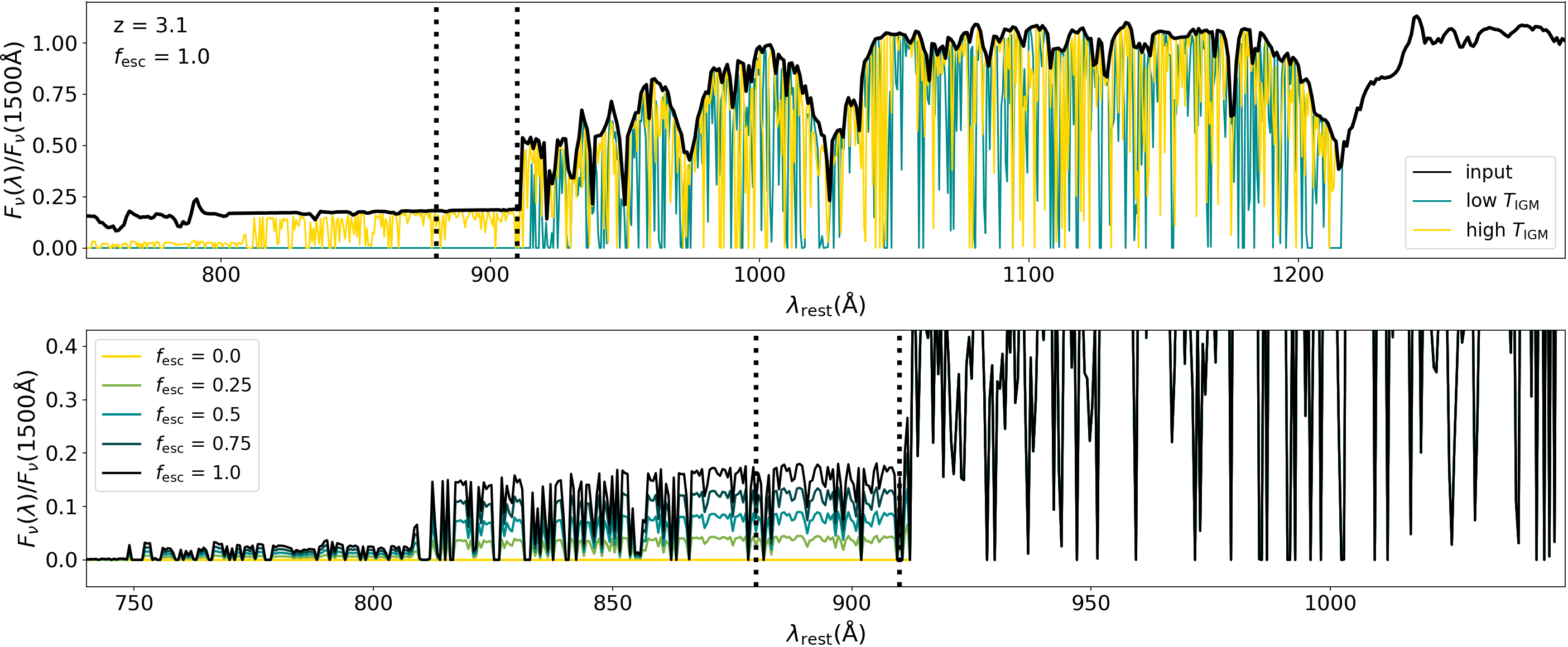}
  \caption{\textit{Top:} Example BPASSv2.1 spectra used in this study. In black is
    shown the input spectrum and in gold and cyan we show the output
    spectra with the high and low IGM transmission curves shown in Figure
    \ref{fig:TMCex}. In this panel, both spectra are shown for the
    {\fesc} = 1.0 case. \textit{Bottom:} The effect of our flat
    treatment of {\fesc} on the output spectra for the high IGM
    transmission spectrum shown in the top panel with {\fesc} varying
    from 0.0 to 1.0. For all spectra in both panels, we show the flux
    in $F_{\nu}$ normalised to the flux at a rest wavelength of 1500 {\AA}.}
  \label{fig:specex}
\end{figure*}

As mentioned above, the process of producing mock galaxy spectra for
our simulations requires three inputs: an underlying SED model, an IGM
attenuation function, and a value
for $f_{\rm esc}(LyC)$. We note that in much of this work we ignore the effects of dust
attenuation (see, however, Section \ref{section:dust}, simply noting
that most LyC detections appear to
originate from relatively dust free galaxies (e.g. S18). Similar to
S18 we construct our SEDs from the BPASSv2.1 \citep{eldridge17} models
with $Z_{*} = 0.001$, IMF slope $\alpha = -2.35$, and stellar mass
limit of 300 $M_{\odot}$. We employ a model with an exponentially
declining SFR with an $e$-folding time of 0.1 Gyr sampled at an age of
$\sim$200 Myr. This provides an input spectrum with an intrinsic LyC
to UV flux ratio, {\Lint}, of 0.18 (e.g. S18). Our SED model corresponds
to a LyC photon production efficiency, $\xi_{ion}$, of
log$_{10}(\xi_{ion})$ = 25.61 Hz erg$^{-1}$, consistent with estimates for high
redshift star-forming galaxies \citep[e.g.][]{bouwens16}. We explore
the effect of altering {\Lint} on our results in Section \ref{section:sedvar}.

Each mock spectrum is scaled such that the non-ionizing UV flux
matches a randomly sampled value characteristic of high redshift,
highly star-forming galaxies. The sampling of UV fluxes is one key
factor in our analysis as this ultimately determines the intrinsic
level of LyC flux from galaxies in our mock samples. In this work we
test samples taken two different UV flux distributions: one based on
the full sample of galaxies observed by the KLCS, which is composed of
a representative subsample of bright Lyman Break Galaxies (LBGs) at 2.9 $<$
$z$ $<$ 3.2 from the flux-limited sample of \citet{reddy12}, and a
second based on $z$ $\sim$ 3.1, narrow-band selected Lyman $\alpha$
emitters (LAEs) characteristic of galaxies targeted by LACES
(F19). For our LBG comparison
UV values used in our work are sampled from measurements of LRIS spectra at 
$\lambda_{\rm rest} = 1500${\AA} taken directly from reported values
of S18. For the comparison with LAEs, UV values are sampled based on
the histograms presented in F19, Figure 15, based on ground based photometry. We compare the
absolute magnitude distributions of the two distributions in Figure
\ref{fig:UVfluxes}, showing LAEs to be significantly fainter than
LBGs\footnote{In both cases sampling of UV fluxes is achieved using the
  {\sc histogram\_oversampler} class of the code {\sc cdf\_sampler.py}
  (https://github.com/robbassett/cdf\_sampler) with spline fitting
  enabled to remove sharp edges of the histogram bins}. We note,
however, that some LBGs have been shown to also exhibit Ly$\alpha$
emission \citep[e.g.][]{shapley03}, thus LBG and LAE classifications
are based on selection methodology. Here, the important distinction is
the relative non-ionizing UV flux with LBGs being significantly brighter.

The sample of S18 covers a redshift range of $z$ $\simeq$
2.8-3.5 and the sample of F19 is at a roughly fixed redshift of
3.1. The mock galaxies in our analysis, however, are produced at 10 discrete redshift values 
with $\Delta z = 0.1$ from $z=2.9$ to $z=3.9$. Thus, we must include a
method to account for cosmological dimming of each of these samples
when considering higher redshifts. In each case, we begin with the
absolute magnitude distributions shown in Figure \ref{fig:UVfluxes}
and assume that this distribution is roughly representative of a
similarly selected sample in each $\Delta z = 0.1$ redshift bin. We
then sample magnitudes from the above distributions then convert each
value to an observed 1500 {\AA} flux at a given redshift. We note that
this is equivalent to a slight increase in depth with redshift,
however we expect this to have a negligible effect on our results as
we are most sensitive to the brightest galaxies at any redshift.

For each of the 10,000 IGM sightlines in a given redshift bin we
produce 100 mock spectra for both the LBG and LAE comparison
samples. For each trial we randomly 
sample a 1500 {\AA} flux (as described above) and a value of {\fesc}, the latter following
Section \ref{section:fesc_PDF}. The current {\Tigm} function is
applied to the input BPASSv2.1 spectrum, then at all wavelengths
shortward of 911.8 {\AA} it is scaled uniformly by the randomly selected {\fesc}
value. The resulting spectrum is then scaled to match the randomly
selected 1500 {\AA} flux. Thus, in each redshift bin we produce one
million galaxy spectra ensuring that the 1500 {\AA} flux and {\fesc}
distributions are well sampled. Example spectra can be seen in Figure
\ref{fig:specex}. 

We can summarise the construction of each individual mock spectrum
with the following equation:
\begin{equation}
  F_{\nu}^{i,j}(\lambda,z) =
  F_{\nu,mod}(\lambda)\frac{F_{1500,obs}^{i}}{F_{1500,mod}}T_{\rm
    IGM}^{j}(\lambda)f_{\rm esc}^{i}(\lambda)
\end{equation}
where $F_{\nu,mod}(\lambda,z)$ is the input BPASS spectrum,
$F_{1500,obs}^{i}$ is the $i$th randomly sampled 1500 {\AA} flux
(noting again that here we have included cosmological dimming),
$F_{1500,mod}$ is the 1500 {\AA} flux of the BPASS model (taken as the mean
value at 1450 $<$ $\lambda$ $<$ 1550 {\AA}), $T_{\rm IGM}^{j}$ is the
current IGM transmission curve (we use the superscript $j$ to indicate
that the same IGM transmission curve will be used 100 times, thus it
is not unique to mock spectrum $i$), and $f_{\rm esc}^{i}(\lambda)$ is
a step function representation of the $i$th randomly sampled {\fesc}
value given as:
\begin{equation}\label{eq:mockspec}
  f_{\rm esc}^{i}(\lambda) =\begin{cases}
      f_{\rm esc}^{i} \quad & \text{if} \, \lambda < 911.8 \\
      1 \quad & \text{if} \, \lambda \geq 911.8
    \end{cases}
\end{equation}

\section{Results}\label{section:results}

The primary results of this paper concern quantifying the observational
bias in IGM transmission for samples of LyC detected galaxies. We reiterate that the
initial results are based on tests performed on a fiducial dust-free,
exponentially declining
SFR SED models at fixed metallicity, IMF slope, and age (see Section
\ref{section:mock_spectra} for a full description). We have
selected our fiducial model to have {\Lint}$\sim$0.18
(comparable to other studies in the literature, e.g. S18, F19), which is
expected to be representative of young, star-forming galaxies
responsible for driving reionization. 

Additionally, as described in Section \ref{section:fesc_PDF}, our fiducial
model assumes the unrealistic 
case of a flat probability distribution for {\fesc} between 0 and
1.0. High {\fesc} will correspond to a bright LyC flux, thus, we
expect a preference towards detections at high {\fesc} in our fiducial
model. {\fesc} for
real galaxies will be, on average, lower than the average of our
fiducial model given the typically low value for observed LyC emitters
(e.g. S18). This means that the level of bias in {\Tigm} for detections
seen in our fiducial model can be seen as a lower limit to the true
bias for observed galaxy samples. 

We explore the quantitative effects of both altering the input PDF of
{\fesc} and changing the value of {\Lint} in Sections
\ref{section:fescdist} and \ref{section:sedvar}, respectively. In the
case of SED variations we test SEDs with $\xi_{ion}$ values covering the range
for exponentially declining SFR models using BPASSv2.1 spectra over
available range of stellar population ages provided.

\subsection{Fiducial IGM Bias}\label{section:IGMbias}

Here we quantify the bias in {\Tigm} affecting samples of LyC
detected galaxies when compared with the average {\Tigm} of all random
sightlines. Formally, we define this bias as: 
\begin{equation}\label{eq:biasdef}
  T_{\rm bias} = \langle T_{\rm det} \rangle - \langle T_{\rm IGM} \rangle
\end{equation}
where $\langle T_{\rm det} \rangle$ is the average {\Tigm} for galaxies
with LyC detected above a specified detection limit and $\langle
T_{\rm IGM}
\rangle$ is the average {\Tigm} for all sightlines. It is worth noting
that transmission values are not inherently additive quantities and it
could be argued that the definition $T_{\rm bias} = \langle T_{\rm
  det} \rangle / \langle T_{\rm IGM} \rangle $ is more sensible, and
possibly more physically motivated as it relates directly to a
difference in optical depth/HI column density. Our choice of
definition is motivated by the fact that the resulting $T_{\rm bias}$
values are roughly redshift independent at fixed observational
detection limit (see, e.g., Section \ref{section:3p1summ}),
providing a simplified framework for applying $T_{\rm bias}$ to a
given set of observations. We also point out that, by definition, such
a correction will never result in an unphysical transmission value for
LyC detections $>$ 1.0. Furthermore, any evolution in $T_{\rm bias}$
with redshift when assuming a fractional definition is primarily reflective
of the redshift evolution of $\langle T_{\rm IGM} \rangle$ as one is
dividing by a value increasingly close to zero. Regardless, either $T_{\rm bias}$ definition
mentioned here will provide an equivalent correction, thus the choice
is somewhat arbitrary. 

In this work, the
calculation of $T_{\rm bias}$ is performed at 11 discrete
redshifts in the range $2.9 \leq z \leq 3.9$ with $\Delta z = 0.1$. We also note that,
similar to {\Tigm} and {\Tm}, $T_{\rm bias}$ can refer to a
wavelength dependent function, a single value at a specified
wavelength, or an average value across a specified wavelength range. 
Due to technical differences between LyC
searches employing spectroscopy (e.g. S18) and photometry
(F19, M20), we present the two cases separately: spectroscopic
biases are presented in Section \ref{section:specdet} and photometric
biases are presented in Section \ref{section:photdet}. In all cases,
we have performed this experiment twice: once for an LBG-like sample
and once for a fainter, LAE-like sample (see Figure
\ref{fig:UVfluxes}). Due to the inherent faintness, our mock LAE samples
are typically only detected deep HST
F336W observations, which compare to F19 (particularly in our higher
redshift bins) who achieve a depth of 30.24 mag. Thus, in most cases we
only provide $T_{\rm bias}$ measurements for this comparison (as
opposed to spectroscopy or CFHT $u$ photometry).  We provide a
summary of our fiducial model in Section \ref{section:3p1summ}. 

\subsubsection{Spectroscopic Detections}\label{section:specdet}

\begin{figure}
  \includegraphics[width=\columnwidth]{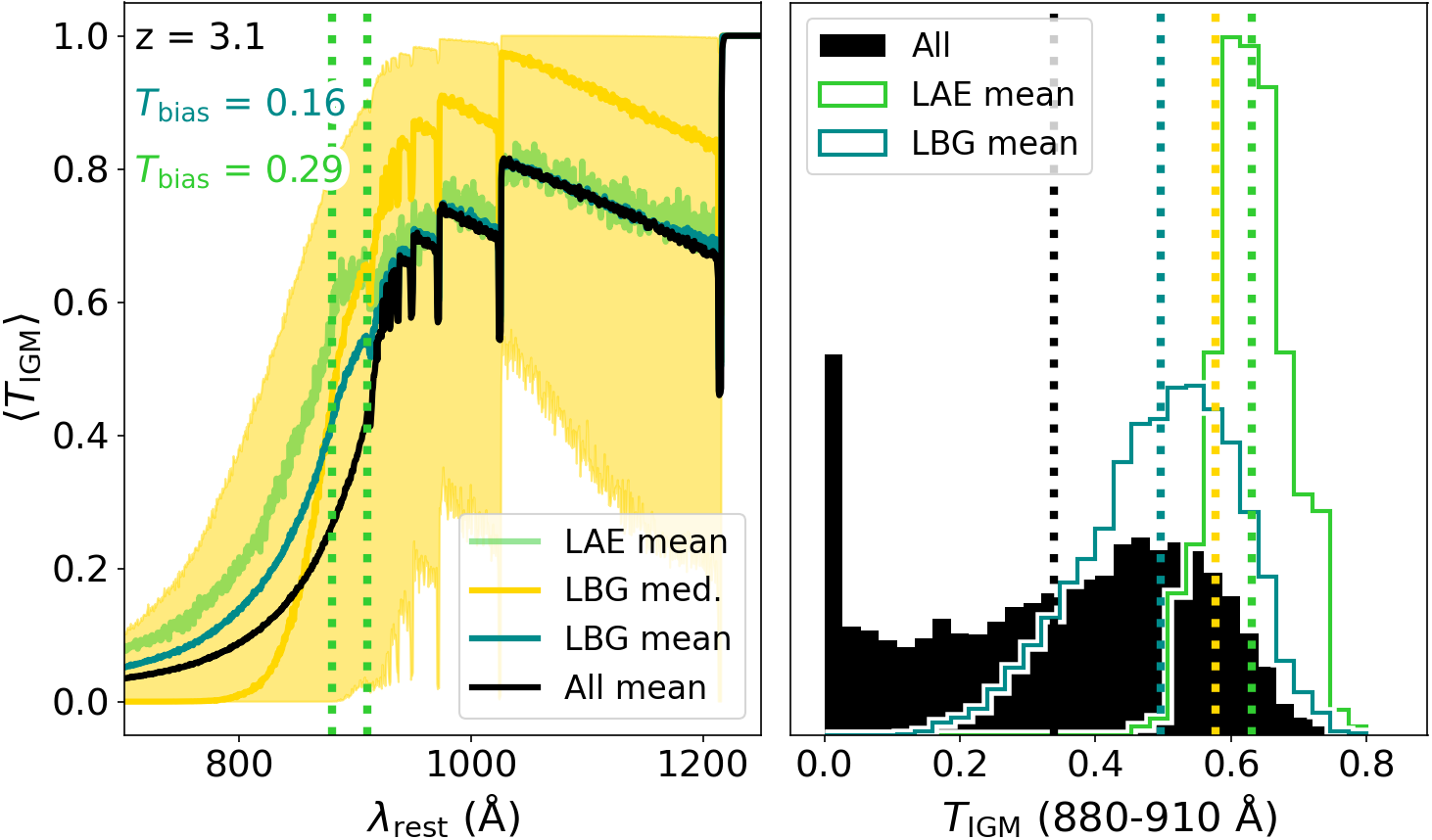}
  \caption{\textit{Left column:} {\Tm} for spectroscopically
    detected LyC emission. Shown are
    results at $z = 3.1$, similar to the average redshift of S18 of
    $\sim$3.05. The Lyman limit is indicated
    with a vertical dotted line, and in all cases the detection limit is fixed at
    0.025 $\mu$Jy ($\sim$27.9 mag), equivalent to a 1-5$\sigma$
    detection, dependent on individual targets, in the sample of
    S18. The mean and median {\Tm} functions for  
    detected LBG-like galaxies are shown with cyan and gold lines with
    the gold shaded area
    enclosing 68\% of {\Tigm} values for detected galaxies at a given
    wavelength. Thus, the lower bound of the gold shaded region is not
    representative of the transmission curve shape for any individual
    sightline. {\Tm} for LyC detected, LAE-like galaxies is shown in
    green, significantly higher than for the LBG-like sample due to
    their relative faintness. The increased dispersion seen for LAE
    samples is a result of a smaller number of detections for such
    galaxies. The mean {\Tigm} for all sightlines is shown in 
    black for comparison. \textit{Right column:} Normalised histograms of
    {\Tigm} for all sightlines (black), detected, LBG-like galaxies
    (cyan), and detected, LAE-like galaxies (green). The mean for all
    galaxies and detections are shown with corresponding vertical,
    dotted lines (matched to corresponding open histograms), and the
    median for LBG-like detections is shown with a gold dotted line,
    noting that this line corresponds to the gold line of the left
    panel and has no matching histogram in the right panel.}
  \label{fig:specdetall}
\end{figure}

Spectroscopic detection of LyC radiation provides a key advantage over
photometric detections in terms of interpretation in the context of
estimating {\fesc}. The reason being that spectroscopy
allows one to probe the same \textit{rest frame} wavelengths just
shortward of the Lyman 
limit, typically probed between 880-910 {\AA}, independent of redshift
in theory. In practice, of course, the redshift range in which LyC can be
probed by a given set of spectroscopic observations is defined by the
wavelength coverage of the instrument used. Furthermore, the detection
limits of a given instrument will be wavelength dependent due to
response variations of the detector. Thus, the experiment presented
here should be considered as a simulation of an idealised
spectroscopic instrument
with uniform sensitivity to LyC radiation at 880-910 {\AA} across the entire redshift
interval from 2.9 $<$ $z$ $<$ 3.9. The
black {\Tm} functions in Figure \ref{fig:specdetall} show that this
wavelength range exhibits the largest {\Tm} values at $\lambda_{\rm rest} <
911.8$ {\AA} meaning that at all redshifts spectroscopic observations
probe LyC
emission at the highest {\Tm} and, thus, the highest
probability of detection (at fixed depth). This is simply due to the
fact that the redshift interval of LyC absorption systems that affect
a given wavelength increases with decreasing
wavelength. This means that at lower wavelengths the probability of
encountering a high column density system in any individual
sightline is higher. 

As we discuss later, this is not the case
for photometric observations which instead probe a fixed
$\lambda_{\rm obs}$ range, thus a decreasing $\lambda_{\rm rest}$ with
increasing redshift. Another important and related point is that
ionizing radiation escaping from galaxies will be completely absorbed
by intervening, high HI column density systems, resulting in rapid
drops in flux based on the redshift of that intervening system
(e.g. the drop at $\sim$810 {\AA} in Figure \ref{fig:TMCex}). This
means that escaping ionizing radiation from high redshift galaxies may
only be visible in a very small wavelength range and this behaviour
will be difficult to capture and interpret from photometric
observations, but will be seen clearly in spectroscopy. As a caveat,
however, we note that, without ancillary, high spatial resolution,
space-based photometric data, it can be difficult to rule out the
possibility of low redshift contamination from ground-based
spectroscopic LyC detections \citep{vanzella10,vanzella12}. 

The fact that spectroscopy probes the most transparent portion of the
emitted spectrum from high redshift galaxies also suggests that spectroscopic
detections of LyC may suffer from relatively low $T_{\rm bias}$ at fixed
detection limits (noting however that photometric detections are
significantly deeper for the same exposure time). We show this in
Figure \ref{fig:specdetall} where we show {\Tm} for all
10,000 sightlines in black and {\Tm} for those where galaxies are detected with a
flux above 0.025 $\mu$Jy, equivalent to $\sim$27.9
mag, at $z=3.1$, with coloured lines. It should be clarified here that
this detection limit is chosen
to be roughly matched to the faintest LyC detection reported in S18
for the galaxy Westphal-MM37 (0.026 $\mu$Jy). Considering the full
parent sample of S18, 0.025 $\mu$Jy corresponds to a 1-5$\sigma$
detection as the observational limits and noise characteristics exhibit
complex dependencies on factors such as observational depth and source
redshift (i.e. the observed wavelength of emitted LyC
radiation). Thus, we reiterate that our results are representative of
an idealised version of the S18 survey as we have not attempted to
simulate the full complexity of their spectroscopic observations.

Returning to Figure \ref{fig:specdetall}, the cyan and gold lines indicate
the mean and median $T_{\rm IGM}$ curves for LyC detected, LBG-like
galaxies (comparable to the S18 sample) while the green line shows the
mean $T_{\rm IGM}$ curve for LAE-like detections (comparable to the
F19 sample). Here we measure 
$\langle T_{\rm IGM} \rangle$ in the rest frame wavelength range 880 $\leq$
$\lambda_{\rm rest}$ $\leq$ 910 {\AA} (indicated in Figure
\ref{fig:specdetall}), also following S18. We note that LAE-like
galaxies are not representative of the S18 sample and are only
detected at these spectroscopic limits in our lowest redshift
bins. In fact, overall detection rates at all redshifts is lower for
the more faint sample of LAEs, which accounts for the increased
dispersion seen in the $\langle T_{\rm det} \rangle$ curve for LyC
detected LAEs. Given this comparison is to S18 who focus on LBGs, we do
not place a large emphasis on this mock sample for spectroscopic
observations. 

Also shown in Figure \ref{fig:specdetall} is the median and 68
percentile range for LyC detected LBGs in gold for comparison. The
median value of {\Tigm}
for detections is seen to be larger in the Ly$\alpha$ forest and lower
beyond the Lyman limit, with a cross-over value around 880 {\AA}. The
significant differences between the mean and median IGM transmission
functions for detected galaxies is a reflection of the non-Gaussian
nature of the underlying {\Tigm} distribution (see Figure
\ref{fig:2dpdf}). Regardless, the median and mean values of {\Tigm}
for detected galaxies are similar and throughout the remainder of this
work we focus on the mean value.

In the right column of Figure \ref{fig:specdetall} we compare the
histograms of {\Tigm} at 880-910 {\AA} between all sightlines
(black) and those associated with LBG-like galaxies detected above 0.025
$\mu$Jy (cyan). We can see that the underlying distribution is bimodal
with the most probable value of {\Tigm} being $\sim$0 while the distribution for
detections is unimodal with the most probably value being close to the
upper mode of the underlying distribution (the distribution for
fainter, LAE-like samples, shown in green, is skewed towards even
higher values). This is not surprising as
for galaxy to be detected at LyC wavelengths the value of {\Tigm}
must not be zero. It is clear that the mean value of {\Tigm} for all
sightlines falls between the peaks of the underlying {\Tigm}
distributions and is thus not among the most probably values for
detected galaxies.

The fact that LyC detections can not occur at $T_{\rm IGM} = 0$ may
occasionally be overlooked in calculations of 
{\fesc} for LyC detected galaxies, and is key to the narrative of this
work. Careful consideration of $T_{\rm IGM}$ variation in the estimate
of {\fesc} for individual detections is common practice
\citep[e.g.][]{shapley06,inoue11,vanzella16}. Ultimately, the goal of
this paper is to provide a clear quantification of this effect. A primary application of
our results will be for estimating $\langle${\fesc}$\rangle$ for larger
samples of LyC detected galaxies that may be returned by future,
extremely deep surveys (see \ref{section:conclusions}). It should also
be mentioned that, when estimating upper limits in {\fesc} for samples
including LyC non-detections, {\Tm} considering all simulated
sightlines is appropriate (i.e. inclusion of $T_{\rm bias}$ is
unnecessary). 

\begin{figure}
  \includegraphics[width=\columnwidth]{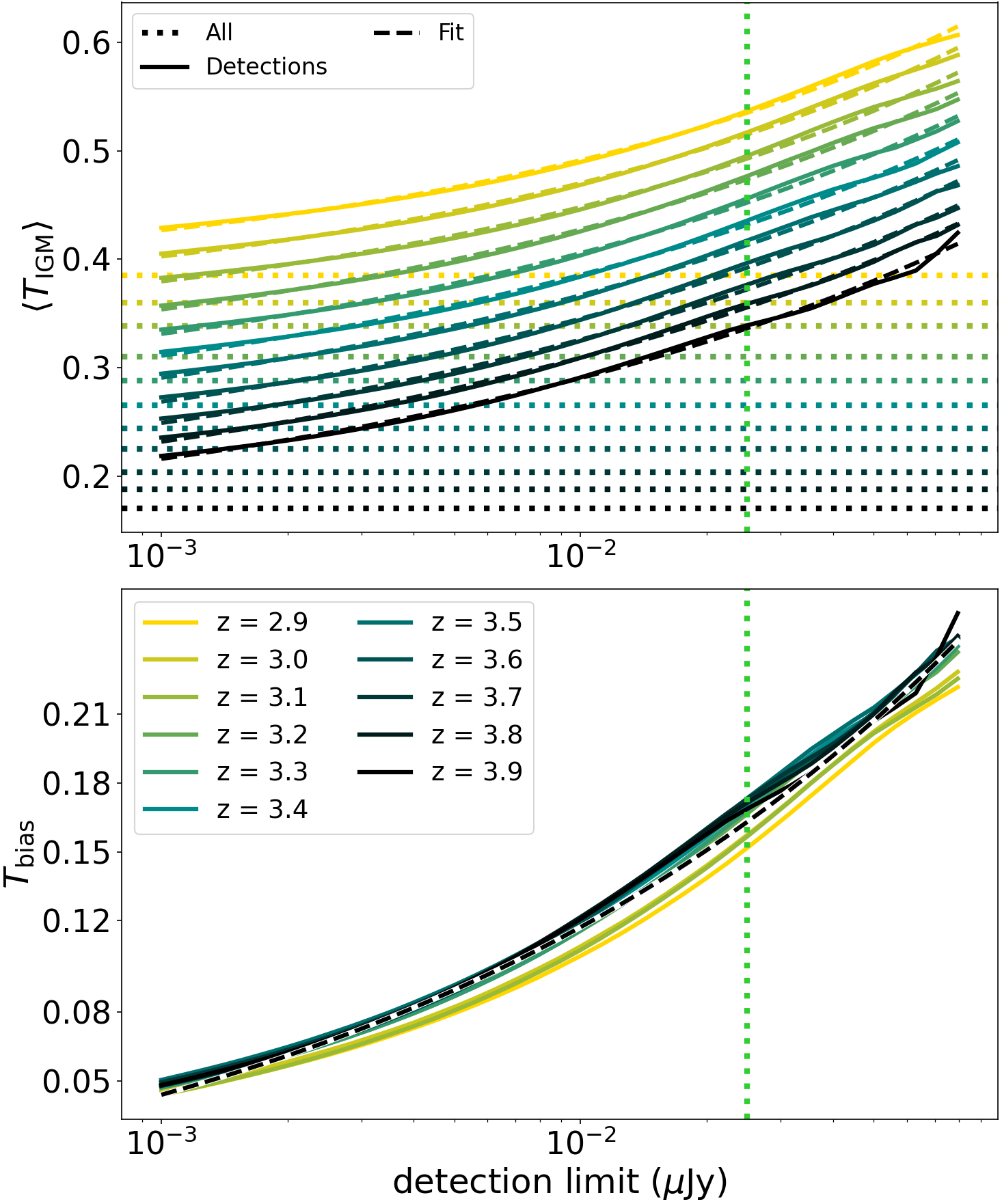}
  \caption{The dependence of $T_{\rm bias}$ on the detection
  limit of spectroscopic observations ($F_{lim}$) for LBG-like detections. \textit{Top:}
  {\Tm} as a
  function of $F_{lim}$ at redshifts between 2.9 and 3.9 (see lower
  panel for legend). Horizontal dotted lines show 
 {\Tm} of all sightlines at a given redshift, and the vertical dotted
line shows the $F_{lim}$ assumed in Figure
\ref{fig:specdetall}. At each redshift we fit the curve of {\Tm} for
detected galaxies with a power law of the form $\langle T_{\rm IGM}
\rangle(F_{lim}) = a F_{lim}^{k} + \epsilon$. \textit{Bottom:} $T_{\rm bias}$ as a
function of $F_{lim}$ for the same redshift interval. We show a
power-law fit, $T_{\rm bias}(F_{lim}) = a F_{lim}^{k} + \epsilon$, to the
combined data for all redshifts as a black dashed line.}
  \label{fig:specdetlim}
\end{figure}

The level of $T_{\rm bias}$ for LyC detections will also be
sensitive to the detection limits, $F_{lim}$, of a given set of observations. We explore the
dependence between $T_{\rm bias}$ and spectroscopic detection limits
in Figure \ref{fig:specdetlim}. In the top panel of Figure
\ref{fig:specdetlim} we show the value of $\langle T_{\rm IGM} \rangle$ for galaxies with
spectroscopically detected LyC emission as a function of detection
limit at redshifts in the range 2.9 $\leq$ $z$ $\leq$ 3.9. For each
redshift, we also show the corresponding $\langle T_{\rm IGM} \rangle$ for all
sightlines with a dotted line of the same colour. The
detection limit assumed in Figure \ref{fig:specdetlim} of 0.025 $\mu$Jy
is shown with a green, vertical, dotted line. We find that at low detection
limits the dependence between {\Tm} and $F_{lim}$ is similar in
all redshift bins apart from the expected vertical offsets due to the
drop in {\Tm} with redshift (reiterating, however, that the definition
$T_{\rm bias} = \langle T_{\rm det} \rangle / \langle T_{\rm IGM}
\rangle$ will result in a clear redshift dependence). At each
redshift the curve can be well fit by a power law of the form $T_{\rm IGM}
\propto F_{lim}^{\beta}$ with $\beta$
in the range $\sim$0.26-0.35. These fits for each redshift are shown in
Figure \ref{fig:specdetlim} with corresponding dashed lines, noting
that these relationships will change for inputs that vary from our
fiducial model (e.g. different values of {\Lint} or a different input
distribution of 1500 {\AA} fluxes). It is also worth reiterating that
sensitivity variations across real spectroscopic detectors will result
in detection limit variation with redshift at fixed exposure time.

In the bottom panel of Figure \ref{fig:specdetlim} we show $T_{\rm bias}$
as a function of $F_{lim}$ at the same discrete $z$ values between 2.9 and
3.9 with $\Delta z = 0.1$. Overall we find a very small scatter in $T_{\rm bias}$ with
the difference between the maximum and minimum $T_{\rm bias}$ at fixed
$F_{lim}$ less that 0.01 at all redshifts in the range considered. Given the smooth curves seen in
Figure \ref{fig:specdetlim}, it is tempting to provide the power law
fits (of the form $\langle T_{\rm IGM} \rangle(F_{lim}) = a
F_{lim}^{k} + \epsilon$, dashed lines in Figure 7, top panel) at each
redshift giving an analytical function for estimating 
$T_{\rm bias}$ as a function of $z$ and $F_{lim}$, however we refrain from
doing so as we would consider any application of such a
function as an overinterpretation of Figure \ref{fig:specdetlim},
which results from our particular implementation for producing {\Tigm} functions as
well as the various inputs of our fiducial model (e.g. here we have
only shown results for LBG-like samples).
For illustrative purposes we have fit a power law to the
combined $T_{\rm bias}$ vs $F_{lim}$ curves $T_{bias} \propto
F_{lim}^{0.29}$. This fit is shown in the bottom panel of Figure
\ref{fig:specdetlim} with a black dashed line. Here, the choice of a power
law is ad hoc, and no specific significance is assigned to the fit
parameters. 

\begin{figure}
  \includegraphics[width=\columnwidth]{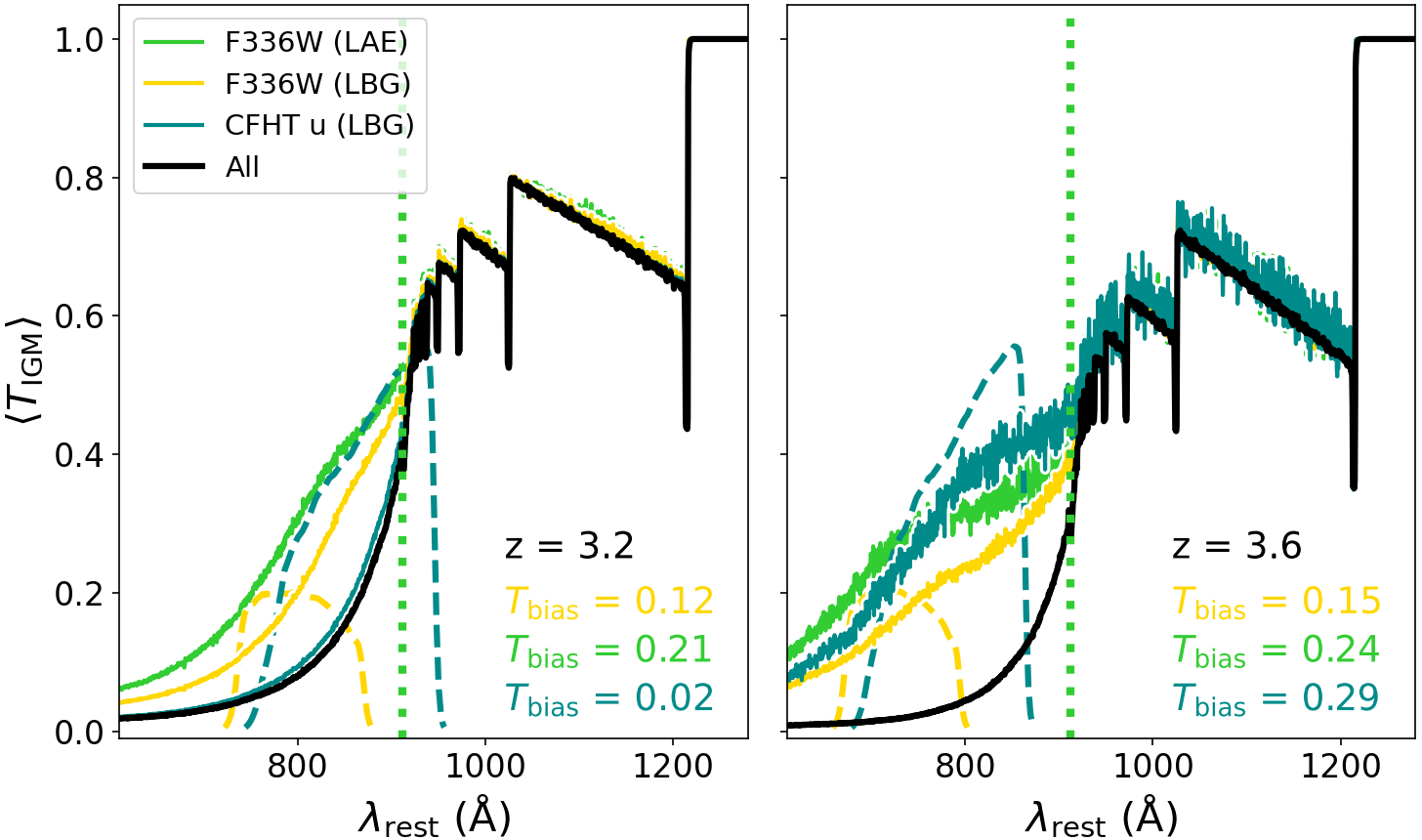}
  \caption{$T_{\rm bias}$ for photometrically detected
    LyC emission in the HST F336W and CFHT $u$ filters. Results are
    shown at $z=3.2$ and $z=3.6$. F336W and CFHT $u$ transmission curves are
    shown as dashed gold and cyan lines, respectively. Detection
    limits are fixed at 30.24 and 27.82 mag for F336W and CFHT $u$,
    respectively (matched to F19 and M20). {\Tm} for F336W and CFHT $u$
    detected LBG-like galaxies are shown in gold and cyan, respectively, and
    {\Tm} for all sightlines is shown in black. The green line
    indicates {\Tm} for LAE-like galaxies detected with the F336W
    filter, more similar to the sample of F19. The increased
    dispersion of the green line relative to the gold line is driven
    by a decrease in the total number of detected galaxies. The Lyman limit is
    indicated in each panel by a vertical dotted line.}
  \label{fig:photdetall}
\end{figure}

It is useful here take a step back and recall two important points:
first there is significant variation in {\Tigm} for individual
sightlines at any redshift (see, e.g., Figure 
\ref{fig:TMCex}) and second the fact that we allow high
\fesc values (up to 1.0) in our 
fiducial model meaning $T_{\rm bias}$ observed in our fiducial model
represents the absolute minimum $T_{\rm bias}$ for a given detection
limit. Thus, we caution the reader from applying values of  
$T_{\rm bias}$ calculated using a similar model to observations of
\textit{individual} galaxies when estimating
{\fesc} without including these caveats. 

\subsubsection{Photometric Detections}\label{section:photdet}

While spectroscopic detection of LyC radiation from galaxies
provides distinct advantages in terms of {\Tm}, achieving this for large
samples of galaxies is inefficient. Photometric surveys have the
potential for detecting large samples of LyC 
emitting galaxies simultaneously. Another important benefit of
photometric surveys when compared to spectroscopy is that photometry
is significantly more sensitive (i.e. deeper) for the same exposure
time. Furthermore, in the case of space-based LyC detections,
ancillary data is not necessary to rule out the possibility of low
redshift contamination. Photometric LyC surveys
must be performed in well studied fields in which targeted galaxies
already have accurate photometric redshift estimates
\citep[e.g. ZFOURGE fields][]{straatman16} or, ideally, secure
spectroscopic redshifts \citep[e.g. 3DHST, DEIMOS10K, VANDELS, MUSE-wide,][]{momcheva16,hasinger18,pentericci18,urrutia18}. In fields such as these, specific redshift windows can
be targeted using photometric bands probing LyC emission such as HST
F336W at $z\sim3.0$ or CFHT $u$ at $z\sim3.4$
(e.g. F19, M20). There are two key drawbacks in the case of
photometric LyC surveys when compared to spectroscopy, however (see
also S18, Section 7.2).

The first drawback in photometric searches for LyC emission when
compared to spectroscopic studies is that the
ionizing radiation may only be observable in a narrow wavelength
range just short of 912 {\AA} as shown in Figure
\ref{fig:specex}. This is due to
intervening, high HI column density 
systems at redshifts corresponding to the Lyman limit occuring at the
wavelength of the drop in flux of our simulated spectra. The fact that
such a drop may occur in the middle of the wavelength sensitivity of a
given filter will result in an underestimation of the flux level of
the emergent LyC radiation. This results from the fact that \textit{calculation}
of the reported photometric flux inherently assumes a flat flux
density across the filter. Of course, the \textit{interpretation} of the
photometric flux can include more complex spectral behaviour,
e.g. extreme [OIII]+H$\beta$ emitters presented in \citet{forrest17}.

The second drawback is that the observed wavelengths of photometric
bands are fixed. This means that the ideal redshift for such
surveys is at the point where the red cutoff
of the filter in question falls just below the Lyman limit (thus
filter dependent). LyC
radiation can be detected to higher redshifts (more likely for 
extremely deep observations), however at high redshift
the filter moves to bluer rest wavelengths where {\Tm}
is significantly lower. This fact causes significant complications
when making comparisons of LyC escape from photometric detections at
different redshifts. We also mention briefly here that some photometric filters
suffer from so-called ``red leak'' with a small amount of radiation at
wavelengths longer than the optimal cutoff of the filter being
transmitted \citep[though this is minimized for the new CFHT $u$
filter used in M20,][]{sawicki19}. As such features will be included in the filter curves
used in our analysis, this effect is implicitly accounted for. 

With these two drawbacks in mind we present the simulated $T_{\rm bias}$ for LyC
detected galaxies for HST F336W and CFHT $u$ detected galaxies in
Figure \ref{fig:photdetall}. Here we use fixed detection limits of
30.24 and 27.82 mag for F336W and CFHT $u$, respectively (matched to
the limits of F19 and M20). The two panels in Figure
\ref{fig:photdetall} show the mean {\Tigm} for all sightlines
(black), for F336W LBG-like detections (gold), CFHT $u$ LBG-like detections
(cyan), and F336W LAE-like detections (green) at redshifts of 3.2 and
3.6 (LAE-like detections for CFHT $u$ are not shown as such detections
are extremely rare due to the relative shallowness of M20
photometry). We find that $T_{\rm bias}$ for LBG-like galaxies is
significantly lower for F336W detections, however this is simply
reflective of the greater depth of our F336W comparison rather than any
intrinsic advantage of HST observations over ground-based for LyC
detections. Comparing F336W LAE-like versus LBG-like detections, we
find that $T_{\rm bias}$ for the former is $\sim$0.1 larger owing to
the relative faintness of LAEs compared to LBGs (see Figure
\ref{fig:UVfluxes}). 

In Figure \ref{fig:photdetall}, F336W and CFHT
$u$ filters are shown with dashed gold and cyan lines, highlighting
the fact that the F336W and $u$ filters
exclusively probe LyC radiation at $z>3.1$ and $z>3.4$,
respectively. This explains why we see a significantly lower $T_{\rm
  bias}$ for the CFHT $u$ filter at $z=3.2$ as the transmission of
this filter peaks redward of the Lyman limit, meaning that it is more
sensitive to non-ionizing radiation at this redshift. In such a case
where a filter straddles the Lyman limit the interpretation of any
observed flux in the context of {\fesc} is significantly complicated
\citep[e.g.][]{bassett19} and such cases should be avoided where
possible.

\begin{figure}
  \includegraphics[width=\columnwidth]{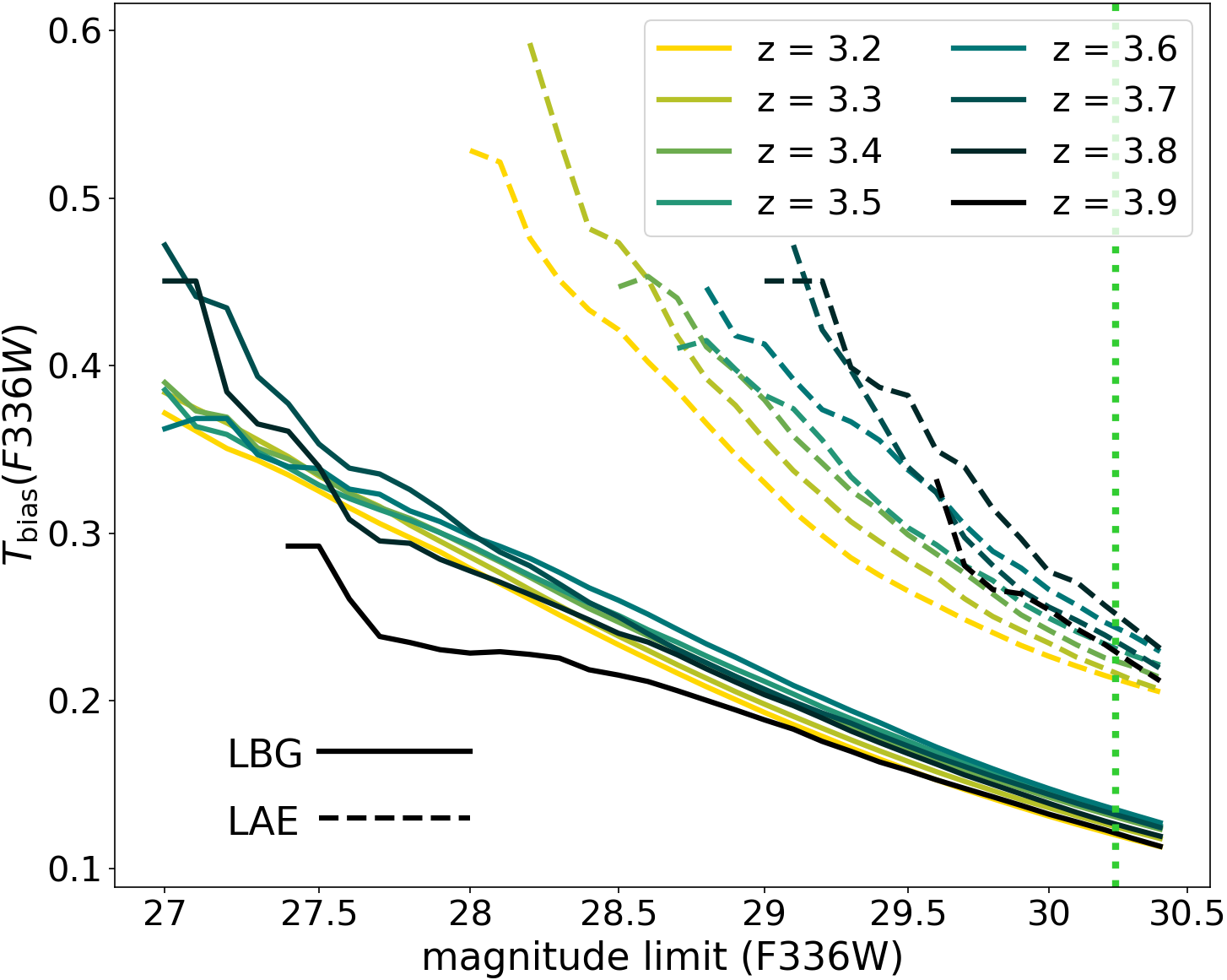}
  \caption{The dependence of $T_{\rm bias}$ on the detection
  limit of F336W observations ($m_{lim}$) at 3.2 $\leq$ $z$ $\leq$ 3.9
  (where F336W probes LyC exclusively). Above $z = 3.2$ the level of
  $T_{\rm bias}$ is relatively constant at fixed $m_{lim}$. The larger redshift
  variation when compared to Figure \ref{fig:specdetlim} and the divergent
  behaviour for shallow observations at high redshift reflect the
  shifting rest wavelengths probed by the F336W with increasing redshift.}
  \label{fig:photdetlim}
\end{figure}

\begin{figure}
  \includegraphics[width=\columnwidth]{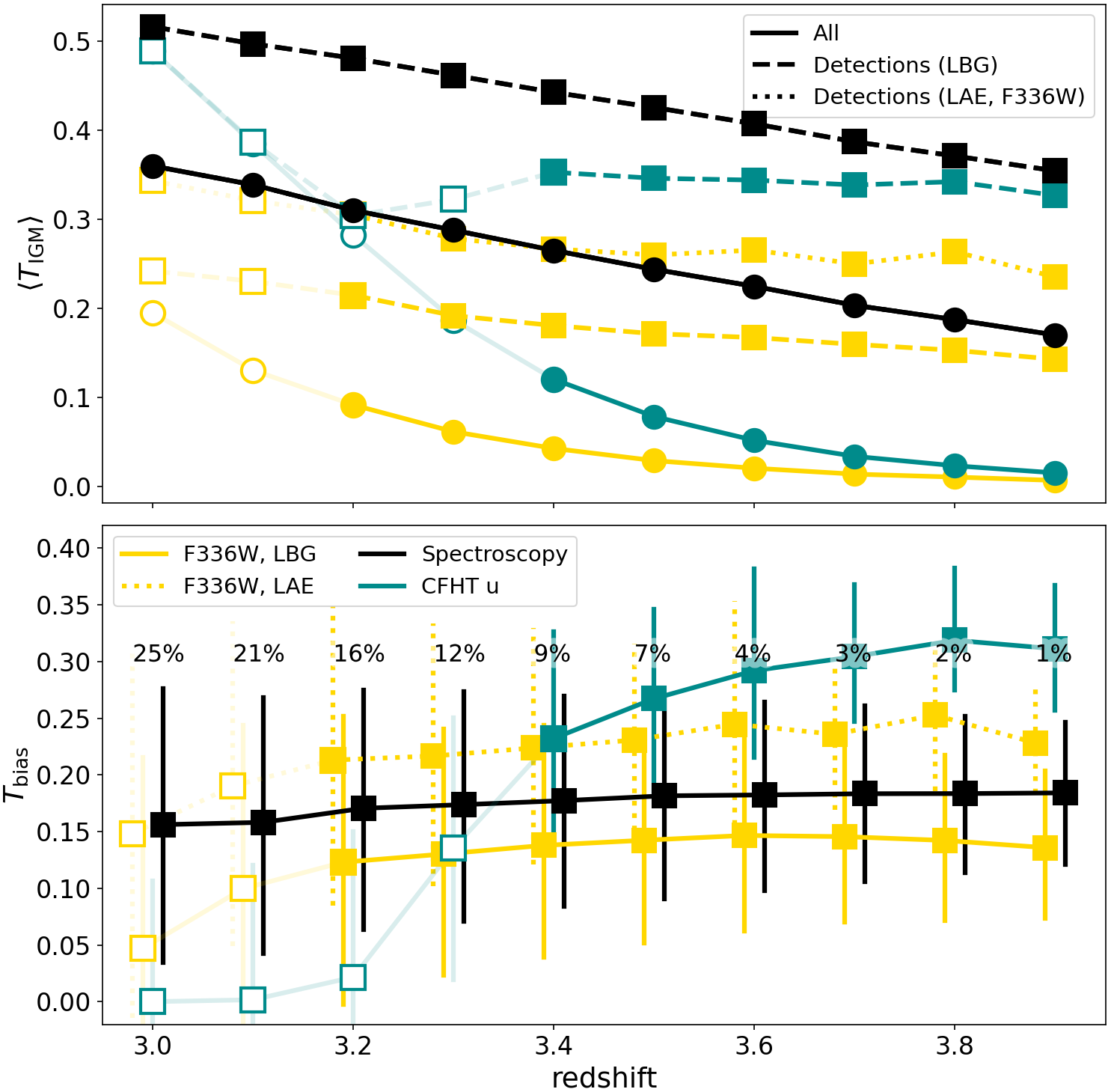}
  \caption{A summary of $T_{\rm bias}$ for our fiducial
    model. \textit{Top:} {\Tm} as a function of redshift for all
    sightlines are shown with solid lines while dashed lines show
    {\Tm} for detected galaxies. Results for spectroscopy, F336W, and
    CFHT $u$ are shown in black, gold, 
    and cyan, respectively. For our fiducial model we assume detection
    limits of 0.025 $\mu$Jy ($\sim$27.9 mag), 30.24 mag, and 27.82 mag for
    spectroscopy, F336W, and CFHT $u$. \textit{Bottom:} $T_{\rm bias}$ as a function
    of redshift for each detection method. Error bars show the 68
    percentile range at each redshift. Values are calculated at fixed
    redshifts between 3.0 and 3.9 with $\Delta z$ = 0.1, slight
    offsets between methods are for clarity only. We also show the
    detection percentage for spectroscopy in black, which decreases
    significantly with redshift, across the top of
    the bottom panel. In both panels, open
  symbols for photometric observations indicate redshifts at which a
  given filter probes (parially or entirely) wavelengths redward of
  the Lyman limit (i.e. non-ionizing photons).}
  \label{fig:biassummaryall}
\end{figure}

$T_{\rm bias}$ for
photometry is also sensitive to observational detection limits. 
The variation in $T_{\rm bias}$ with detection limit (in magnitudes,
$m_{lim}$) is demonstrated in  
Figure \ref{fig:photdetlim} for the HST F336W filter. Solid lines show
results for LBG-like detections and dashed lines for LAE-like
detections. Similar to
spectroscopic results presented in Figure \ref{fig:specdetlim}, we
find that, at fixed $z$, $T_{\rm bias}$ decreases linearly with an
increasing magnitude limit. When compared to the spectroscopic
results of Figure \ref{fig:specdetlim}, with $\Delta T_{bias} \lesssim
0.01$ for all redshifts, we find more variation with
redshift. This is due to the changing rest-frame wavelengths
probed by the F336W filter with redshift. Again, a more significant
redshift evolution will be observed assuming the definition $T_{\rm
  bias} = \langle T_{\rm det} \rangle / \langle T_{\rm IGM} \rangle
$. For LAE-like
detections, the fact that very few LyC fluxes reach magnitudes
brighter than 28.5 (and only in the lowest redshift bins) means that
detections occur in only those sightlines with the highest $T_{\rm
  IGM}$(F336W). Thus the trends shown for LAE samples in Figure
\ref{fig:photdetlim} exhibit more scatter due to an increased
sensitivity to the stochasticity of our IGM transmission
functions. The dashed lines in Figure \ref{fig:photdetlim} also
demonstrate why we find so few LyC detected LAEs for our mock
spectroscopic and CFHT $u$ observations given the depth of these two
comparisons are fixed at $\sim$27.9 and 27.82 mag, respectively.

\subsubsection{Fiducial Model Summary}\label{section:3p1summ}

The results of our fiducial model for fixed detection limits of 0.025
$\mu$Jy ($\sim$27.9 mag), 30.24 mag, and 27.82 mag for spectroscopy,
F336W, and CFHT $u$, respectively, are 
summarised in Figure \ref{fig:biassummaryall} for mock observations of
galaxies with 1500 {\AA} flux distributions characteristic of LBGs
(F336W results for fainter, LAE-like galaxies are also shown with
dotted lines). As described in
Sections \ref{section:specdet} and \ref{section:photdet},
spectroscopic detections at this depth (targeting a fixed rest wavelength window
at 880 $<$ $\lambda_{\rm rest}$ $<$ 910 {\AA}) experience a roughly
constant $T_{\rm bias}$ of $\sim$0.15-0.17 ($\sim$0.32 for fainter,
LAE-like samples). We find a
slight redshift dependence on $T_{\rm bias}$, 
which increases from 0.157 at $z=2.9$ to 0.173 at $z=3.7$ then
decreases slightly to 0.169 at $z=3.9$. This change in $T_{\rm bias}$
of less than 2\% is significantly smaller than the variance seen at
any given redshift and is driven entirely by our cosmological dimming
(see Equation \ref{eq:mockspec}). Thus, we conclude that $T_{\rm
  bias}$ is effectively constant at 3.0 $<$ $z$ $<$ 3.9 for our chosen
definition.

The fact that $T_{\rm bias}$ is found to be constant with redshift is
somewhat counterintuitive. Instead, one may expect a
monotonic increase in $T_{\rm bias}$ with redshift due to the fixed
detection limit and linear decrease in {\Tm}. For our additive
definition of $T_{\rm bias}$, the constant $T_{\rm
  bias}$ observed can be explained by a decrease in detection rate
with redshift where only the brightest galaxies contribute to $T_{\rm
  bias}$ at the high $z$ end. This is illustrated in Figure
\ref{fig:biassummaryall} with the detection percentages for
spectroscopy at each redshift indicated in black. 

Condsidering photometric
detections, $T_{\rm bias}$ is seen to increase while the Lyman limit passes through
the filter in question. At redshifts where a given filter has passed
fully blueward of the Lyman limit, the level of $T_{\rm bias}$ is seen to
level off (within errors) at a value dependent on the photometric depth. For our
fiducial depths, this plateau level is $\sim$0.11-0.14 and $\sim$0.22-0.31 for
the F336W (magnitude limit = 30.24) and CFHT $u$ filters (magnitude
limit = 27.82), respectively. In the case of LAE-like
1500 {\AA} flux distributions, we show results only for F336W as this
comparison has significantly deeper flux limits compared with
spectroscopy and CFHT $u$ (where detections of LAE-like samples are
vanishingly rare). In the case of LAEs, we find that $T_{\rm bias}$ is
roughly 0.1 higher than for LBGs at fixed redshift, with values in the
range $\sim$0.21-0.24 across the redshift range sampled.

We also observe a
slight dip in $T_{\rm bias}$ for photometric detections at the highest
redshifts in the bottom panel of Figure
\ref{fig:biassummaryall}. Unlike spectroscopic detections, by $z \sim
3.8$ our photometric filters are probing very blue $\lambda_{\rm rest}$
where {\Tm} is near zero. Furthermore, as seen in Figure
\ref{fig:2dpdf}, the {\Tigm} distribution at these wavelengths is a
skewed, unimodal distribution peaked at {\Tigm} = 0. This
means that the probability of finding a sightline with {\Tigm} much
higher than zero is very low. This could explain why $T_{\rm bias}$ for
photometry dips at high $z$, as even those small number of detected
galaxies will be found in sightlines approaching zero transmission at
wavelengths probed by a given filter. This means that the level of
$T_{\rm bias}$ seen at lower redshifts simply can not be maintained
given the underlying $T_{\rm IGM}$ distribution for the wavelengths
probed. At higher redshifts the {\Tigm}
distribution becomes so strongly peaked at {\Tigm} = 0.0 that no
detections are expected, thus we do not expect the results presented
here for 2.9 $<$ $z$ $<$ 3.9 to be generalizable towards higher
redshifts. This is not necessarily the case for spectroscopic 
detections as the {\Tigm} distribution at 880 $<$ $\lambda_{\rm rest}$
$<$ 910 {\AA} remains bimodal even at high redshift, thus no obvious
dip in $T_{\rm bias}$ is seen. Regardless, in all cases {\Tm} is
decreasing with redshift, thus detections become rarer. This
manifests as a decreasing 68 percentile range for $T_{\rm bias}$, a
reflection of the drop in the numbers of detected galaxies with
redshift. 

Finally, as mentioned at the
start of Section \ref{section:IGMbias}, the 
definition $T_{\rm bias} = \langle T_{\rm det} \rangle / \langle
T_{\rm IGM} \rangle$ is equally valid to the definition adopted in
this work. Under this alternative definition, a very clear trend
between $T_{\rm bias}$ and redshift is apparent increasing from
$\sim$1.3 to $\sim$1.8 for spectroscopic observations and from $\sim$3
to $\sim$35 for F336W observations for LAE samples. We reiterate that,
although a fractional definition may be more physical (in the sense
that it relates directly to a ratio of HI column densities), the
redshift evolution of $T_{\rm bias}$ in this case reflects primarily
the fact that $\langle T_{\rm IGM} \rangle$ moves increasingly close
to zero with redshift while the actual difference in the mean IGM
transmission between detections and all sightlines is roughly
constant, as our chosen definition illustrates. Thus, our definition
provides a simplified correction when calculating {\fesc} for LyC
detected samples from an observational point of view.

\subsection{An Alternative {\fesc} Distribution}\label{section:fescdist}

\begin{figure}
  \includegraphics[width=\columnwidth]{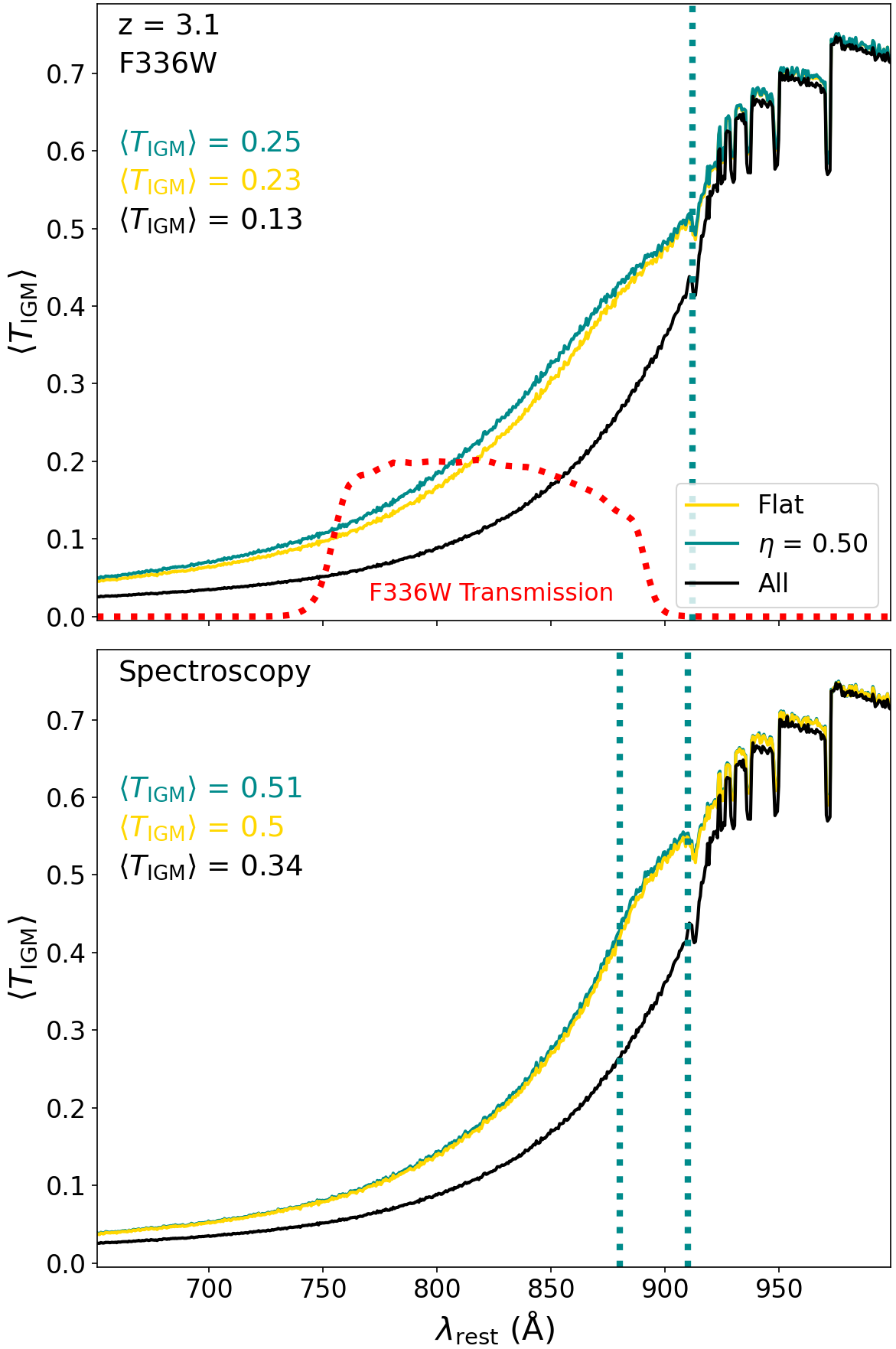}
  \caption{{\Tigm} curves for galaxies
    detected with F336W (top) and spectroscopically,
    (bottom). Both panels show the results at $z=3.1$ with
    detection limits of 30.24 mag for F336W and 0.025 $\mu$Jy
    ($\sim$27.9 mag) for spectroscopy. In
    both cases, the $\eta$ = 0.30 model exhibits only slightly higher {\Tigm}
    than the fiducial model. Spectroscopic (F336W) values of $T_{\rm bias}$ increase
    modestly from 0.16 (0.10) for the fiducial model to 0.17 (0.13)
    for the $\eta$ = 0.50 model.}
  \label{fig:3dex}
\end{figure}

It is expected that if LyC emission is detected
from a given galaxy, it must have a high {\fesc} and/or a high
{\Tigm}. From current observations of LyC emitters (particularly
considering the large number of non-detections), it seems
that {\fesc} values, i.e. 0.0-0.2, are most common
\citep[e.g.][]{boutsia11,grazian16,smith18}. The results presented for our
fiducial model
in Section \ref{section:IGMbias}, however, allow for {\fesc} values from 0.0 to
1.0 with no preference. This means that a large number of detections
from our fiducial model exhibit a large {\fesc} and are detected in
sightlines with relatively low {\Tigm}. If we instead choose an underlying
{\fesc} distribution skewed towards low {\fesc}, we might expect
that the average {\Tigm} for detections will increase, thus
increasing $T_{\rm bias}$. 

In this Section, we explore how altering the PDF of selected {\fesc}
values affects the level of $T_{\rm bias}$ and the distributions of
{\fesc} for LyC detections. For comparison, the fiducial model
can be treated as a flat PDF between 0 and 1. Here we test an
alternative {\fesc} PDF model
designed to give more weight to lower {\fesc} values. In this cases we
choose an exponentially declining {\fesc} PDF of the form:
\begin{equation}\label{eq:expdec}
  PDF(f_{\rm esc}) \propto e^{-f_{\rm esc}/\eta}
\end{equation}
where $\eta$ represents an exponential cut off in {\fesc}. Here we
test the value $\eta$ = 0.50 (see Figure \ref{fig:fesc_pdfs}) motivated by LyC
detection rates from S18 (see Section
\ref{section:detrate}). For brevity our fiducial model will be described
as ``flat'' and our alternative model will be referred to as
$\eta$ = 0.50. As with our fiducial model, for our $\eta$ = 0.50
model we recreate 100 mock spectra for each of our 10,000 IGM
transmission functions at each discrete redshift value as described in
Section \ref{section:mock_spectra}.

We show example {\Tm} curves at $z=3.1$ for detected LBG-like galaxies in each
of our two models in Figure \ref{fig:3dex}. Though not shown here,
results for LAE-like galaxies are qualitatively similar.
Here we see that the flat {\fesc} PDF exhibits a lower $T_{\rm bias}$ than
the $\eta$ = 0.50 as expected. The increases in $T_{\rm bias}$ for
both spectroscopic and photometric detections are found to be only
0.01 and 0.02, respectively. These increases in $T_{\rm bias}$ are
essentially negligible considering the spread in {\Tm} for LyC
detections seen in Figure \ref{fig:biassummaryall}. Thus, in the case
of our $\eta = 0.5$ model, we find no significant difference in
$T_{\rm bias}$ when compared to the fiducial, flat {\fesc} PDF and
note that this behaviour is the same in all redshift bins. In
cases where the underlying {\fesc} is more strongly skewed towards
{\fesc} = 0 (i.e. smaller values of $\eta$), the difference in $T_{\rm
  bias}$ when compared to a flat PDF is certain to increase. Such
a low $\eta$ model (or any other similarly skewed {\fesc} PDF) may be
appropriate for galaxy samples with selection biases different from
the LBG and LAE samples considered here if $\langle$\fesc$\rangle$ does indeed vary
with galaxy properties (see Section \ref{section:detrate} for more
discussion). 

\begin{figure}
  \includegraphics[width=\columnwidth]{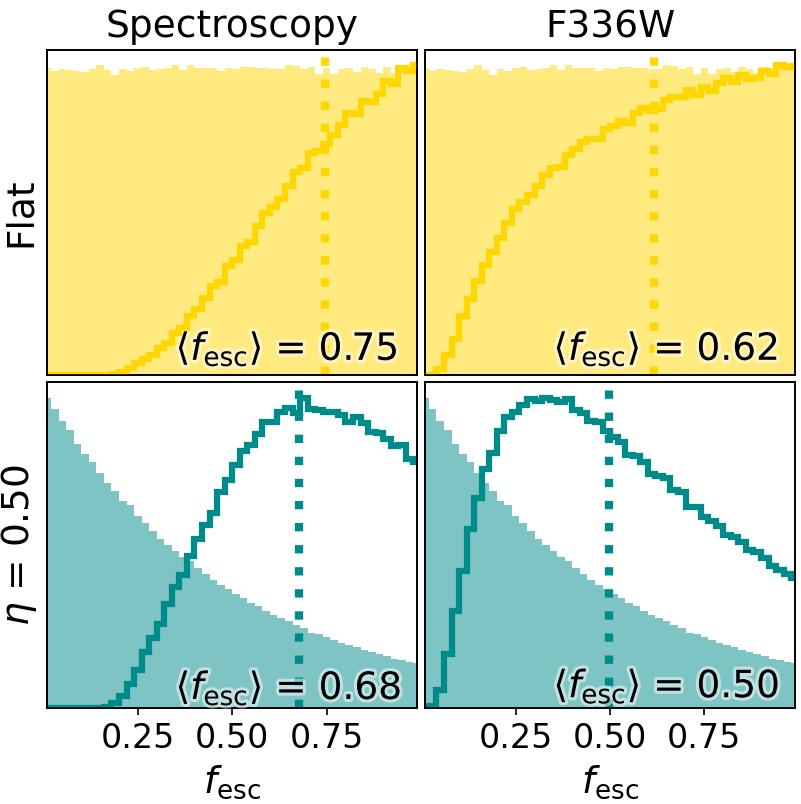}
  \caption{Histograms of {\fesc} for our two {\fesc} PDF models:
    flat in the top row and $\eta = 0.50$ on the bottom. The
    left column shows results for spectroscopy and the right for
    F336W. In each panel the underlying {\fesc} distribution is shown
    with a filled histogram, the {\fesc} distribution of detections
    with an open histogram, and the mean value for detections is shown
  with a vertical dotted line and indicated in the top left of each
  panel. Note that each histogram has been normalised by the maximum
  value for ease of comparison.}
  \label{fig:fescdethist}
\end{figure}

The fact that $T_{\rm bias}$ for our $\eta = 0.5$ model is only
negligibly larger than our flat {\fesc} PDF does not mean the two
models are interchangeable in regards to estimates of {\fesc} from
observed samples.
To illustrate this, we show in Figure \ref{fig:fescdethist} the histograms of
{\fesc} for detections only vs all trials at $z=3.1$ for spectroscopy (left)
and F336W (right). Filled histograms show the underlying {\fesc}
distributions and open histograms show the {\fesc} distribution for
LyC detections. We find that the {\fesc} distributions of LyC
detections (i.e. the posterior) for both observational methods is
skewed towards {\fesc} = 1.0, inconsistent with the low values
typically seen in observations. The posterior for the $\eta = 0.5$
model, on the other hand peaks at lower values, more consistent with
estimates in the literature. We show the mean values for posterior
distributions in each panel with a vertical dotted line. 
When assuming an $\eta=0.5$ model, the inferred average {\fesc} value
is lower by 0.07 and 0.12 for 
spectroscopy and F336W detections, respectively. Thus, although $T_{\rm bias}$
is roughly the same between the two {\fesc} PDF models, the
differences when considering the inferred {\fesc} for galaxy samples
is significant.

We note that the posterior distribution
for the $\eta = 0.5$ model for a given detection method is equivalent
to the posterior for the flat {\fesc} PDF model multiplied by the
input $\eta = 0.5$ distribution (the prior) in line with the framework
of Bayesian statistics. This is true in general, thus one can simply
determine the posterior distribution for any arbitrarily defined
{\fesc} PDF once the posterior for a flat distribution is determined
for a given observational method and detection limit without the need
to run a separate analysis. We stress again that the actual distribution of {\fesc}
is essentially unknown, however we discuss possibilities for placing
some constraints on this in Sections \ref{section:detrate} and
\ref{section:3dv2d}.  

\begin{figure*}
  \includegraphics[width=\textwidth]{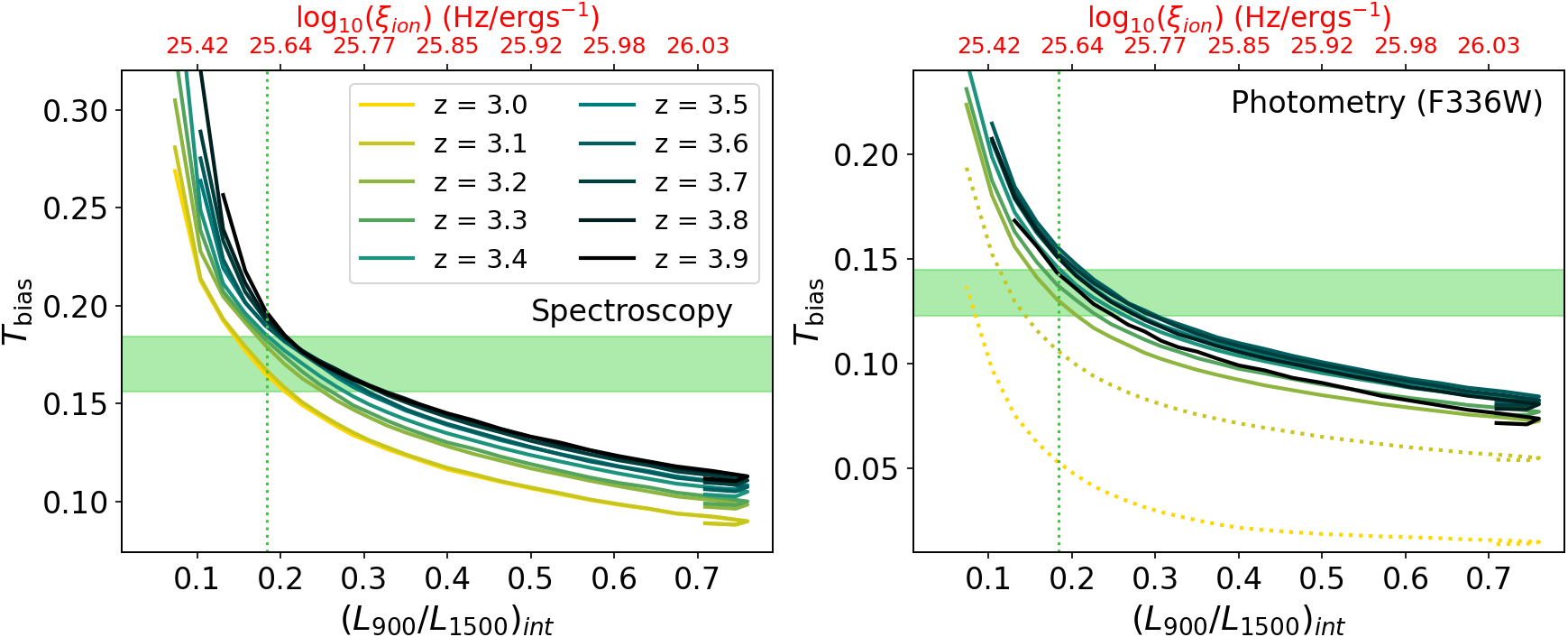}
  \caption{$T_{\rm bias}$ as a function of intrinsic ratio of LyC (at
    900{\AA}) to UV (at 1500{\AA}) luminosities for spectroscopy
    (left) and photometry (right). Solid lines show
    results in each redshift bin as indicated in each legend while
    dotted lines in the right panel indicate redshifts at which the
    F336W filter contains contamination from Ly$\alpha$ forest photons
    as it has not passed fully into the LyC portion of the spectrum. The
    location corresponding to intrinsic ratios 
    presented in Sections \ref{section:IGMbias} and
    \ref{section:fescdist} are indicated with dotted green
    lines and shaded regions. The top axis of both panels indicates $\xi_{ion}$
    for the BPASSv2.1 model at a given {\Lint}.
     }
  \label{fig:intrat}
\end{figure*}

\subsection{Dependence on SED Variations}\label{section:sedvar}

In this Section we briefly explore the effects that varying the SED
shape will have on our estimates of $T_{\rm bias}$ presented in Sections
\ref{section:IGMbias} and \ref{section:fescdist}. In regards to
detecting LyC from a given galaxy above a specified limit, the key
difference resulting from a change in SED shape will be a change in
the flux ratio of the LyC and UV ($\lambda_{\rm rest}$ $\sim$ 1500
{\AA}) portions of the observed spectrum, {\Frat}, at a fixed
{\Tigm}. The factor that will affect
{\Frat} (in addition to {\Tigm}) considered here is
variation in the \textit{intrinsic} ratio of LyC and UV emission,
{\Lint}. 

To test the effect of altering {\Lint} on our results we rerun the
analysis described in Section \ref{section:mock_spectra} for each age of our
exponentially declining BPASSv2.1 models (with $e$-folding timescale
of 0.1 Gyr) in the range 6.0 $<$ log(age) $<$ 9.0 in steps of
$\Delta$log(age) = 0.1. The models produced exhibit {\Lint} in the
range $\sim$0.07-0.77, with corresponding values of $\xi_{ion}$ from
$\sim$25.4-26.0. For each aged model we again create 100 mock spectra
for each of the 10,000 IGM transmission functions produced at each
redshift (2.9 $<$ $z$ $<$ 3.9, $\Delta z$ = 0.1) with 1500 {\AA}
fluxes sampled from an LBG-like distribution. We then repeat our
measurements of LyC flux as in previous sections and adopt the flux
limits of our fiducial model: $F_{lim}$(spectroscopy) = 0.025 $\mu$Jy
($\sim$27.9 mag) and $m_{lim}$(F336W) = 30.24. The CFHT $u$ comparison
is not considered here as the relatively shallow nature of these
observations results in prohibitively few detections at low
{\Lint}. For a similar reason, we also do not consider LAE-like
samples in this section.

We show the resulting {\Lint}
versus $T_{\rm bias}$ for spectroscopic, LBG-like LyC 
detections in the left panel of Figure \ref{fig:intrat}. Similar to the results for
our test on detection limits we find only slight variation in
$T_{\rm bias}$ with redshift with a total spread in values of $\sim$0.03
for all redshifts at a fixed {\Lint} above {\Lint} = 0.15 (again, a
fractional definition of $T_{\rm bias}$ will result in significant
redshift variation). For
reference, we show the location of 
the fiducial model presented in Section \ref{section:specdet} with the dotted green
line and shaded regions. Slight differences can be attributed to
stochasticity as the analysis here represents and independent sample
of 1500 {\AA} fluxes and {\fesc} values at the same {\Lint}
value. Regardless, the results of Figure \ref{fig:intrat} are
consistent with those of \ref{section:specdet} within errors.

 Results for F336W detections are shown similarly in the right 
panel of Figure \ref{fig:intrat}. We show results at redshifts where
F336W partially probes non-ionizing photons with dashed lines
(i.e. $z < 3.2$). Qualitatively the curves are similar
to those in the left panel, with the difference in $T_{\rm bias}$
again attributed to the increased depth of the F336W observational
comparison. 

Finally, we note that none of the models presented to this point have
considered the effects of dust attenuation on the observed LyC flux
from mock galaxies. The effect that dust will have on LyC will be to
further reduce the observed value of $F(LyC)/F(UV)$ relative to
{\Lint}. In this way, dust attenuation is a third level of degeneracy
between {\fesc} and {\Tigm}. Given the low attenuation for LyC
detections (e.g. S18), we ignore the effects of dust simply noting
that detections should be biased towards galaxies with low
dust attenuation \citep[or even none in the case of LAEs,
e.g.][]{fletcher19,nakajima20}. 

\section{Discussion}\label{section:discussion}

\subsection{Correlation between $T_{\rm IGM}(LyC)$ and
  $T_{\rm IGM}(Ly\alpha)$}

One major difficulty in accurately measuring {\fesc} from high
redshift galaxies is the unknown value of {\Tigm}. So far, there is
no clear observational indicator of $T_{\rm IGM}($LyC$)$, which has
necessitated statistical methods such as those explored in this
paper. In the work of \citet{inoue08}, however, it was argued that the
{\Tigm} at Ly$\alpha$ wavelengths may correlate with $T_{\rm IGM}($LyC$)$
(their Section 4.4, Figure 10). This claim is in direct contrast with
previous results of \citet{shapley06} who found no such correlation at
$z=3.06$. \citet{inoue08} suggest that the lack of correlation seen in
\citet{shapley06} was due to those authors exploring {\Tigm} at only
one redshift. Here we test for a correlation between $T_{\rm IGM}($LyC$)$
and $T_{\rm IGM}($Ly$\alpha)$ for our simulated IGM transmission
functions, noting that our simulations differ from those of
\citet{shapley06} and \citet{inoue08} in that we include a CGM
component to our HI column density distributions following the work of
S18 and \citet{rudie13}. 

To perform this test, we assess all one million IGM sightlines we have
produced in Section \ref{section:TMC}, measuring {\Tigm} for LyC at 880 $<$
$\lambda_{\rm rest}$ $<$ 910 {\AA} and for Ly$\alpha$ at 1050 $<$
$\lambda_{\rm rest}$ $<$ 1170 {\AA} following \citet{shapley06} and
\citet{inoue08}. We note, however, that the LyC and Ly$\alpha$
wavelength ranges used in these works are not probing the same
redshift range. Thus, we also measure an alternative Ly$\alpha$
wavelength range 1173 $<$ $\lambda_{\rm rest}$ $<$ 1213 {\AA}, matched to
the redshift of LyC in the specified range. In each redshift bin, we
measure the spearman rank-order correlation coefficient between
$T_{\rm IGM}($LyC$)$ and $T_{\rm IGM}($Ly$\alpha)$ in both wavelength ranges
and also the correlation coefficient of the combined data from all
redshift bins. 

\begin{figure}
  \includegraphics[width=\columnwidth]{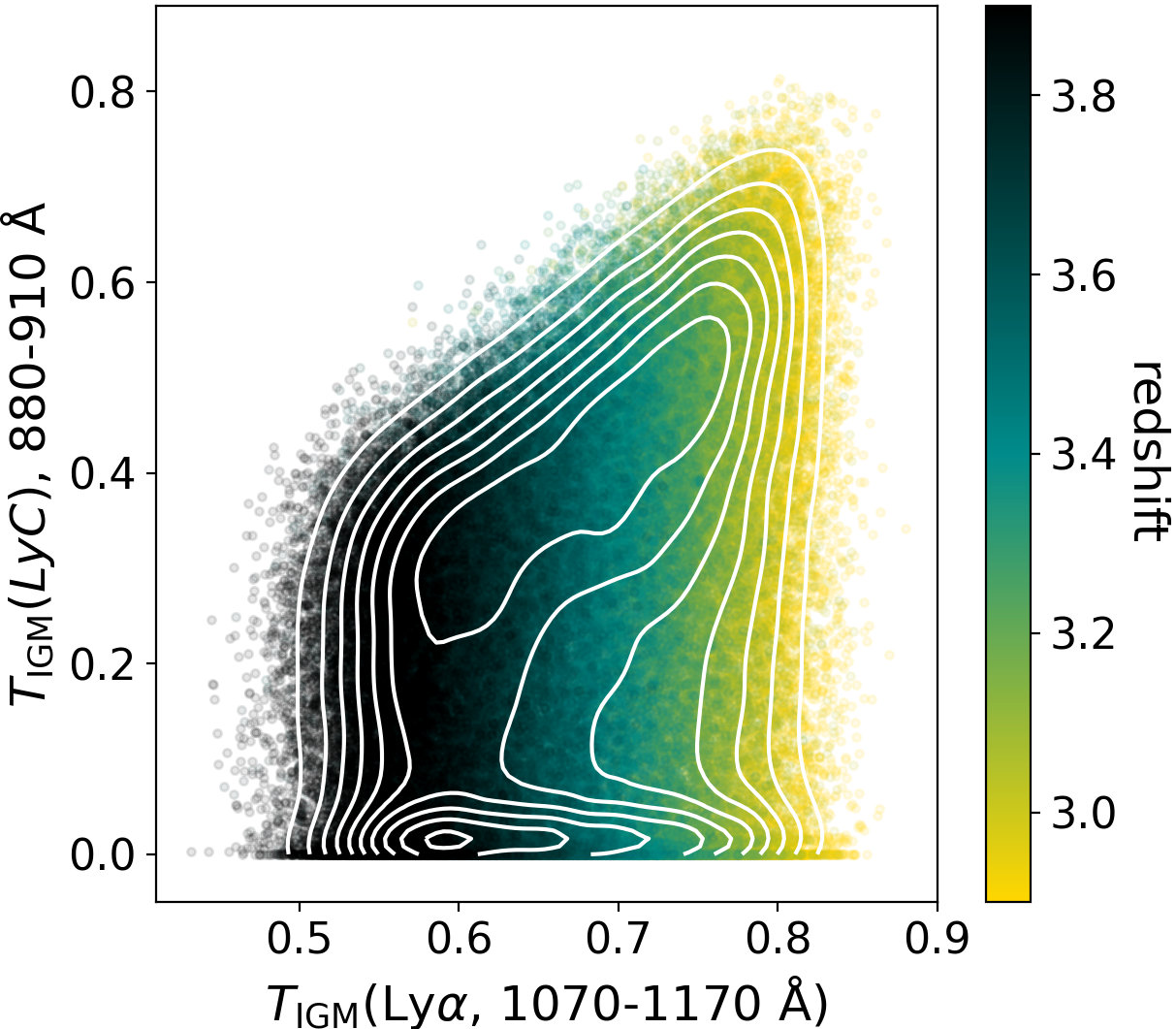}
  \caption{A comparison of {\Tigm} for LyC (880-910 {\AA}) and
    Ly$\alpha$ (1070-1170 {\AA}) for all one million simulated
    sightlines. Points are coloured based on their source
    redshift. We find that the apparent correlation seen between the IGM
    transmission of LyC and Ly$\alpha$ radiation is driven by the fact
  that both values exhibit individual redshift dependencies rather
  than any correlation between these two values. Indeed, there is no
  apparent correlation between $T_{\rm IGM}$(LyC) and $T_{\rm
    IGM}$(Ly$\alpha$) at fixed redshift.}
  \label{fig:tlacorr}
\end{figure}

Figure \ref{fig:tlacorr} shows $T_{\rm IGM}($LyC$)$ vs
$T_{\rm IGM}($Ly$\alpha)$ for all one million sightlines colored by their
redshift. Overall there appears to be a correlation between the two
values (albeit with large scatter), however at any individual redshift
such a correlation is less apparent. For $T_{\rm IGM}($Ly$\alpha)$ at
1050-1170 {\AA} we measure correlation coefficients at individual
redshifts finding values in the range 0.05-0.08 indicating no
correlation with $T_{\rm IGM}($LyC$)$ at fixed redshift consistent with
\citet{shapley06}. Considering all redshift bins together we find a
drastic increase in the correlation coefficient to 0.34. This is still
lower than the correlation of 0.86 quoted by \citet{inoue08}, however,
in this work the authors tested a much wider redshift range from 0.2
to 6.0. Our results combined with those of \citet{inoue08} suggest
that any apparent correlation between {\Tigm} at LyC and Ly$\alpha$
wavelengths is driven only by the fact that both values correlate
similarly with redshift (e.g. Figure \ref{fig:TMCex}). For any individual galaxy (or sample) at a
given redshift, however, $T_{\rm IGM}($Ly$\alpha)$ provides no useful
prediction for $T_{\rm IGM}($LyC$)$. Indeed, this is apparent from the
contours shown in \citet{inoue08} Figure 10.

As we have pointed out, however, the Ly$\alpha$ wavelength range
considered in \citet{shapley06} and \citet{inoue08} is not well
matched to the LyC wavelength range they considered. If we instead use
our alternative Ly$\alpha$ range, 1173-1213 {\AA}, we find a significant increase
in the correlation coefficient at fixed redshift range to
0.32-0.37. Combining the values for all bins we find a modest increase
to 0.45. Thus, we find a weak correlation between {\Tigm} for LyC
and Ly$\alpha$ at fixed redshift where the wavelength ranges for these
two are well matched. We note that a direct comparison to the results
of \citet{inoue08} and \citet{shapley06} may be slightly tenuous as
the IGM transmission curves produced there do not include a CGM
component while our models do. Indeed, this may be the reason that we
find such a large increase in the correlation coefficient at fixed
redshift when the wavelength ranges of LyC and Ly$\alpha$ are properly
matched. Given the large scatter and
the fact that $T_{\rm IGM}($LyC$)$ is found to be 0 for a range of
$T_{\rm IGM}($Ly$\alpha)$ at fixed redshift, however, we would be
hesitant to try and estimate one from the other regardless of the
apparent weak correlation. 

\subsection{Effects of $T_{\rm bias}$ on {\fesc} Estimates for Samples}\label{section:biasdisc}

The analysis presented in Section \ref{section:IGMbias} was designed
to predict the average bias for a sample of LyC detected galaxies,
which in turn can be used to estimate the average {\fesc} of the
sample (e.g. S18, F19). Thus, it may not be appropriate
to blindly apply values measured here to individual galaxies. Here we
test the discrepancy
between the average value of {\fesc} for a sample of LyC detections
estimated with and without including $T_{\rm bias}$ when compared to the
true average {\fesc}. This should be seen as a highly simplified test
as all mock galaxies represent dust-free BPASSv2.1 models with a fixed
{\Lint} of 0.18 (log$_{10}(\xi_{ion}/[$Hz erg$^{-1}]) = 26.51$). Real
galaxy samples are likely to exhibit a range of {\Lint}, and will thus
decrease the accuracy of {\fesc} estimates when compared to this test.

\begin{figure}
  \includegraphics[width=\columnwidth]{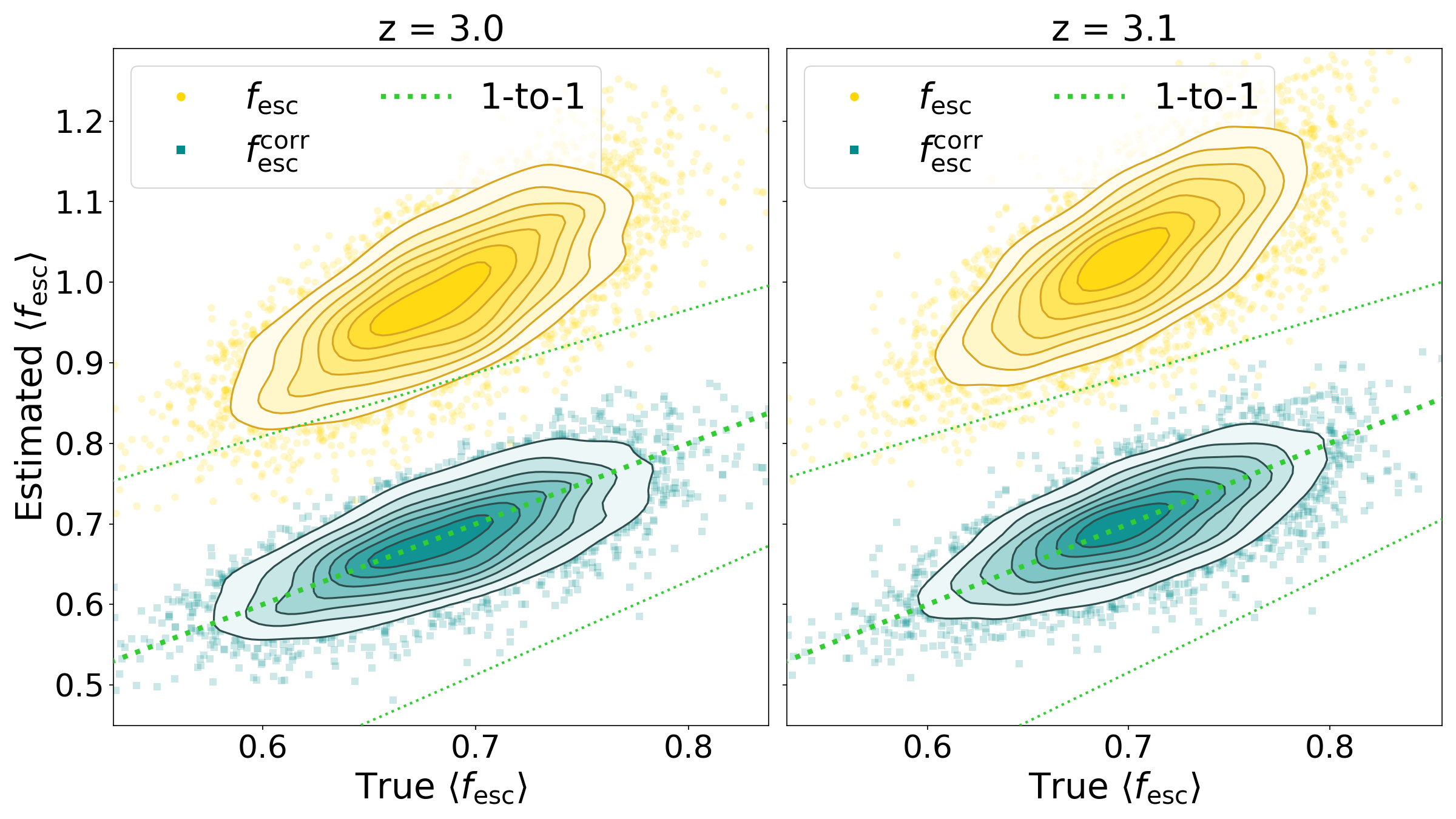}
  \caption{Comparison of $\langle${\fesc}$\rangle$ computed via
    Equation \ref{eq:fescnorm} (gold) and Equation \ref{eq:fescbias}
    (cyan) compared to the true $\langle${\fesc}$\rangle$ at $z=3.0$
    (left) and $z=3.1$ (right). Here we
    perform 5000 trials in which 15 LyC detected galaxies are
    selected at random (comparable to the number detected in S18) from
    among the one million mock galaxies produced at each redshift as
    decribed in Section \ref{section:mock_spectra}. We calculate the
    average {\Frat} among the 15 galaxies and use this value to
    estimate $\langle${\fesc}$\rangle$ using Equations
    \ref{eq:fescnorm} and \ref{eq:fescbias}. The 1-to-1 relation is shown
    with the thick dotted line while the thin dotted lines represent
    the average 68 percentile spread of the 15 galaxies selected in
    individual trials (more description in text), which we find to
    decrease roughly linearly with increasing the mean {\fesc} for a given
  trial.}
  \label{fig:stacktest}
\end{figure}

The typical method of estimating {\fesc} is to employ an equation of
the form (or similar to):
\begin{equation}\label{eq:fescnorm}
  f_{\rm esc} =
  \frac{(F_{900}/F_{1500})_{obs}}{(L_{900}/L_{1500})_{int}} \times
  \frac{1}{\langle T_{\rm IGM} \rangle} 
\end{equation}
noting that the effects of dust attenuation are ignored here. In order to
estimate {\fesc} for a given level of $T_{\rm bias}$, Equation
\ref{eq:fescnorm} must be modified in the following way:
\begin{equation}\label{eq:fescbias}
  f_{\rm esc}^{corr} =
  \frac{(F_{900}/F_{1500})_{obs}}{(L_{900}/L_{1500})_{int}} \times
  \frac{1}{\langle T_{\rm IGM} \rangle + T_{\rm bias}}
\end{equation}
Our test of the recovery of $\langle${\fesc}$\rangle$ for a sample of
spectroscopically LyC detected galaxies is performed on the mock
observations described
in Section \ref{section:mock_spectra}. We first select those mock galaxies
with output 880 $<$ $\lambda_{\rm rest}$ $<$ 910 {\AA} fluxes above the
detection limit of 0.025 $\mu$Jy. We then perform 5000 trials in which
we randomly select 15 mock LyC detections (matched to the number of
detections in S18) and measure $\langle${\Frat}$\rangle$ of this
subsample. For each trial we calculate the average {\fesc} using Equations
\ref{eq:fescnorm} and \ref{eq:fescbias} and compare this with the true
$\langle${\fesc}$\rangle$ for the 15 selected detections.

The results of this test at $z=3.0$ and $z=3.1$ for our $\eta$ = 0.5 {\fesc}
PDF model are shown in Figure \ref{fig:stacktest}, though we find
similar results for the flat {\fesc} PDF of our fiducial model.
Here we plot the true $\langle${\fesc}$\rangle$ versus two estimated
values. Gold and cyan contours show the distribution for
$\langle${\fesc}$\rangle$ estimated using Equations \ref{eq:fescnorm}
and \ref{eq:fescbias}, respectively. The thick dotted green line shows the
1-to-1 relation. The thin green lines are meant to be representative
of the average 68 percentile spread of the 15 galaxies from any
individual trial. To produce these lines we measure the 68 percentile
lower and upper bounds and the average values of {\fesc} for the 15 galaxies from
each of the 5000 trials. We find that the upper and lower bounds
for a given trial decrease roughly linearly with increasing mean
{\fesc} (albiet with significant scatter), thus we fit each bound with
a straight line as a function of mean {\fesc}. In this way, we are
attempting to illustrate, roughly, the expected speard in {\fesc}
values for a random selection of 15 LyC detected galaxies having a
given mean {\fesc} value. 

At both redshifts there is good agreement between
$f_{\rm esc}^{\rm corr}$ and the true value, while failing to account
for $T_{\rm bias}$ results in an overestimate of the average
{\fesc}. The level of overestimation is lower at $z=3.0$ due to the
fact that {\fesc} in Equations \ref{eq:fescnorm} and \ref{eq:fescbias}
depends on the reciprocal of $T_{\rm IGM}$, which is decreasing with
redshift towards 0. At $z=3.1$, Only $\sim$1.4\% of the
estimated $\langle${\fesc}$\rangle$ values calculated using Equation
\ref{eq:fescnorm} fall within the range of typical ``true'' {\fesc} values for
our detected sample (noting this percentage is stochastic). At higher
redshifts this falls to 0\%. Considering $f_{\rm esc}^{\rm corr}$, we
find that, typically, less than 1\% of trials fall outside of the
rough confidence intervals presented in Figure \ref{fig:stacktest}. This
test illustrates that not accounting 
for $T_{\rm bias}$ when estimating the stacked {\fesc} for detected
galaxies can result in a significant overestimate of the true value. 

Of course, as has been repeated throughout this work, the absolute
differences between $f_{\rm esc}$ and $f_{\rm esc}^{corr}$ (as well as the
fractional decrease) will have some dependence on the details of
our method for producing IGM transmission curves (e.g. $N_{\rm HI}$
distributions), the assumed value(s) of {\Lint}, the assumed
{\fesc} PDF, the input distribution of 1500 {\AA} fluxes, etc. In
addition, the inclusion of dust, choice of dust curve, and any assumed
dependence between E(B-V) and {\fesc} will further affect these
results. Although not shown, we also performed the test presented here with dust
attenuation included following the method outlined in Section
\ref{section:dust} (where E(B-V) values are sampled from a
distribution characterised by the observed values from S18) and find
similar results with a similar level of scatter. This of course
assumes that both the average $E(B-V)$ for detected galaxies as well
as the exact form of the attenuation curve is precisely
known. Inevitably, these values will be highly uncertain for real
observations resulting in a higher level of scatter. Providing
more realistic tests of the associated 
effects on our stacking, while possible, would be highly model
dependent, thus not particularly useful.

The fact that {\Tigm} for LyC detected galaxies is
expected to be larger than {\Tm} at a given redshift will be
true regardless of the exact implementations, however. Thus, the purpose of the
illustration presented here is simply to highlight the fact that the
assumption that {\Tm} is representative of IGM sightlines towards LyC
detected galaxies will result in an overestimation of {\fesc}. Given
that $T_{\rm bias}$ increases with decreasing observational depth, the
overestimation of {\fesc} will be the higher for shallower LyC
surveys.

\subsection{Survey Detection Rate vs {\fesc} PDF}\label{section:detrate}

As we have shown in Section \ref{section:fescdist}, the value of
$\langle${\fesc}$\rangle$ inferred for stacked samples of LyC
detections will depend on the PDF assumed for {\fesc}. It has been
repeated throughout this work that, observationally, there appears to
be a preference for low (or zero) {\fesc} from high redshift galaxies
\citep[e.g.][]{japelj17,smith18,bian20}.
This creates a chain of circular reasoning, however, as accurate measurements of
{\fesc} thus requires knowledge of the PDF of {\fesc}, which seemingly
requires accurate measurements of {\fesc} to determine. The way
forward is to determine an observational metric that can help to
determine the PDF {\fesc} that is independent of the individual values
of {\fesc}. 

In this Section we propose that the detection rates of LyC from
dedicated surveys can be used to probe the parameters of a given
{\fesc} PDF. We construct two mock versions of the surveys of S18, F19,
and M20, one with a flat {\fesc} PDF and one with an exponentially
declining {\fesc} PDF. In the latter case we tune the value of
$\eta$ (see Equation \ref{eq:expdec}) to match the observed detection
rate of a given survey (more description to follow). In each case, we again use the same fixed input
BPASSv2.1 SED model as our fiducial model with {\Lint} = 0.18. 
In all cases the input 1500 {\AA} fluxes are sampled
from a distribution matched to the fluxes reported by each of those
surveys. Thus, in this case, the F19 sample, which is made up of LAEs,
have a characteristic 1500 {\AA} flux that is lower than that of the
LBG samle of S18 at the same redshift resulting in a lower
relative LyC flux (see Figure \ref{fig:UVfluxes}). This is important as the selection method of a
given sample strongly influences the distribution of galaxy properties
included (e.g. typical 1500 {\AA} flux, {\Lint}, among others), which
ultimately determine the output LyC fluxes, and thus the detectability
of a given galaxy. Therefore, the toy model presented here is primarily
for illustrative purposes.

The set of inputs described above is combined with our {\Tigm}
sightlines to produce an output sample of LyC fluxes. In
the case of S18 and F19 we simply use the {\Tigm} functions already
produced, noting that this requires all of our mock galaxies to be at
discrete redshifts with $\Delta z$ = 0.1. For S18 we match the
observed redshift distribution in each bin from that work and for F19
all galaxies are simulated at $z=3.1$. As M20 explores significantly
higher redshifts we simply randomly sample values across the full
redshift range (matched to the observed distribution from that work)
and create new {\Tigm} functions each time. For each survey we
produce 10,000 mock observations. We then randomly draw
subsamples from these mock observations with sizes matched to the
observed sample sizes in each paper and measure the detection rates of
LyC for each subsample with detection rates of 0.025 $\mu$Jy
($\sim$27.9 mag), 30.24
mag, and 27.82 mag for S18, F19, and M20, respectively. We perform
this test with a few different $\eta$ values for the exponential
{\fesc} PDF, and coarsely tune the model such that the average
detection rate falls within the range quoted for each survey.

\begin{figure}
  \includegraphics[width=\columnwidth]{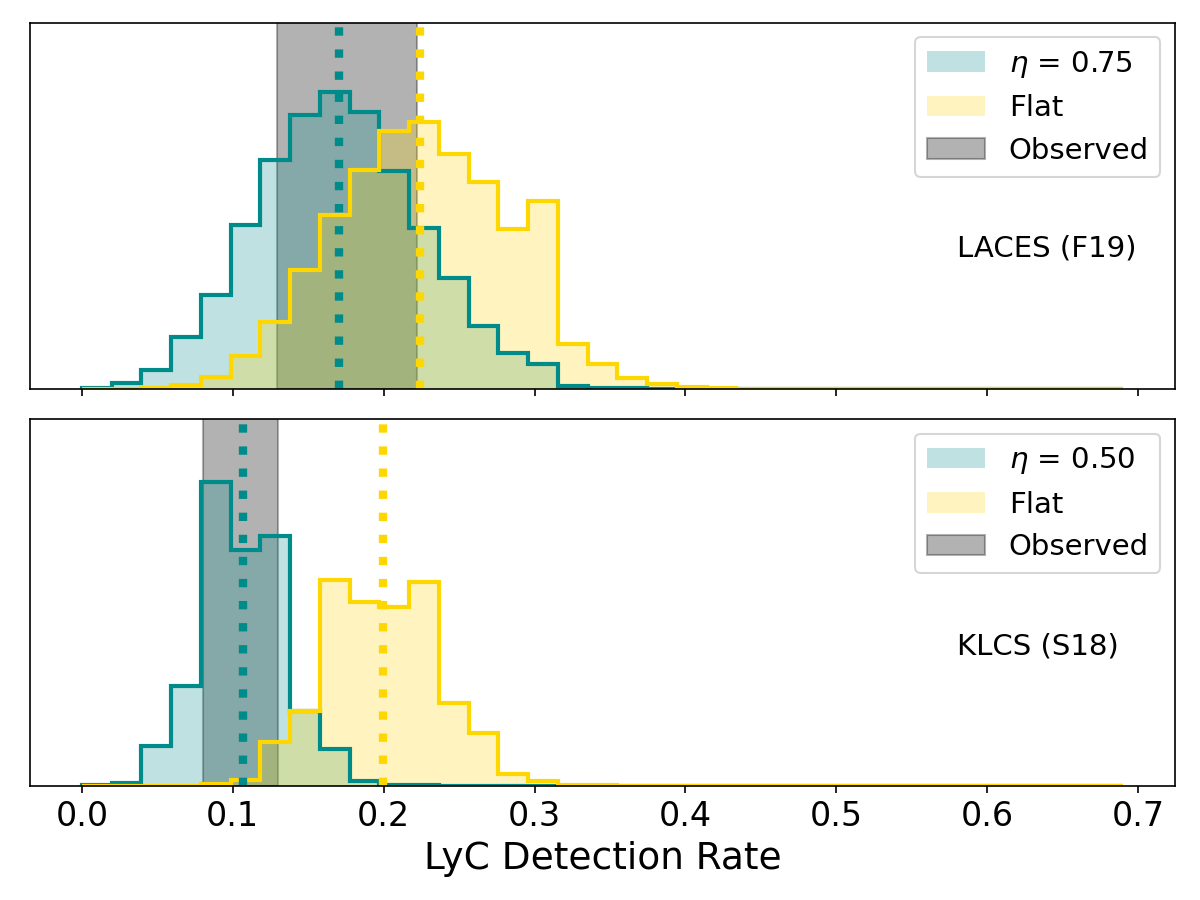}
  \caption{Detection rate distributions for our mock LACES (top) and
    KLCS (bottom) surveys. In each case we create 50 thousand mock
    spectra at the respective survey redshifts following Section
    \ref{section:mock_spectra}, however we now include the effects of dust
    attenuation for the S18 comparison (see text). The observed detection rates are shown in gray while
    the detection rates assuming a flat and exponentially declining
    {\fesc} PDF are shown in gold and cyan, respectively. In each
    case, the value of $\eta$ for the exponentially declining PDF is
  coarsely tuned to match the detection rate of a given survey.}
  \label{fig:detrateplot}
\end{figure}

The results of our test for KLCS and LACES are shown in Figure
\ref{fig:detrateplot}. In each panel the detection rate distribution
for the tuned, exponentially declining model is shown in cyan (with
the tuned $\eta$ value in the legend) and the distribution for the
flat model is shown in gold. For LACES the observed detection rate
range is defined by either only considering their
``gold'' sample (low) or  considering the ``gold'' plus ``silver'' samples (high)
and for KLCS we take their detection rate of 15/124 $\sim$ 0.12. In
the case of KLCS, the flat {\fesc} PDF model is seen to predict a detection
rate that is too large to reproduce the survey in question. While the
$\eta$ = 0.5 model is well matched to the observed detection rate. In
the case of LACES, however, though the average detection rate for the
flat model is close to the upper limit for the detection rate of that
suvey, it is difficult to rule out a flat PDF. The coarsely matched
exponential model requires a relatively high value of $\eta$ =
0.75. From Figure \ref{fig:fescdethist} we expect that the inferred {\fesc}
values from this model will not differ significantly from a flat
distribution. In the case of our mock M20 test, we were unable to reproduce the high
detection rate reported in the paper, which falls in the range
0.02-0.11 (1-5 out of 44) depending on the reliability cut for the LyC
emitting galaxy candidates from that work. For our mocks we find
an overall detection rate from the flat {\fesc} PDF (which will give
the highest detection rate) of 0.005, thus only a small fraction of
random selections of 44 galaxies will even contain one detection. 

Taken together, this toy model test for our three comparison samples
provides strong evidence that the underlying {\fesc} PDFs will be
sensitive to the selection bias of the galaxy sample in question. In
the case of KLCS and LACES, the former probes the bright end of the
UV luminosity function characterised by LBGs while the latter
significantly fainter LAEs. The fact that the detection rate of KLCS
requires the {\fesc} PDF to be skewed towards 0 while LACES is not
inconsistent with a flat {\fesc} PDF points towards a scenario in
which faint galaxies, on average, have a {\fesc} PDF less biased
towards 0 \citep[similar
to the results of][]{finkelstein19}. Our inability to reproduce the
high detection rate of M20, even employing a flat {\fesc} PDF,
suggests that this sample may be biased towards high values of {\fesc}
(though we have not tested such a model here). Interestingly, the goal
of M20 was to provide a methodology for preferentially selecting high
{\fesc} galaxies, consistent with the toy model presented here. The
key point highlighted here is that we have shown the calculation of
{\fesc} to be sensitive to the underlying PDF (e.g. Figure
\ref{fig:fescdethist}), which is in turn appears to depend on sample
selection. Thus, a consideration of the {\fesc} PDF should be
considered in particular when comparing inferred {\fesc} values
between disparate samples (e.g. LAEs vs LBGs). 

We reiterate that the exponentially declining {\fesc} PDF favoured here
is simply an ad-hoc solution chosen for its bias towards low {\fesc}
values and a preference for {\fesc} = 0. This selection was motivated
by the low detection rate of such emission and the, generally, low
estimates of the average {\fesc} for large galaxy samples
\citep[e.g.][]{vanzella10,grazian16,smith18}. The
true functional form of the PDF of
{\fesc} is very likely more complex and may include dependencies on
galaxy properties such as, e.g., stellar mass
\citep{finkelstein19,naidu20}. Further clarification of this issue
will require larger samples of LyC detections at $z>3$ which would be
greatly aided by more sensitive instrumentation at $u$-band
wavelengths. It is also likely that
inputs from high-resolution, hydrodynamics simulations of high
redshift galaxies that include full radiative transfer can help
greatly with the interpretation of detections (and non-detections),
though running such simulations is computationally
expensive. Regardless, we show here evidence that the most
likely PDF for {\fesc} for LBG like galaxies favours a model with a
reasonable bias towards low {\fesc}. 

\subsection{Dust Attenuation}\label{section:dust}

To this point, we have avoided one key topic in the study of optical
and UV radiation from star-forming galaxies: dust
attenuation. In general, the level of attenuation at fixed E(B-V)
increases with decreasing $\lambda_{\rm rest}$ such that UV
wavelengths experience the highest levels of attenuation (i.e. lowest
transmission) irrespective of the functional form of the assumed
attenuation curve \citep[e.g.][etc]{gordon98,calz00,reddy16}. This
statement, of course, assumes that extending the chosen attenuation
curve to short wavelengths ($\lesssim$1500 {\AA}) is reasonable. We
acknowledge that \citet{buat02} have investigated dust attenuation at
900 {\AA} in a handful of local star-forming galaxies and
\citet{weingartner01} have explored theoretical models of dust
attenuation in a similar regime (based on the
Magellanic clouds), however there applicability to high redshift
galaxies is also uncertain. Regardless, the expected high level of
attenuation at LyC wavelengths may lead one to expect that galaxies
with a high enough LyC flux to be detected in current surveys should
be biased towards low attenuation. Indeed, LyC detections from KLCS
all have $\langle E(B-V) \rangle$ = 0.045 (and 0.129 for full LBG parent
sample, S18), and those of LACES all exhibit
negligible attenuation \citep[$E(B-V)$ $<$ 0.07, $\langle E(B-V) \rangle
$ $\simeq$ 0.01-0.03, F19,][]{nakajima20}. 

Regardless, to expect all LyC emitting galaxies to contain negligible
amounts of dust is likely too simplistic. Thus, some consideration of
the effects of dust in the interpretive framework for {\fesc}
calculations outlined in this paper is warranted. We advocate a
methodology similar to that outlined in F19. First, a determination of
the stellar E(B-V) value should be 
computed based on the available photometric data for a given sample of
objects. This can be achieved through full SED fitting or through
calibrations such as those based on the UV slope, $\beta$
\citep[e.g.][]{meurer99}. In the case of SED fitting, we advocate a
method only incorporating bands redward of Ly$\alpha$, as shorter
wavelengths are strongly affected by IGM attenuation (e.g. effects not
intrinsic to the galaxy) that should be treated independently to avoid
added degeneracy in the SED model. The effects of including or
omitting flux with wavelengths shortward of Ly$\alpha$ during the SED
fitting process will be tested in future work (Bassett et al., in
prep). The computed E(B-V) is combined with a
choice of dust attenuation curve, $k(\lambda)$, to correct the
observed 1500 {\AA} flux to the ``intrinsic'', dust-free,
value. Finally, the chosen input SED template for a given sample
(either computed through SED fitting or simply selecting a template
with a reasonable value of {\Lint}) is scaled to match the corrected
1500 {\AA} flux. Through this process, the intrinsic LyC flux can be
determined, noting this value will be dependent on the selection of
the intrinsic SED. 

From the intrinsic LyC flux calculated in this manner, one can then
determine the expected value of {\fesc} by comparing with the observed
value. In this way, any attenuation of LyC flux due to dust is
incorporated into the definition of {\fesc}, as pointed out by F19
(i.e. there is no distinction between dust attenuation and absorption
of LyC by neutral hydrogen). We also follow the methodology of F19 who
allow {\fesc} for a given value of $E(B-V)$ to only be as large as the
transmission allowed by the extrapolated dust attenuation curve. This
is reasonable as, in the case of a galaxy with relatively large
$E(B-V)$, a value of {\fesc} = 1.0 would imply zero
dust attenuation for LyC and high attenuation at 1500 {\AA}. We note,
however, that such a case is not entirely impossible given LyC
emission is often dominated by stellar populations with ages $<$ 10
Myr while stellar populations as old as a few hundred Myr can provide
significant flux at 1500 {\AA} \citep[e.g.][]{eldridge17}, thus the
emission at each wavelength may originate from different locations
within a given galaxy.

We do note,
however, that this maximum {\fesc} allowed by the assumed attenuation
curve is highly model dependent. For example a Small
Magellanic Cloud attenuation curve \citep[e.g.][]{gordon98} will have
a much higher attenuation at LyC wavelengths when compared to
either a \citet{calz00} or \citet{reddy16} attenuation curve with the
same 1500 {\AA} attenuation. This results from the fact that
the extension of the functional form of either a \citet{calz00} or
\citet{reddy16} $k(\lambda)$ is significantly flatter at $\lambda$ $<$
1500 {\AA} than that of \citet{gordon98}. This
caveat is important to keep in mind when considering the possible
``maximum'' {\fesc} allowable for a given value of $E(B-V)$. 

\subsection{LyC Detections at Other Redshifts}\label{section:lowz}

The study of LyC escape from galaxies in ground-based studies is
limited to redshifts $\gtrsim$ 2.8 due to the low atmospheric
transmission of UV photons. This limitation is not suffered by
space-based instrumentation, thus, studies of LyC at lower redshifts
can be performed at significantly lower redshifts with satellite 
instrumentation. In particular LyC has been detected at $z\sim2.5$ by
\citet{bian17} using the HST-WFC3 F275W filter, and recently at
$z=1.42$ by \citet{saha20} with the Ultra-Violet-Imaging Telescope
(UVIT) on board AstroSat. We also note that there has been significant
activity in spectroscopic detection of LyC from green pea galaxies at
$z\sim0.3-4$ with HST COS \citep[e.g.][]{Izotov16,izotov18b}. At such low redshifts, however,
IGM transmission should be negligible, thus these studies are not of
particular relevance to the study of $T_{\rm bias}$. 

\begin{figure}
  \includegraphics[width=\columnwidth]{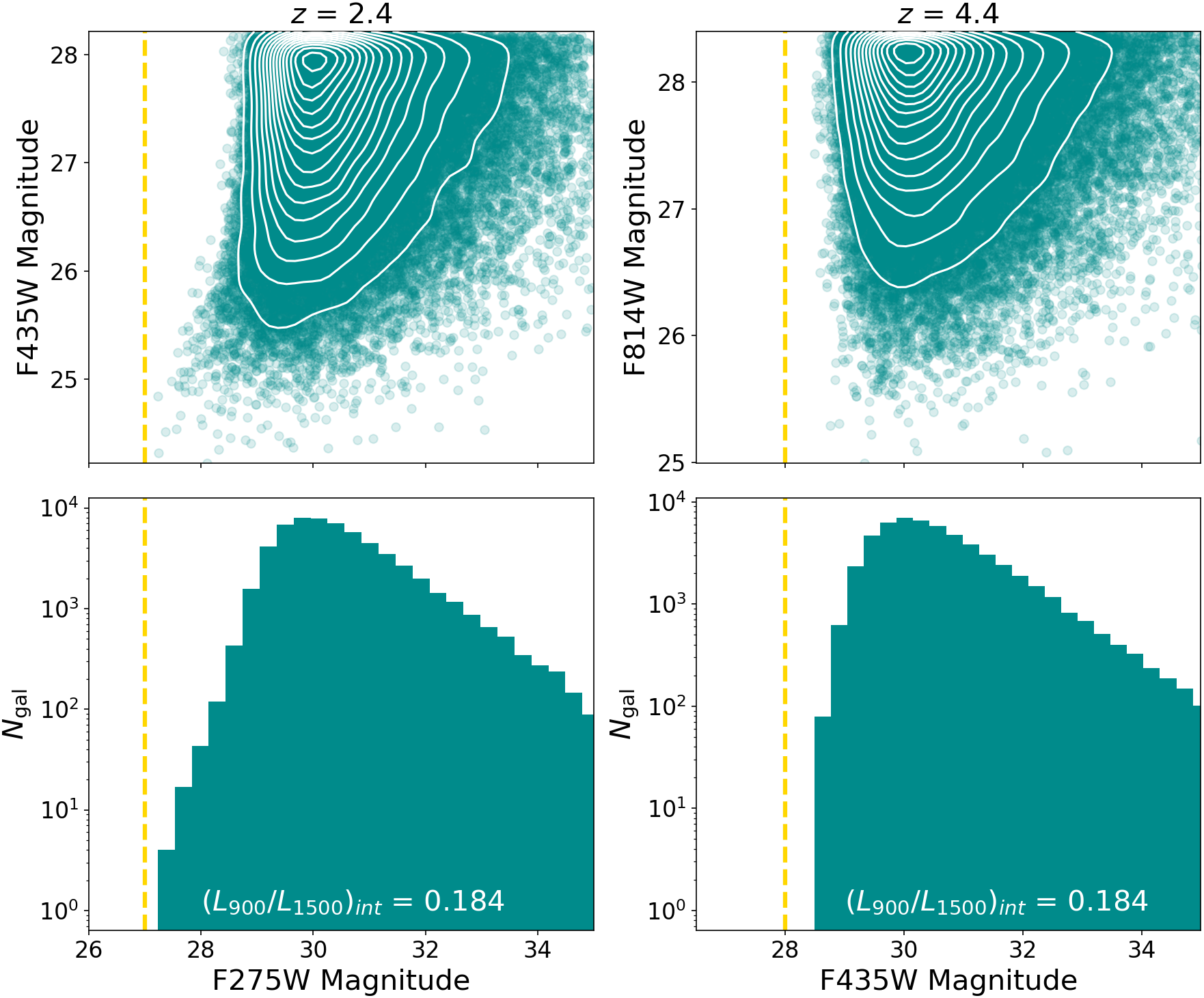}
  \caption{UV magnitudes of mock star-forming galaxies at $z=2.4$ and
    $z=4.4$ as
    observed by the UVCANDELS survey. Here 1500 {\AA} fluxes are
    sampled from the $z\sim2.5$ UV luminosity function of
    \citet{moutard20} and from the $z\sim4.0$ UV luminosity function
    of \citet{bouwens15} with depths matched to UVCANDELS. \textit{Top
      row:} LyC vs
    non-ionizing UV magnitudes of mock observations. \textit{Bottom row:} histograms
    of magnitudes for LyC probing bands. In all panels the vertical dashed line
    indicates the depths of UVCANDELS observations. Here mock galaxies are
    produced as dust-free, exponentially declining SFR BPASSv2.1 SEDs
    with {\Lint} = 0.184, and assuming a flat {\fesc} PDF. We note
    that a larger value of {\Lint} can produce a handful of individual LyC detected galaxies.}
  \label{fig:f275w}
\end{figure}

Here we focus on providing predictions for detection of LyC within the
Ultraviolet Imaging of the Cosmic Assembly Near-infrared Deep
Extragalactic Legacy Survey Fields (UVCANDELS; PI: Teplitz, PID
15647), a $\sim$430 arcmin$^{2}$, 164-orbit Cycle 26 UV HST
program. UVCANDELS will provide 3-orbit depth of WFC3/275W and
parallel ACS/F435W in four CANDELS fields: GOODS-N, GOODS-S, EGS, and
COSMOS. 

For our predictions we
follow a similar procedure outlined in Section \ref{section:detrate},
however here we have produced 10,000 IGM transmission curves at
both $z=2.4$ and $z=4.4$ for the purpose of providing mock observations of LyC using
the WFC3/F275W and ACS/F435W filters, respectively. To sample the input 1500 {\AA} fluxes or
this comparison we sample from UV luminosity functions of
\citet{moutard20} derived from the CLAUDS survey for $z=2.4$ and from
fits to B-band dropouts ($z \gtrsim 3.8$) from \citet{bouwens15} for
$z=4.4$. In both cases we use luminosity functions described by a Schechter
function with $\alpha$ = -1.4, $\phi^{*}$ = 2.708$\times10^{-3}$, and
$M^{*} = -20.623$ at $z=2.2$ and $\alpha$ = -1.64, $\phi^{*}$ = 1.97$\times10^{-3}$, and
$M^{*} = -20.88$ at $z=4.4$. For all mock galaxies we assume a value of {\Lint}
of 0.18. 

To sample the $\lambda_{\rm rest} \sim 1500$ {\AA} flux for
mock galaxies we measure our SED with the ACS/F435W and WFC3/F814W for
$z=2.4$ and $z=4.4$ respectively. The 5$\sigma$ depths of each filter are
matched to the observations at 27.0 mag for F275W, 28.0 for F435W, and
28.4 for F814W. For the best chance of
detecting LyC emission we produce roughly 100,000 mock observations of
galaxies at each redshift, significantly more than should be expected
in the UVCANDELS volume. We also assume a flat {\fesc} PDF, to further
increase the possibility of producing galaxies with very bright LyC flux.

The results of this test are shown in Figure \ref{fig:f275w}. The top
row shows the non-ionizing UV versus LyC magnitudes (observational
band is redshift dependent) and the bottom row
show the histograms of F275W and F435W magnitudes in logscale. In both panels we
show the magnitude limits of UVCANDELS for respective LyC probing
bands with a vertical dashed
line. For galaxies with {\Lint} = 0.18, the 5$\sigma$ limits of
UVCANDELS are too shallow to detect individual galaxies within the
UVCANDELS footprint as the brightest. In the event that UV
bright galaxies with significantly higher {\Lint} exist within the
UVCANDELS footprint, it may be possible that one or two individual
detections will be found. We conclude that pushing observations to a 
depth of 30 mag and beyond in small, targeted fields
\citep[i.e. similar to the $z=3.1$ observations of][]{fletcher19} is
likely to be more fruitful than shallow searches over large areas such as
UVCANDELS.

Given the expected faintness of LyC emission, wide area surveys such
as UVCANDELS will likely rely on stacking
analysis in order to estimate the average {\fesc} for galaxy
subsamples. In this scenario, prior information regarding the
likelihood of escaping LyC emission (e.g. evidence of a hard ionizing
spectrum or high Ly$\alpha$ escape, if available) will be
useful. This is due to the fact that, although high LyC flux is more
common for UV bright galaxies, galaxies with the same UV brightness
are also commonly found with relatively low LyC flux (as shown in Figure
\ref{fig:f275w}). Similarly, we find galaxies with relatively faint
non-ionizing UV flux with relatively high LyC flux. Thus, simply
stacking the galaxies with the highest 1500 {\AA} flux does not
guarantee that the galaxies with the highest LyC flux have been chosen.

Finally, we note a few caveats to this analysis. First, this analysis
has been performed using $T_{\rm 
  IGM}$ curves produced independently, while UVCANDELS covers 4 individual
fields. In the event that there is strong correlation in $T_{\rm IGM}$
across the field, the resulting LyC fluxes may be systematically
higher (in the case of high $T_{\rm IGM}$) or lower (in the case of
low $T_{\rm IGM}$) for one particular field. Consideration of the correlation of $T_{\rm IGM}$
between sources in individual fields of a given area may require
dedicated analysis of large scale simulations of the HI distributionns
and is beyond the scope of this work. And second, this analysis has ignored
variation in {\Lint} (as noted), assumed no dust attenuation, and
employed a flat {\fesc} PDF,  all three of which will affect our
resulting LyC fluxes.

\subsection{Observed vs True {\fesc}}\label{section:3dv2d}

Ultimately the ongoing search for LyC emission from high redshift
galaxies is closely connected with our understanding what types of
galaxies are responsible for reionizing the universe. Characterising
the population of strong LyC emitters will be key to informing our
picture of the topological evolution of ionized regions during the
EoR \citep{seiler18}. There exists, however, an inherent difficulty
regarding the interpretation of {\fesc} values measured
observationally due to the complex geometry of LyC escape from
galaxies, independent of $T_{\rm IGM}$. Indeed various models have been hypothesised that may
provide slightly different interpretations of the detected LyC flux in
the context of measuring {\fesc}. A detailed
discussion of various LyC escape models can be found in Section 9.4 of
S18. 

Crucially, it has been pointed out \citep[e.g.][]{bassett19, barrow20} that the detection
of LyC from any individual galaxy is reflective of only the fraction
of LyC that is able to escape into our single line-of-sight. There is
still no reliable way of inferring if the observed {\fesc} value is
reflective of {\fesc} in all directions, i.e. the 3D {\fesc}
\citep[though intriguing indirect measurement techniques for the 3D {\fesc} have been
proposed, which warrant further exploration within an anisotropic
LyC escape scenario, e.g.][]{zackrisson13,yamanaka20}. Similarly, the
lack of LyC emission from any individual galaxy is not evidence of
{\fesc} = 0.0 as large quantities of LyC photons could be escaping in
directions other than our line-of-sight. As we have shown, the value
of {\fesc} is dependent on the assumed underlying PDF and current
detection rates may disfavour a flat distribution. One way to provide
a theoretically sound basis for our assumptions on the PDF of {\fesc}
for galaxies or galaxy samples is through the careful consideration of
high resolution hydrodynamical simulations.

LyC escape can be measured in such simulations by applying full
radiative transfer, then measuring {\fesc} from a large number of
sight lines towards the galaxy. This method provides the full three
dimensional {\fesc} at a given time and has shown that even for
individual galaxies {\fesc} is highly variable and can swing from 0 to
1 within 100 Myr \citep{paardekooper15,trebitsch17,rosdahl18} though
there may be some mass dependence
on the 3D {\fesc} PDF. It has been shown, however, that for a galaxy
with given 3D {\fesc} value the value of {\fesc} in any particular
sightline may vary from 0 to values larger than the true 3D value
\citep[e.g.][Figure 13]{paardekooper15}. Thus, to construct the
underlying {\fesc} PDF for a given sample of galaxies may require the
combination of the 3D {\fesc} of galaxies (with possible dependencies
on mass or other properties) with the probability distribution of 2D
{\fesc} (line-of-sight) for a given 3D {\fesc} value. Disentangling
the various dependencies on these underlying PDFs will require suites
of high resolution simulations with full radiative transfer, but is of
the utmost importance in interpreting the 2D {\fesc} values from
observations with the true 3D {\fesc} distributions. Ultimately it is
the full 3D {\fesc} values from galaxies that are of interest in the
context of the EoR, which can only be connected to our 2D
observational results through such a complex line of reasoning as is
described here. 

\section{Summary and Conclusions}\label{section:conclusions}

In this paper we have explored the level of bias in the IGM
transmission, {\Tigm}, for galaxies with LyC detections at $z$=3-4
under the observational limits imposed by current instruments and
surveys. Our tests were performed by simulating one million IGM
transmission functions in our redshift range of interest and applying
these to empirically motivated mock galaxy spectra constructed from
the BPASSv2.1 models \citep{eldridge17}. We have also tested how the
level of IGM transmission bias, $T_{\rm bias}$,
depends on both the assumed probability distribution function, PDF, of {\fesc}
and SED shape (which controls {\Lint}, a key value for measuring
{\fesc}). Our analysis has included modeling designed to approximate
both spectroscopic and photometric LyC detections from recent surveys
of \citet{steidel18}, \citet{fletcher19}, and
\citet{mestric20}.

Broadly, we find that, in all cases the average value of {\Tigm} at
LyC wavelengths for galaxies with LyC detections is found to be larger
than the average {\Tigm} for all simulated sightlines at the same
redshift. This results from the fact that the
underlying {\Tigm} distribution at 880 {\AA} $<$ $\lambda_{\rm rest}$
$<$ 910 {\AA}
is bimodal with the stronger peak at {\Tigm} = 0, but the simple
fact that the galaxy has been detected means that {\Tigm} $\neq$
0. Thus, the {\Tigm} distribution for LyC \textit{detected} galaxies
is unimodal with a peak at relatively high {\Tigm}, while the mean
for all sightlines falls below this due to the inclusion of the
{\Tigm} = 0 peak. The result is that the assumption of a mean
{\Tigm} for all sightlines when calculating $\langle${\fesc}$\rangle$
for a sample of LyC detected galaxies results in an
overestimate of the true value. This result is similar to the recent
results of \citet{byrohl20} for Ly$\alpha$ transmission. Thus, it is
becoming clear that, while tempting, using a single statistic (e.g. median or
mean) when considering {\Tigm} for individual objects provides
misleading results for LyC detected samples. Considering samples which include (are
composed entirely of) LyC non-detected galaxies, the use of {\Tm} when
calculating upper limits on $\langle${\fesc}$\rangle$ is
appropriate, however. The remainder of our conclusions can be
summarised as follows:

\begin{itemize}
  \item Assuming the an LBG-like UV flux distribution and applying
    detection limits of \citet{steidel18}, \citet{fletcher19}, and
    \citet{mestric20} we estimate minimum levels of $T_{\rm bias}$ to be
    $\sim$0.15, $\sim$0.11, and $\sim$0.22 for each survey,
    respectively.
  \item In the case of a UV flux distribution more characteristic of
    LAE sample \citep[e.g. those of][]{fletcher19} a higher
    $T_{\rm bias}$ should be expected. In this case, mock HST F336W
    observations similar to \citet{fletcher19}, the minimum $T_{\rm
      bias}$ increases to $\sim$0.21.
  \item We have shown in Section \ref{section:fescdist} that, although
    $T_{\rm bias}$ does not increase significantly assuming an {\fesc}
    PDF mildly biased towards 0, there may be a slight decrease in the
    recovered {\fesc} value in such a model.
  \item We have also demonstrated that the current detection rates of
    LyC radiation from surveys may reflect information regarding the
    underlying {\fesc} PDF. Our simplified model presented in Section
    \ref{section:detrate}, for example, appears to slightly disfavour a flat
    {\fesc} PDF for LBGs \citep[e.g.][]{steidel18}, though this may
    not be the case for LAE samples \citep[e.g.][]{fletcher19}.
\end{itemize}

This final point may suggest that fainter galaxies, represented by LAE
samples, are more likely to exhibit a higher {\fesc} than bright
galaxies, represented by LBGs. Such a scenario is in agreement with
other recent studies \citep[e.g.][]{finkelstein19}. Our comparisons in
this context in Section \ref{section:detrate} with the detection rates
of \citet{steidel18} and \citet{fletcher19} are still in the realm of
low statistical significance. Thus, confirmation of these results will
require larger samples of LyC detected galaxies on which to perform a
similar analysis.

Of course, all of our results will depend on the various input parameters of
our models including the assumed distribution of 1500 {\AA}
(rest-frame) fluxes, our treatment (or lack thereof) of dust
attenuation, our assumptions regarding the intrinsic luminosity ratio
({\Lint}) of galaxies, and even the details of our methods for
producing {\Tigm} functions (e.g. HI distribution functions). Thus,
we do not claim that the absolute values of $T_{\rm bias}$ from this work
to be in any way definitive. The purpose of this work is to highlight
the ways in which different assumptions regarding the underlying
distributions of {\Tigm} and {\fesc} affect our attempts to estimate
{\fesc} from galaxies. It is clear that significant theoretical work is still
required to better understand these PDFs that are critical to our
interpretation of LyC detections from observations.

From an observational point of view, it is also clear that larger
samples of LyC detections will be essential in disentangling the
various dependencies on {\fesc} (e.g. stellar mass, SFR, etc). It is
possible that more efficient searches can be conducted in the near
future with a focus on increasing both depth and
field-of-view (FOV). Indeed, we find the highest detection rates among our
mock surveys for our mock LACES survey \citep[][Section \ref{section:detrate}]{fletcher19}, primarily due to
those observations reaching 30.24 mag. The drawback is that this study
is performed with WFC3, an instrument with a relatively small
FOV. One possible future instrument that may push LyC surveys to the
next level is the Keck Wide Field Imager \citep[KWFI,][]{gillingham20} that is expected to
achieve a signal to noise of $\sim$2 at 30th magnitude in the $u$-band
across a 1 degree diameter FOV in just under 8 hours of exposures
(private communication). From our mock LACES survey, we estimate that
$\sim$50\% of all simulated galaxies fall in the magnitude range
between 30 and 32. Thus, the era of large samples of known LyC
emitting galaxies may be near.

\section*{Data Availability}

Simulated data used in this work is produced primarily using publicly
available codes found at https://github.com/robbassett as well as
publicly available galaxy SED models from the BPASS collaboration
\citep{eldridge17}. Observational data used for comparison is
available from publications associated with those surveys.

\section*{Acknowledgements}

This research was conducted by the
Australian Research Council Centre of Excellence for All Sky
Astrophysics in 3 Dimensions (ASTRO 3D), through project
number CE170100013.  The authors wish to thank Chris Blake, Adam
Batten, and Katinka Ger\'{e}b for useful and illuminating discussions.
We also wish to thank our referee, Akio K. Inoue, for careful
consideration of the manuscript, which has resulted in an improved
focus within the context of current studies exploring LyC emission
from galaxies at high redshift. Results
presented in this work have made extensive use of the python3
programming language \citep{python3} and, in particular, the authors wish to
acknowledge the the numpy \citep{numpy}, matplotlib \citep{matplotlib},
and scipy \citep{scipy} packages. MR
and LP acknowledge support from HST programs 15100 and 15647. Support
for Program numbers 15100 and 15647 were provided by NASA through a
grant from the Space Telescope Science Institute, which is operated by
the Association of Universities for Research in Astronomy,
Incorporated, under NASA contract NAS5-26555.

%%%%%%%%%%%%%%%%%%%%%%%%%%%%%%%%%%%%%%%%%%%%%%%%%%



\bibliographystyle{mnras}
\bibliography{refs} % if your bibtex file is called example.bib

\begin{thebibliography}{}
\makeatletter
\relax
\def\mn@urlcharsother{\let\do\@makeother \do\$\do\&\do\#\do\^\do\_\do\%\do\~}
\def\mn@doi{\begingroup\mn@urlcharsother \@ifnextchar [ {\mn@doi@}
  {\mn@doi@[]}}
\def\mn@doi@[#1]#2{\def\@tempa{#1}\ifx\@tempa\@empty \href
  {http://dx.doi.org/#2} {doi:#2}\else \href {http://dx.doi.org/#2} {#1}\fi
  \endgroup}
\def\mn@eprint#1#2{\mn@eprint@#1:#2::\@nil}
\def\mn@eprint@arXiv#1{\href {http://arxiv.org/abs/#1} {{\tt arXiv:#1}}}
\def\mn@eprint@dblp#1{\href {http://dblp.uni-trier.de/rec/bibtex/#1.xml}
  {dblp:#1}}
\def\mn@eprint@#1:#2:#3:#4\@nil{\def\@tempa {#1}\def\@tempb {#2}\def\@tempc
  {#3}\ifx \@tempc \@empty \let \@tempc \@tempb \let \@tempb \@tempa \fi \ifx
  \@tempb \@empty \def\@tempb {arXiv}\fi \@ifundefined
  {mn@eprint@\@tempb}{\@tempb:\@tempc}{\expandafter \expandafter \csname
  mn@eprint@\@tempb\endcsname \expandafter{\@tempc}}}

\bibitem[\protect\citeauthoryear{{Barrow}, {Robertson}, {Ellis}, {Nakajima},
  {Saxena}, {Stark}  \& {Tang}}{{Barrow} et~al.}{2020}]{barrow20}
{Barrow} K. S.~S.,  {Robertson} B.~E.,  {Ellis} R.~S.,  {Nakajima} K.,
  {Saxena} A.,  {Stark} D.~P.,   {Tang} M.,  2020, arXiv e-prints, \href
  {https://ui.adsabs.harvard.edu/abs/2020arXiv201000592B} {p. arXiv:2010.00592}

\bibitem[\protect\citeauthoryear{{Bassett} et~al.,}{{Bassett}
  et~al.}{2019}]{bassett19}
{Bassett} R.,  et~al., 2019, \mn@doi [\mnras] {10.1093/mnras/sty3320}, \href
  {https://ui.adsabs.harvard.edu/abs/2019MNRAS.483.5223B} {483, 5223}

\bibitem[\protect\citeauthoryear{{Becker}, {Hewett}, {Worseck}  \&
  {Prochaska}}{{Becker} et~al.}{2013}]{becker13}
{Becker} G.~D.,  {Hewett} P.~C.,  {Worseck} G.,   {Prochaska} J.~X.,  2013,
  \mn@doi [\mnras] {10.1093/mnras/stt031}, \href
  {https://ui.adsabs.harvard.edu/abs/2013MNRAS.430.2067B} {430, 2067}

\bibitem[\protect\citeauthoryear{{Bershady}, {Charlton}  \&
  {Geoffroy}}{{Bershady} et~al.}{1999}]{bershady99}
{Bershady} M.~A.,  {Charlton} J.~C.,   {Geoffroy} J.~M.,  1999, \mn@doi [\apj]
  {10.1086/307257}, \href
  {https://ui.adsabs.harvard.edu/abs/1999ApJ...518..103B} {518, 103}

\bibitem[\protect\citeauthoryear{{Bian} \& {Fan}}{{Bian} \&
  {Fan}}{2020}]{bian20}
{Bian} F.,  {Fan} X.,  2020, \mn@doi [\mnras] {10.1093/mnrasl/slaa007}, \href
  {https://ui.adsabs.harvard.edu/abs/2020MNRAS.493L..65B} {493, L65}

\bibitem[\protect\citeauthoryear{{Bian}, {Fan}, {McGreer}, {Cai}  \&
  {Jiang}}{{Bian} et~al.}{2017}]{bian17}
{Bian} F.,  {Fan} X.,  {McGreer} I.,  {Cai} Z.,   {Jiang} L.,  2017, \mn@doi
  [\apjl] {10.3847/2041-8213/aa5ff7}, \href
  {http://adsabs.harvard.edu/abs/2017ApJ...837L..12B} {837, L12}

\bibitem[\protect\citeauthoryear{{Boutsia} et~al.,}{{Boutsia}
  et~al.}{2011}]{boutsia11}
{Boutsia} K.,  et~al., 2011, \mn@doi [\apj] {10.1088/0004-637X/736/1/41}, \href
  {https://ui.adsabs.harvard.edu/abs/2011ApJ...736...41B} {736, 41}

\bibitem[\protect\citeauthoryear{{Bouwens}, {Illingworth}, {Oesch}, {Caruana},
  {Holwerda}, {Smit}  \& {Wilkins}}{{Bouwens} et~al.}{2015}]{bouwens15}
{Bouwens} R.~J.,  {Illingworth} G.~D.,  {Oesch} P.~A.,  {Caruana} J.,
  {Holwerda} B.,  {Smit} R.,   {Wilkins} S.,  2015, \mn@doi [\apj]
  {10.1088/0004-637X/811/2/140}, \href
  {http://adsabs.harvard.edu/abs/2015ApJ...811..140B} {811, 140}

\bibitem[\protect\citeauthoryear{{Bouwens}, {Smit}, {Labb{\'e}}, {Franx},
  {Caruana}, {Oesch}, {Stefanon}  \& {Rasappu}}{{Bouwens}
  et~al.}{2016}]{bouwens16}
{Bouwens} R.~J.,  {Smit} R.,  {Labb{\'e}} I.,  {Franx} M.,  {Caruana} J.,
  {Oesch} P.,  {Stefanon} M.,   {Rasappu} N.,  2016, \mn@doi [\apj]
  {10.3847/0004-637X/831/2/176}, \href
  {http://adsabs.harvard.edu/abs/2016ApJ...831..176B} {831, 176}

\bibitem[\protect\citeauthoryear{{Buat}, {Burgarella}, {Deharveng}  \&
  {Kunth}}{{Buat} et~al.}{2002}]{buat02}
{Buat} V.,  {Burgarella} D.,  {Deharveng} J.~M.,   {Kunth} D.,  2002, \mn@doi
  [\aap] {10.1051/0004-6361:20020925}, \href
  {https://ui.adsabs.harvard.edu/abs/2002A&A...393...33B} {393, 33}

\bibitem[\protect\citeauthoryear{{Byrohl} \& {Gronke}}{{Byrohl} \&
  {Gronke}}{2020}]{byrohl20}
{Byrohl} C.,  {Gronke} M.,  2020, arXiv e-prints, \href
  {https://ui.adsabs.harvard.edu/abs/2020arXiv200610041B} {p. arXiv:2006.10041}

\bibitem[\protect\citeauthoryear{{Calzetti}, {Armus}, {Bohlin}, {Kinney},
  {Koornneef}  \& {Storchi-Bergmann}}{{Calzetti} et~al.}{2000}]{calz00}
{Calzetti} D.,  {Armus} L.,  {Bohlin} R.~C.,  {Kinney} A.~L.,  {Koornneef} J.,
   {Storchi-Bergmann} T.,  2000, \mn@doi [\apj] {10.1086/308692}, \href
  {http://adsabs.harvard.edu/abs/2000ApJ...533..682C} {533, 682}

\bibitem[\protect\citeauthoryear{{Carswell} \& {Webb}}{{Carswell} \&
  {Webb}}{2014}]{carswell14}
{Carswell} R.~F.,  {Webb} J.~K.,  2014, {VPFIT: Voigt profile fitting program}
  (\mn@eprint {ascl} {1408.015})

\bibitem[\protect\citeauthoryear{{Eldridge}, {Stanway}, {Xiao}, {McClelland},
  {Taylor}, {Ng}, {Greis}  \& {Bray}}{{Eldridge} et~al.}{2017}]{eldridge17}
{Eldridge} J.~J.,  {Stanway} E.~R.,  {Xiao} L.,  {McClelland} L.~A.~S.,
  {Taylor} G.,  {Ng} M.,  {Greis} S.~M.~L.,   {Bray} J.~C.,  2017, \mn@doi
  [\pasa] {10.1017/pasa.2017.51}, \href
  {http://adsabs.harvard.edu/abs/2017PASA...34...58E} {34, e058}

\bibitem[\protect\citeauthoryear{{Fan}, {Carilli}  \& {Keating}}{{Fan}
  et~al.}{2006}]{fan06}
{Fan} X.,  {Carilli} C.~L.,   {Keating} B.,  2006, \mn@doi [\araa]
  {10.1146/annurev.astro.44.051905.092514}, \href
  {http://adsabs.harvard.edu/abs/2006ARA%26A..44..415F} {44, 415}

\bibitem[\protect\citeauthoryear{{Fern{\'a}ndez-Soto}, {Lanzetta}  \&
  {Chen}}{{Fern{\'a}ndez-Soto} et~al.}{2003}]{fernandez-soto03}
{Fern{\'a}ndez-Soto} A.,  {Lanzetta} K.~M.,   {Chen} H.~W.,  2003, \mn@doi
  [\mnras] {10.1046/j.1365-8711.2003.06622.x}, \href
  {https://ui.adsabs.harvard.edu/abs/2003MNRAS.342.1215F} {342, 1215}

\bibitem[\protect\citeauthoryear{{Finkelstein} et~al.,}{{Finkelstein}
  et~al.}{2019}]{finkelstein19}
{Finkelstein} S.~L.,  et~al., 2019, \mn@doi [\apj] {10.3847/1538-4357/ab1ea8},
  \href {https://ui.adsabs.harvard.edu/abs/2019ApJ...879...36F} {879, 36}

\bibitem[\protect\citeauthoryear{{Fletcher}, {Tang}, {Robertson}, {Nakajima},
  {Ellis}, {Stark}  \& {Inoue}}{{Fletcher} et~al.}{2019}]{fletcher19}
{Fletcher} T.~J.,  {Tang} M.,  {Robertson} B.~E.,  {Nakajima} K.,  {Ellis}
  R.~S.,  {Stark} D.~P.,   {Inoue} A.,  2019, \mn@doi [\apj]
  {10.3847/1538-4357/ab2045}, \href
  {https://ui.adsabs.harvard.edu/abs/2019ApJ...878...87F} {878, 87}

\bibitem[\protect\citeauthoryear{{Forrest} et~al.,}{{Forrest}
  et~al.}{2017}]{forrest17}
{Forrest} B.,  et~al., 2017, \mn@doi [\apjl] {10.3847/2041-8213/aa653b}, \href
  {http://adsabs.harvard.edu/abs/2017ApJ...838L..12F} {838, L12}

\bibitem[\protect\citeauthoryear{{Giallongo}, {Cristiani}, {D'Odorico}  \&
  {Fontana}}{{Giallongo} et~al.}{2002}]{giallongo02}
{Giallongo} E.,  {Cristiani} S.,  {D'Odorico} S.,   {Fontana} A.,  2002,
  \mn@doi [\apjl] {10.1086/340254}, \href
  {https://ui.adsabs.harvard.edu/abs/2002ApJ...568L...9G} {568, L9}

\bibitem[\protect\citeauthoryear{{Gillingham}, {Cooke}, {Glazebrook}, {Mould},
  {Smith}  \& {Steidel}}{{Gillingham} et~al.}{2020}]{gillingham20}
{Gillingham} P.,  {Cooke} J.,  {Glazebrook} K.,  {Mould} J.,  {Smith} R.,
  {Steidel} C.,  2020, in \procspie. p. 112030F, \mn@doi{10.1117/12.2540717}

\bibitem[\protect\citeauthoryear{{Gordon} \& {Clayton}}{{Gordon} \&
  {Clayton}}{1998}]{gordon98}
{Gordon} K.~D.,  {Clayton} G.~C.,  1998, \mn@doi [\apj] {10.1086/305774}, \href
  {http://adsabs.harvard.edu/abs/1998ApJ...500..816G} {500, 816}

\bibitem[\protect\citeauthoryear{{Grazian} et~al.,}{{Grazian}
  et~al.}{2016}]{grazian16}
{Grazian} A.,  et~al., 2016, \mn@doi [\aap] {10.1051/0004-6361/201526396},
  \href {http://adsabs.harvard.edu/abs/2016A%26A...585A..48G} {585, A48}

\bibitem[\protect\citeauthoryear{{Greig} \& {Mesinger}}{{Greig} \&
  {Mesinger}}{2017}]{greig17}
{Greig} B.,  {Mesinger} A.,  2017, \mn@doi [\mnras] {10.1093/mnras/stx2118},
  \href {https://ui.adsabs.harvard.edu/abs/2017MNRAS.472.2651G} {472, 2651}

\bibitem[\protect\citeauthoryear{{Hasinger} et~al.,}{{Hasinger}
  et~al.}{2018}]{hasinger18}
{Hasinger} G.,  et~al., 2018, \mn@doi [\apj] {10.3847/1538-4357/aabacf}, \href
  {https://ui.adsabs.harvard.edu/abs/2018ApJ...858...77H} {858, 77}

\bibitem[\protect\citeauthoryear{{Hopkins}, {Richards}  \&
  {Hernquist}}{{Hopkins} et~al.}{2007}]{hopkins07}
{Hopkins} P.~F.,  {Richards} G.~T.,   {Hernquist} L.,  2007, \mn@doi [\apj]
  {10.1086/509629}, \href {http://adsabs.harvard.edu/abs/2007ApJ...654..731H}
  {654, 731}

\bibitem[\protect\citeauthoryear{{Hui} \& {Rutledge}}{{Hui} \&
  {Rutledge}}{1999}]{hui99}
{Hui} L.,  {Rutledge} R.~E.,  1999, \mn@doi [\apj] {10.1086/307202}, \href
  {https://ui.adsabs.harvard.edu/abs/1999ApJ...517..541H} {517, 541}

\bibitem[\protect\citeauthoryear{Hunter}{Hunter}{2007}]{matplotlib}
Hunter J.~D.,  2007, \mn@doi [Computing in Science \& Engineering]
  {10.1109/MCSE.2007.55}, 9, 90

\bibitem[\protect\citeauthoryear{{Inoue} \& {Iwata}}{{Inoue} \&
  {Iwata}}{2008}]{inoue08}
{Inoue} A.~K.,  {Iwata} I.,  2008, \mn@doi [\mnras]
  {10.1111/j.1365-2966.2008.13350.x}, \href
  {https://ui.adsabs.harvard.edu/abs/2008MNRAS.387.1681I} {387, 1681}

\bibitem[\protect\citeauthoryear{{Inoue}, {Iwata}, {Deharveng}, {Buat}  \&
  {Burgarella}}{{Inoue} et~al.}{2005}]{inoue05}
{Inoue} A.~K.,  {Iwata} I.,  {Deharveng} J.-M.,  {Buat} V.,   {Burgarella} D.,
  2005, \mn@doi [\aap] {10.1051/0004-6361:20041769}, \href
  {http://adsabs.harvard.edu/abs/2005A%26A...435..471I} {435, 471}

\bibitem[\protect\citeauthoryear{{Inoue}, {Iwata}  \& {Deharveng}}{{Inoue}
  et~al.}{2006}]{inoue06}
{Inoue} A.~K.,  {Iwata} I.,   {Deharveng} J.-M.,  2006, \mn@doi [\mnras]
  {10.1111/j.1745-3933.2006.00195.x}, \href
  {https://ui.adsabs.harvard.edu/abs/2006MNRAS.371L...1I} {371, L1}

\bibitem[\protect\citeauthoryear{{Inoue} et~al.,}{{Inoue}
  et~al.}{2011}]{inoue11}
{Inoue} A.~K.,  et~al., 2011, \mn@doi [\mnras]
  {10.1111/j.1365-2966.2010.17851.x}, \href
  {https://ui.adsabs.harvard.edu/abs/2011MNRAS.411.2336I} {411, 2336}

\bibitem[\protect\citeauthoryear{{Inoue}, {Shimizu}, {Iwata}  \&
  {Tanaka}}{{Inoue} et~al.}{2014}]{inoue14}
{Inoue} A.~K.,  {Shimizu} I.,  {Iwata} I.,   {Tanaka} M.,  2014, \mn@doi
  [\mnras] {10.1093/mnras/stu936}, \href
  {http://adsabs.harvard.edu/abs/2014MNRAS.442.1805I} {442, 1805}

\bibitem[\protect\citeauthoryear{{Iwata} et~al.,}{{Iwata}
  et~al.}{2009}]{iwata09}
{Iwata} I.,  et~al., 2009, \mn@doi [\apj] {10.1088/0004-637X/692/2/1287}, \href
  {https://ui.adsabs.harvard.edu/abs/2009ApJ...692.1287I} {692, 1287}

\bibitem[\protect\citeauthoryear{{Izotov}, {Schaerer}, {Thuan}, {Worseck},
  {Guseva}, {Orlitov{\'a}}  \& {Verhamme}}{{Izotov} et~al.}{2016}]{Izotov16}
{Izotov} Y.~I.,  {Schaerer} D.,  {Thuan} T.~X.,  {Worseck} G.,  {Guseva} N.~G.,
   {Orlitov{\'a}} I.,   {Verhamme} A.,  2016, \mn@doi [\mnras]
  {10.1093/mnras/stw1205}, \href
  {http://adsabs.harvard.edu/abs/2016MNRAS.461.3683I} {461, 3683}

\bibitem[\protect\citeauthoryear{{Izotov}, {Worseck}, {Schaerer}, {Guseva},
  {Thuan}, {Fricke}, {Verhamme}  \& {Orlitov{\'a}}}{{Izotov}
  et~al.}{2018}]{izotov18b}
{Izotov} Y.~I.,  {Worseck} G.,  {Schaerer} D.,  {Guseva} N.~G.,  {Thuan} T.~X.,
   {Fricke} K.~J.,  {Verhamme} A.,   {Orlitov{\'a}} I.,  2018, \mn@doi [\mnras]
  {10.1093/mnras/sty1378}, \href
  {http://adsabs.harvard.edu/abs/2018MNRAS.tmp.1318I} {}

\bibitem[\protect\citeauthoryear{{Janknecht}, {Reimers}, {Lopez}  \&
  {Tytler}}{{Janknecht} et~al.}{2006}]{janknecht06}
{Janknecht} E.,  {Reimers} D.,  {Lopez} S.,   {Tytler} D.,  2006, \mn@doi
  [\aap] {10.1051/0004-6361:20065372}, \href
  {https://ui.adsabs.harvard.edu/abs/2006A&A...458..427J} {458, 427}

\bibitem[\protect\citeauthoryear{{Japelj} et~al.,}{{Japelj}
  et~al.}{2017}]{japelj17}
{Japelj} J.,  et~al., 2017, \mn@doi [\mnras] {10.1093/mnras/stx477}, \href
  {https://ui.adsabs.harvard.edu/abs/2017MNRAS.468..389J} {468, 389}

\bibitem[\protect\citeauthoryear{{Kakiichi} \& {Dijkstra}}{{Kakiichi} \&
  {Dijkstra}}{2018}]{kakiichi18b}
{Kakiichi} K.,  {Dijkstra} M.,  2018, \mn@doi [\mnras] {10.1093/mnras/sty2214},
  \href {https://ui.adsabs.harvard.edu/abs/2018MNRAS.480.5140K} {480, 5140}

\bibitem[\protect\citeauthoryear{{Kakiichi} et~al.,}{{Kakiichi}
  et~al.}{2018}]{kakiichi18a}
{Kakiichi} K.,  et~al., 2018, \mn@doi [\mnras] {10.1093/mnras/sty1318}, \href
  {https://ui.adsabs.harvard.edu/abs/2018MNRAS.479...43K} {479, 43}

\bibitem[\protect\citeauthoryear{{Kimm} \& {Cen}}{{Kimm} \&
  {Cen}}{2014}]{kimm14}
{Kimm} T.,  {Cen} R.,  2014, \mn@doi [\apj] {10.1088/0004-637X/788/2/121},
  \href {https://ui.adsabs.harvard.edu/abs/2014ApJ...788..121K} {788, 121}

\bibitem[\protect\citeauthoryear{{Ma}, {Quataert}, {Wetzel}, {Hopkins},
  {Faucher-Gigu{\`e}re}  \& {Kere{\v{s}}}}{{Ma} et~al.}{2020}]{ma20}
{Ma} X.,  {Quataert} E.,  {Wetzel} A.,  {Hopkins} P.~F.,  {Faucher-Gigu{\`e}re}
  C.-A.,   {Kere{\v{s}}} D.,  2020, arXiv e-prints, \href
  {https://ui.adsabs.harvard.edu/abs/2020arXiv200305945M} {p. arXiv:2003.05945}

\bibitem[\protect\citeauthoryear{{Mason}, {Treu}, {Dijkstra}, {Mesinger},
  {Trenti}, {Pentericci}, {de Barros}  \& {Vanzella}}{{Mason}
  et~al.}{2018}]{mason18}
{Mason} C.~A.,  {Treu} T.,  {Dijkstra} M.,  {Mesinger} A.,  {Trenti} M.,
  {Pentericci} L.,  {de Barros} S.,   {Vanzella} E.,  2018, \mn@doi [\apj]
  {10.3847/1538-4357/aab0a7}, \href
  {https://ui.adsabs.harvard.edu/abs/2018ApJ...856....2M} {856, 2}

\bibitem[\protect\citeauthoryear{{Meiksin}}{{Meiksin}}{2006}]{meiksin06}
{Meiksin} A.,  2006, \mn@doi [\mnras] {10.1111/j.1365-2966.2005.09756.x}, \href
  {https://ui.adsabs.harvard.edu/abs/2006MNRAS.365..807M} {365, 807}

\bibitem[\protect\citeauthoryear{{Meurer}, {Heckman}  \& {Calzetti}}{{Meurer}
  et~al.}{1999}]{meurer99}
{Meurer} G.~R.,  {Heckman} T.~M.,   {Calzetti} D.,  1999, \mn@doi [\apj]
  {10.1086/307523}, \href {http://adsabs.harvard.edu/abs/1999ApJ...521...64M}
  {521, 64}

\bibitem[\protect\citeauthoryear{{Me{\v{s}}tri{\'c}}
  et~al.,}{{Me{\v{s}}tri{\'c}} et~al.}{2020}]{mestric20}
{Me{\v{s}}tri{\'c}} U.,  et~al., 2020, \mn@doi [\mnras]
  {10.1093/mnras/staa920}, \href
  {https://ui.adsabs.harvard.edu/abs/2020MNRAS.494.4986M} {494, 4986}

\bibitem[\protect\citeauthoryear{{Micheva}, {Iwata}, {Inoue}, {Matsuda},
  {Yamada}  \& {Hayashino}}{{Micheva} et~al.}{2017}]{micheva17}
{Micheva} G.,  {Iwata} I.,  {Inoue} A.~K.,  {Matsuda} Y.,  {Yamada} T.,
  {Hayashino} T.,  2017, \mn@doi [\mnras] {10.1093/mnras/stw2700}, \href
  {https://ui.adsabs.harvard.edu/abs/2017MNRAS.465..316M} {465, 316}

\bibitem[\protect\citeauthoryear{{M{\o}ller} \& {Jakobsen}}{{M{\o}ller} \&
  {Jakobsen}}{1990}]{moller90}
{M{\o}ller} P.,  {Jakobsen} P.,  1990, \aap, \href
  {https://ui.adsabs.harvard.edu/abs/1990A&A...228..299M} {228, 299}

\bibitem[\protect\citeauthoryear{{Momcheva} et~al.,}{{Momcheva}
  et~al.}{2016}]{momcheva16}
{Momcheva} I.~G.,  et~al., 2016, \mn@doi [\apjs] {10.3847/0067-0049/225/2/27},
  \href {https://ui.adsabs.harvard.edu/abs/2016ApJS..225...27M} {225, 27}

\bibitem[\protect\citeauthoryear{{Moutard}, {Sawicki}, {Arnouts}, {Golob},
  {Coupon}, {Ilbert}, {Yang}  \& {Gwyn}}{{Moutard} et~al.}{2020}]{moutard20}
{Moutard} T.,  {Sawicki} M.,  {Arnouts} S.,  {Golob} A.,  {Coupon} J.,
  {Ilbert} O.,  {Yang} X.,   {Gwyn} S.,  2020, \mn@doi [\mnras]
  {10.1093/mnras/staa706}, \href
  {https://ui.adsabs.harvard.edu/abs/2020MNRAS.494.1894M} {494, 1894}

\bibitem[\protect\citeauthoryear{{Naidu}, {Tacchella}, {Mason}, {Bose}, {Oesch}
   \& {Conroy}}{{Naidu} et~al.}{2020}]{naidu20}
{Naidu} R.~P.,  {Tacchella} S.,  {Mason} C.~A.,  {Bose} S.,  {Oesch} P.~A.,
  {Conroy} C.,  2020, \mn@doi [\apj] {10.3847/1538-4357/ab7cc9}, \href
  {https://ui.adsabs.harvard.edu/abs/2020ApJ...892..109N} {892, 109}

\bibitem[\protect\citeauthoryear{{Nakajima}, {Ellis}, {Robertson}, {Tang}  \&
  {Stark}}{{Nakajima} et~al.}{2020}]{nakajima20}
{Nakajima} K.,  {Ellis} R.~S.,  {Robertson} B.~E.,  {Tang} M.,   {Stark} D.~P.,
   2020, \mn@doi [\apj] {10.3847/1538-4357/ab6604}, \href
  {https://ui.adsabs.harvard.edu/abs/2020ApJ...889..161N} {889, 161}

\bibitem[\protect\citeauthoryear{{Nestor}, {Shapley}, {Steidel}  \&
  {Siana}}{{Nestor} et~al.}{2011}]{nestor11}
{Nestor} D.~B.,  {Shapley} A.~E.,  {Steidel} C.~C.,   {Siana} B.,  2011,
  \mn@doi [\apj] {10.1088/0004-637X/736/1/18}, \href
  {http://adsabs.harvard.edu/abs/2011ApJ...736...18N} {736, 18}

\bibitem[\protect\citeauthoryear{Oliphant}{Oliphant}{2006}]{numpy}
Oliphant T.,  2006, Guide to NumPy

\bibitem[\protect\citeauthoryear{{Osterbrock}}{{Osterbrock}}{1989}]{osterbrock89}
{Osterbrock} D.~E.,  1989, {Astrophysics of gaseous nebulae and active galactic
  nuclei}

\bibitem[\protect\citeauthoryear{{Ouchi} et~al.,}{{Ouchi}
  et~al.}{2009}]{ouchi09}
{Ouchi} M.,  et~al., 2009, \mn@doi [\apj] {10.1088/0004-637X/706/2/1136}, \href
  {http://adsabs.harvard.edu/abs/2009ApJ...706.1136O} {706, 1136}

\bibitem[\protect\citeauthoryear{{Paardekooper}, {Khochfar}  \& {Dalla
  Vecchia}}{{Paardekooper} et~al.}{2015}]{paardekooper15}
{Paardekooper} J.-P.,  {Khochfar} S.,   {Dalla Vecchia} C.,  2015, \mn@doi
  [\mnras] {10.1093/mnras/stv1114}, \href
  {http://adsabs.harvard.edu/abs/2015MNRAS.451.2544P} {451, 2544}

\bibitem[\protect\citeauthoryear{{Parsa}, {Dunlop}  \& {McLure}}{{Parsa}
  et~al.}{2018}]{parsa18}
{Parsa} S.,  {Dunlop} J.~S.,   {McLure} R.~J.,  2018, \mn@doi [\mnras]
  {10.1093/mnras/stx2887}, \href
  {https://ui.adsabs.harvard.edu/abs/2018MNRAS.474.2904P} {474, 2904}

\bibitem[\protect\citeauthoryear{{Pentericci} et~al.,}{{Pentericci}
  et~al.}{2018}]{pentericci18}
{Pentericci} L.,  et~al., 2018, \mn@doi [\aap] {10.1051/0004-6361/201833047},
  \href {https://ui.adsabs.harvard.edu/abs/2018A&A...616A.174P} {616, A174}

\bibitem[\protect\citeauthoryear{{Planck Collaboration} et~al.,}{{Planck
  Collaboration} et~al.}{2016}]{planck16}
{Planck Collaboration} et~al., 2016, \mn@doi [\aap]
  {10.1051/0004-6361/201628890}, \href
  {https://ui.adsabs.harvard.edu/abs/2016A&A...596A.107P} {596, A107}

\bibitem[\protect\citeauthoryear{{Reddy}, {Pettini}, {Steidel}, {Shapley},
  {Erb}  \& {Law}}{{Reddy} et~al.}{2012}]{reddy12}
{Reddy} N.~A.,  {Pettini} M.,  {Steidel} C.~C.,  {Shapley} A.~E.,  {Erb} D.~K.,
    {Law} D.~R.,  2012, \mn@doi [\apj] {10.1088/0004-637X/754/1/25}, \href
  {https://ui.adsabs.harvard.edu/abs/2012ApJ...754...25R} {754, 25}

\bibitem[\protect\citeauthoryear{{Reddy}, {Steidel}, {Pettini},
  {Bogosavljevi{\'c}}  \& {Shapley}}{{Reddy} et~al.}{2016}]{reddy16}
{Reddy} N.~A.,  {Steidel} C.~C.,  {Pettini} M.,  {Bogosavljevi{\'c}} M.,
  {Shapley} A.~E.,  2016, \mn@doi [\apj] {10.3847/0004-637X/828/2/108}, \href
  {http://adsabs.harvard.edu/abs/2016ApJ...828..108R} {828, 108}

\bibitem[\protect\citeauthoryear{{Rivera-Thorsen} et~al.,}{{Rivera-Thorsen}
  et~al.}{2019}]{rivera-thorsen19}
{Rivera-Thorsen} T.~E.,  et~al., 2019, \mn@doi [Science]
  {10.1126/science.aaw0978}, \href
  {https://ui.adsabs.harvard.edu/abs/2019Sci...366..738R} {366, 738}

\bibitem[\protect\citeauthoryear{{Robertson}, {Ellis}, {Furlanetto}  \&
  {Dunlop}}{{Robertson} et~al.}{2015}]{robertson15}
{Robertson} B.~E.,  {Ellis} R.~S.,  {Furlanetto} S.~R.,   {Dunlop} J.~S.,
  2015, \mn@doi [\apjl] {10.1088/2041-8205/802/2/L19}, \href
  {http://adsabs.harvard.edu/abs/2015ApJ...802L..19R} {802, L19}

\bibitem[\protect\citeauthoryear{{Rosdahl} et~al.,}{{Rosdahl}
  et~al.}{2018}]{rosdahl18}
{Rosdahl} J.,  et~al., 2018, preprint, \href
  {http://adsabs.harvard.edu/abs/2018arXiv180107259R} {} (\mn@eprint {arXiv}
  {1801.07259})

\bibitem[\protect\citeauthoryear{{Rudie}, {Steidel}, {Shapley}  \&
  {Pettini}}{{Rudie} et~al.}{2013}]{rudie13}
{Rudie} G.~C.,  {Steidel} C.~C.,  {Shapley} A.~E.,   {Pettini} M.,  2013,
  \mn@doi [\apj] {10.1088/0004-637X/769/2/146}, \href
  {https://ui.adsabs.harvard.edu/abs/2013ApJ...769..146R} {769, 146}

\bibitem[\protect\citeauthoryear{{Saha} et~al.,}{{Saha} et~al.}{2020}]{saha20}
{Saha} K.,  et~al., 2020, \mn@doi [Nature Astronomy]
  {10.1038/s41550-020-1173-5}, \href
  {https://ui.adsabs.harvard.edu/abs/2020NatAs.tmp..164S} {}

\bibitem[\protect\citeauthoryear{{Sawicki} et~al.,}{{Sawicki}
  et~al.}{2019}]{sawicki19}
{Sawicki} M.,  et~al., 2019, \mn@doi [\mnras] {10.1093/mnras/stz2522}, \href
  {https://ui.adsabs.harvard.edu/abs/2019MNRAS.489.5202S} {489, 5202}

\bibitem[\protect\citeauthoryear{{Seiler}, {Hutter}, {Sinha}  \&
  {Croton}}{{Seiler} et~al.}{2018}]{seiler18}
{Seiler} J.,  {Hutter} A.,  {Sinha} M.,   {Croton} D.,  2018, \mn@doi [\mnras]
  {10.1093/mnrasl/sly122}, \href
  {https://ui.adsabs.harvard.edu/abs/2018MNRAS.480L..33S} {480, L33}

\bibitem[\protect\citeauthoryear{{Shapley}, {Steidel}, {Pettini}  \&
  {Adelberger}}{{Shapley} et~al.}{2003}]{shapley03}
{Shapley} A.~E.,  {Steidel} C.~C.,  {Pettini} M.,   {Adelberger} K.~L.,  2003,
  \mn@doi [\apj] {10.1086/373922}, \href
  {http://adsabs.harvard.edu/abs/2003ApJ...588...65S} {588, 65}

\bibitem[\protect\citeauthoryear{{Shapley}, {Steidel}, {Pettini}, {Adelberger}
  \& {Erb}}{{Shapley} et~al.}{2006}]{shapley06}
{Shapley} A.~E.,  {Steidel} C.~C.,  {Pettini} M.,  {Adelberger} K.~L.,   {Erb}
  D.~K.,  2006, \mn@doi [\apj] {10.1086/507511}, \href
  {http://adsabs.harvard.edu/abs/2006ApJ...651..688S} {651, 688}

\bibitem[\protect\citeauthoryear{{Shapley}, {Steidel}, {Strom},
  {Bogosavljevi{\'c}}, {Reddy}, {Siana}, {Mostardi}  \& {Rudie}}{{Shapley}
  et~al.}{2016}]{shapley16}
{Shapley} A.~E.,  {Steidel} C.~C.,  {Strom} A.~L.,  {Bogosavljevi{\'c}} M.,
  {Reddy} N.~A.,  {Siana} B.,  {Mostardi} R.~E.,   {Rudie} G.~C.,  2016,
  \mn@doi [\apjl] {10.3847/2041-8205/826/2/L24}, \href
  {http://adsabs.harvard.edu/abs/2016ApJ...826L..24S} {826, L24}

\bibitem[\protect\citeauthoryear{{Siana} et~al.,}{{Siana}
  et~al.}{2007}]{siana07}
{Siana} B.,  et~al., 2007, \mn@doi [\apj] {10.1086/521185}, \href
  {http://adsabs.harvard.edu/abs/2007ApJ...668...62S} {668, 62}

\bibitem[\protect\citeauthoryear{{Siana} et~al.,}{{Siana}
  et~al.}{2015}]{siana15}
{Siana} B.,  et~al., 2015, \mn@doi [\apj] {10.1088/0004-637X/804/1/17}, \href
  {https://ui.adsabs.harvard.edu/abs/2015ApJ...804...17S} {804, 17}

\bibitem[\protect\citeauthoryear{{Smith} et~al.,}{{Smith}
  et~al.}{2018}]{smith18}
{Smith} B.~M.,  et~al., 2018, \mn@doi [\apj] {10.3847/1538-4357/aaa3dc}, \href
  {https://ui.adsabs.harvard.edu/abs/2018ApJ...853..191S} {853, 191}

\bibitem[\protect\citeauthoryear{{Steidel}, {Pettini}  \&
  {Adelberger}}{{Steidel} et~al.}{2001}]{steidel01}
{Steidel} C.~C.,  {Pettini} M.,   {Adelberger} K.~L.,  2001, \mn@doi [\apj]
  {10.1086/318323}, \href {http://adsabs.harvard.edu/abs/2001ApJ...546..665S}
  {546, 665}

\bibitem[\protect\citeauthoryear{{Steidel}, {Bogosavlevic}, {Shapley}, {Reddy},
  {Rudie}, {Pettini}, {Trainor}  \& {Strom}}{{Steidel}
  et~al.}{2018}]{steidel18}
{Steidel} C.~C.,  {Bogosavlevic} M.,  {Shapley} A.~E.,  {Reddy} N.~A.,  {Rudie}
  G.~C.,  {Pettini} M.,  {Trainor} R.~F.,   {Strom} A.~L.,  2018, preprint,
  \href {http://adsabs.harvard.edu/abs/2018arXiv180506071S} {} (\mn@eprint
  {arXiv} {1805.06071})

\bibitem[\protect\citeauthoryear{{Straatman} et~al.,}{{Straatman}
  et~al.}{2016}]{straatman16}
{Straatman} C.~M.~S.,  et~al., 2016, \mn@doi [\apj]
  {10.3847/0004-637X/830/1/51}, \href
  {http://adsabs.harvard.edu/abs/2016ApJ...830...51S} {830, 51}

\bibitem[\protect\citeauthoryear{{Tepper-Garc{\'\i}a}}{{Tepper-Garc{\'\i}a}}{2006}]{tepper-garcia06}
{Tepper-Garc{\'\i}a} T.,  2006, \mn@doi [\mnras]
  {10.1111/j.1365-2966.2006.10450.x}, \href
  {https://ui.adsabs.harvard.edu/abs/2006MNRAS.369.2025T} {369, 2025}

\bibitem[\protect\citeauthoryear{{Trebitsch}, {Blaizot}, {Rosdahl}, {Devriendt}
   \& {Slyz}}{{Trebitsch} et~al.}{2017}]{trebitsch17}
{Trebitsch} M.,  {Blaizot} J.,  {Rosdahl} J.,  {Devriendt} J.,   {Slyz} A.,
  2017, \mn@doi [\mnras] {10.1093/mnras/stx1060}, \href
  {http://adsabs.harvard.edu/abs/2017MNRAS.470..224T} {470, 224}

\bibitem[\protect\citeauthoryear{{Urrutia} et~al.,}{{Urrutia}
  et~al.}{2019}]{urrutia18}
{Urrutia} T.,  et~al., 2019, \mn@doi [\aap] {10.1051/0004-6361/201834656},
  \href {https://ui.adsabs.harvard.edu/abs/2019A&A...624A.141U} {624, A141}

\bibitem[\protect\citeauthoryear{Van~Rossum \& Drake}{Van~Rossum \&
  Drake}{2009}]{python3}
Van~Rossum G.,  Drake F.~L.,  2009, Python 3 Reference Manual.
CreateSpace, Scotts Valley, CA

\bibitem[\protect\citeauthoryear{{Vanzella}, {Siana}, {Cristiani}  \&
  {Nonino}}{{Vanzella} et~al.}{2010}]{vanzella10}
{Vanzella} E.,  {Siana} B.,  {Cristiani} S.,   {Nonino} M.,  2010, \mn@doi
  [\mnras] {10.1111/j.1365-2966.2010.16408.x}, \href
  {http://adsabs.harvard.edu/abs/2010MNRAS.404.1672V} {404, 1672}

\bibitem[\protect\citeauthoryear{{Vanzella} et~al.,}{{Vanzella}
  et~al.}{2012}]{vanzella12}
{Vanzella} E.,  et~al., 2012, \mn@doi [\apj] {10.1088/0004-637X/751/1/70},
  \href {http://adsabs.harvard.edu/abs/2012ApJ...751...70V} {751, 70}

\bibitem[\protect\citeauthoryear{{Vanzella} et~al.,}{{Vanzella}
  et~al.}{2016}]{vanzella16}
{Vanzella} E.,  et~al., 2016, \mn@doi [\apj] {10.3847/0004-637X/825/1/41},
  \href {http://adsabs.harvard.edu/abs/2016ApJ...825...41V} {825, 41}

\bibitem[\protect\citeauthoryear{{Vanzella} et~al.,}{{Vanzella}
  et~al.}{2018}]{vanzella18}
{Vanzella} E.,  et~al., 2018, \mn@doi [\mnras] {10.1093/mnrasl/sly023}, \href
  {http://adsabs.harvard.edu/abs/2018MNRAS.476L..15V} {476, L15}

\bibitem[\protect\citeauthoryear{{Virtanen} et~al.,}{{Virtanen}
  et~al.}{2020}]{scipy}
{Virtanen} P.,  et~al., 2020, \mn@doi [Nature Methods]
  {https://doi.org/10.1038/s41592-019-0686-2}, \href {https://rdcu.be/b08Wh} {}

\bibitem[\protect\citeauthoryear{{Weingartner} \& {Draine}}{{Weingartner} \&
  {Draine}}{2001}]{weingartner01}
{Weingartner} J.~C.,  {Draine} B.~T.,  2001, \mn@doi [\apj] {10.1086/318651},
  \href {https://ui.adsabs.harvard.edu/abs/2001ApJ...548..296W} {548, 296}

\bibitem[\protect\citeauthoryear{{Wise} \& {Cen}}{{Wise} \&
  {Cen}}{2009}]{wise09}
{Wise} J.~H.,  {Cen} R.,  2009, \mn@doi [\apj] {10.1088/0004-637X/693/1/984},
  \href {http://adsabs.harvard.edu/abs/2009ApJ...693..984W} {693, 984}

\bibitem[\protect\citeauthoryear{{Yajima}, {Choi}  \& {Nagamine}}{{Yajima}
  et~al.}{2011}]{yajima11}
{Yajima} H.,  {Choi} J.-H.,   {Nagamine} K.,  2011, \mn@doi [\mnras]
  {10.1111/j.1365-2966.2010.17920.x}, \href
  {http://adsabs.harvard.edu/abs/2011MNRAS.412..411Y} {412, 411}

\bibitem[\protect\citeauthoryear{{Yamanaka} et~al.,}{{Yamanaka}
  et~al.}{2020}]{yamanaka20}
{Yamanaka} S.,  et~al., 2020, \mn@doi [\mnras] {10.1093/mnras/staa2507}, \href
  {https://ui.adsabs.harvard.edu/abs/2020MNRAS.498.3095Y} {498, 3095}

\bibitem[\protect\citeauthoryear{{Zackrisson}, {Inoue}  \&
  {Jensen}}{{Zackrisson} et~al.}{2013}]{zackrisson13}
{Zackrisson} E.,  {Inoue} A.~K.,   {Jensen} H.,  2013, \mn@doi [\apj]
  {10.1088/0004-637X/777/1/39}, \href
  {http://adsabs.harvard.edu/abs/2013ApJ...777...39Z} {777, 39}

\makeatother
\end{thebibliography}

%%%%%%%%%%%%%%%%%%%%%%%%%%%%%%%%%%%%%%%%%%%%%%%%%%

%%%%%%%%%%%%%%%%% APPENDICES %%%%%%%%%%%%%%%%%%%%%

%\appendix

%%%%%%%%%%%%%%%%%%%%%%%%%%%%%%%%%%%%%%%%%%%%%%%%%%

% Don't change these lines
\bsp	% typesetting comment
\label{lastpage}
\end{document}